	\newcommand{\manuscripttitle}{%
	Flux-driven integrated modelling of main ion pressure and trace tungsten transport in ASDEX Upgrade
	}

	\newcommand{\manuscriptauthors}{%
	O. Linder$^{1,2,3}$, 
	J. Citrin$^1$, 
	G.M.D. Hogeweij$^1$, 
	C. Angioni$^3$,
	C. Bourdelle$^4$,
	F.J. Casson$^5$,
	E. Fable$^3$, 
	A. Ho$^1$,
	F. Koechl$^6$,
	M. Sertoli$^3$, 
	the EUROfusion MST1 Team\footnote{See author list of H. Meyer et al. \href{https://doi.org/10.1088/1741-4326/aa6084}{\textit{Nucl. Fusion}} \textbf{57}, 102014 (2017)}
	\hspace*{2pt}and the ASDEX Upgrade Team\footnote{See author list of A. Kallenbach et al. \href{https://doi.org/10.1088/1741-4326/aa64f6}{\textit{Nucl. Fusion}} \textbf{57}, 102015 (2017)}
	}

	\newcommand{\manuscriptinstitutes}{%
	$^1$DIFFER - Dutch Institute for Fundamental Energy Research, De Zaale 20, 5612 AJ Eindhoven, the Netherlands\\
	$^2$Eindhoven University of Technology, De Zaale, 5612 AJ Eindhoven, the Netherlands\\
	$^3$Max-Planck-Institut f\"ur Plasmaphysik, Boltzmannstra\ss e 2, D-85748 Garching, Germany\\	
	$^4$CEA, IRFM, F-13108 Saint-Paul-lez-Durance, France\\
	$^5$CCFE, Culham Science Centre, Abingdon, Oxon, OX14 3DB, UK\\
	$^6$\"OAW/ATI, Atominstitut, TU Wien, 1020 Vienna, Austria
	}

	\newcommand{\manuscriptabstract}{%
	Neoclassical and turbulent heavy impurity transport in tokamak core plasmas are determined by main ion temperature, density and toroidal rotation profiles. Thus, in order to understand and prevent experimental behaviour of W accumulation, flux-driven integrated modelling of main ion heat and particle transport over multiple confinement times is a vital prerequisite. For the first time, the quasilinear gyrokinetic code QuaLiKiz has been applied for successful predictions of core kinetic profiles in an ASDEX Upgrade H-mode discharge in the turbulence dominated region within the integrated modelling suite JETTO. Neoclassical contributions are calculated by NCLASS; auxiliary heat and particle deposition profiles due to NBI and ECRH prescribed from previous analysis with TRANSP. Turbulent and neoclassical contributions are insufficient in explaining main ion heat and particle transport inside the $q=1$ surface, necessitating the prescription of further transport coefficients to mimic the impact of MHD activity on central transport. The ion to electron temperature ratio at the simulation boundary at $\rho_\mathrm{tor} = 0.85$ stabilizes ion scale modes while destabilizing ETG modes when significantly exceeding unity. Careful analysis of experimental measurements using Gaussian process regression techniques is carried out to explore reasonable uncertainties. In following trace W impurity transport simulations performed with additionally NEO, neoclassical transport under consideration of poloidal asymmetries alone is found to be insufficient to establish hollow central W density profiles. Reproduction of these conditions measured experimentally is found possible only when assuming the direct impact of a saturated $(m,n)=(1,1)$ MHD mode on heavy impurity transport.
	}

	\documentclass[10pt,a4paper,fleqn,twocolumn]{scrartcl}
	
	\usepackage[utf8]{inputenc}
	\usepackage[T1]{fontenc}
	\usepackage{graphicx}
	\usepackage{color}

	\usepackage{tgtermes}
	\usepackage[left=2cm,right=2cm,top=2.5cm,bottom=2cm]{geometry}
	\usepackage[nouppercase]{scrpage2}
		\cofoot{}
		\rohead{\pagemark}
		\lohead{Flux-driven integrated modelling of main ion pressure and trace tungsten transport in ASDEX Upgrade}
		\setheadsepline{0.2pt}
		
	\usepackage{setspace}
	\usepackage{titlesec, blindtext}
		\titleformat*{\section}{\large\bfseries}
		\titleformat*{\subsection}{\normalsize\itshape}
		\titleformat*{\subsubsection}{\normalsize\itshape}

	\usepackage[pdftex,
		pdffitwindow	=true,
		pdfsubject 		={High-temperature plasma physics},
		pdfcreator 		={O. Linder},
		pdfproducer 	={O. Linder},
		pdftitle		={Flux-driven integrated modelling of main ion pressure and trace tungsten transport in ASDEX Upgrade},
		pdfauthor		={O. Linder, J. Citrin, G.M.D. Hogeweij, C. Angioni, C. Bourdelle, F.J. Casson, E. Fable, A. Ho, F. Koechl, M. Sertoli, the EUROfusion MST1 Team, the ASDEX Upgrade Team},
		pdfkeywords		={tokamaks, impurities, modelling, transport, turbulence, MHD},
		]{hyperref}
	\usepackage{dblfloatfix}

	\usepackage[format=plain,labelfont=bf]{caption}
		
	\usepackage{booktabs}
	\usepackage{colortbl}
	
	\usepackage{amssymb}
	
	\usepackage{mathptmx}
	
	\usepackage{empheq}
	\usepackage{mathtools}
	\usepackage[integrals]{wasysym}
	
	\usepackage[nottoc]{tocbibind}
	\let\OLDthebibliography\thebibliography
	\renewcommand\thebibliography[1]{
  		\OLDthebibliography{#1}
  		\setlength{\parskip}{0pt}
  		\setlength{\itemsep}{0pt}
		}
	\usepackage[superscript,biblabel]{cite}
	
	\usepackage{dcolumn}

	\usepackage{bm}

	\usepackage{soul}
	
	\newcommand{\Fst}{1$^{\text{st}}$ }	
	\newcommand{\Snd}{2$^{\text{nd}}$ }	
	\newcommand{\TiTe}{$T_\mathrm{i}/T_\mathrm{e}|_\mathrm{bc}$ }
	\newcommand{\TiTeE}{$T_\mathrm{i}/T_\mathrm{e}|_\mathrm{bc}$}
	\newcommand{\stdcap}[2]{%
		\centering
		\captionsetup{width=1.0\linewidth,font=small}
		\caption[\normalsize{#1}]{#2}
		}
		
	\definecolor{Cmap100}{rgb}{0.6470588235294118, 0.0, 0.14901960784313725}
	\definecolor{Cmap95}{rgb}{0.73933102652825844, 0.088581314878892731, 0.15086505190311419}
	\definecolor{Cmap83}{rgb}{0.91672433679354093, 0.34302191464821219, 0.22399077277970011}
	\definecolor{Cmap75}{rgb}{0.97347174163783168, 0.54740484429065761, 0.31810841983852373}
	\definecolor{Cmap67}{rgb}{0.99346405228758172, 0.74771241830065382, 0.44183006535947733}
	\definecolor{Cmap50}{rgb}{0.99992310649750094, 0.99761630142252977, 0.74540561322568244}
	\definecolor{Cmap33}{rgb}{0.73986928104575156, 0.88496732026143787, 0.93333333333333324}
	\definecolor{Cmap25}{rgb}{0.56485966935793919, 0.76639753940791999, 0.86758938869665503}
	\definecolor{Cmap17}{rgb}{0.38985005767012687, 0.60092272202998842, 0.77946943483275666}
	\definecolor{Cmap0}{rgb}{0.19215686274509805, 0.21176470588235294, 0.58431372549019611}
	
	\title{%
		\vspace*{4.5mm}\LARGE{\textnormal{\textbf{%
			\manuscripttitle
		}}}}

	\author{%
		\hspace*{-2mm}\normalsize{\parbox{\linewidth}{%
		\begin{flushright}
		\parbox{0.85\linewidth}{%
			\large
			\manuscriptauthors
			}
		\end{flushright}
		\begin{flushright}
		\parbox{0.85\linewidth}{\small{
			\manuscriptinstitutes
			}}
		\end{flushright}
		}}}
	
	\date{%
		\hspace*{0mm}\vspace*{-5mm}
		\begin{flushright}\parbox{0.85\linewidth}{\normalsize{%
			\textbf{Abstract.}\quad %
			\manuscriptabstract
			}}
		\end{flushright}
		}
\begin{document}
	\onecolumn
	\begingroup\let\center\flushleft\let\endcenter\endflushleft
	\maketitle\endgroup\thispagestyle{empty}

	\vspace*{5cm}
	\begin{center}	
		\parbox{.6\linewidth}{%
		\begin{tabular}{ll}
			\parbox{2cm}{\includegraphics[scale=.64]{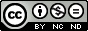}}
			&
			\parbox{.9\linewidth}{%
			\textbf{This work is licensed under a \href{http://creativecommons.org/licenses/by-nc-nd/3.0/}{Creative Commons Attribution-NonCommercial-NoDerivs 3.0 Unported License}.}}
		\end{tabular}
		}
	\end{center}
	
	\clearpage
	\twocolumn
	
\section{Introduction}
In a future commercial fusion reactor, key requirements for the materials used in plasma-facing components are low tritium retention, as well as low erosion under high heat and particle fluxes. Tungsten (W) has been identified as a promising candidate\cite{Bolt04, Causey02, Philipps11} and is therefore presently used as plasma facing material in various devices such as in the ASDEX Upgrade (AUG) tokamak\cite{Neu07}, in the JET ITER-like wall\cite{Matthews10} and in the ITER-like divertor in WEST\cite{Bucalossi14}. Yet, strong line radiation of non-fully ionized tungsten at fusion relevant temperatures \cite{Post77} can significantly cool the central plasma, deteriorating fusion performance. Consequently, central W accumulation has to be avoided to keep concentrations in the core plasma below $10^{-5}-10^{-4}$ \cite{Neu97,Philipps11}. In AUG, central wave heating is applied regularly for W impurity density control \cite{Neu02, Dux03, Putterich13}. Yet, complete understanding of all W transport mechanisms involved is still an outstanding issue. 

In present day devices, such as JET and AUG, heavy impurity transport is dominated by neoclassical transport in the inner half radius \cite{Dux03, Putterich13, Angioni14, Casson15, Angioni15, Angioni17_2}. Density peaking of the main ion species drives inward convection, whereas neoclassical temperature screening due to gradients of the main ion temperature gives rise to outward transport. However, poloidal asymmetries in the heavy impurity density distribution can enhance neoclassical transport of these species by up to an order of magnitude\cite{Angioni14,Casson15}. The large mass of high $Z$ impurities causes localization on the low field side in rotating plasmas due to centrifugal forces, whereas the high charge leads to sensitivity of these impurities to variations in the background electrostatic potential, for example due to minority heating by ion cyclotron resonance heating (ICRH)\cite{Reinke12}. Since the impact of poloidal asymmetries on neoclassical heavy impurity transport is strongly dependent on collisionality and plasma gradients \cite{Casson15}, both central W accumulation \cite{Angioni14} and enhanced outward W transport \cite{Casson15} can be observed due to poloidal asymmetries under different conditions.

In AUG discharges heated by electron cyclotron resonance heating (ECRH) and neutral beam injection (NBI), hollow W density profiles are regularly measured in the presence of saturated $(m,n) = (1,1)$ magnetohydrodynamic (MHD) modes\cite{Angioni17, Sertoli17}. Neoclassical and turbulent transport are insufficient as explanation. Instead, good temporal correlation between changes in W transport and saturated MHD activity is observed.\cite{Putterich13, Sertoli15, Sertoli15_2} However, the exact impact of MHD activity on heavy impurity transport is yet to be explained, giving rise to the possibility of directly or indirectly MHD driven transport\cite{Gunter99, Nave03, Hender16, Goniche17}, the latter through e.g. geometrical effects or modifications of neoclassical transport. Similarly, heavy impurity transport is enhanced by central wave heating \cite{Dux03, Angioni17}. Application of ECRH can lead to a suppression of the neoclassical pinch\cite{Sertoli11, Sertoli15}, possibly in parts due to density profile flattening close to the magnetic axis\cite{Angioni14}. Furthermore, greatly increased anomalous diffusive transport is observed with on-axis ECRH\cite{Dux03,Sertoli11}. The emergence of a saturated $(1,1)$ MHD mode may have a similar impact, yet the exact mechanism responsible for mitigation of central W accumulation under these conditions is not understood completely.

Since neoclassical contributions depend strongly on the main ion density and temperature profiles\cite{Angioni15, Casson15}, accurate modelling of main ion transport is a vital prerequisite to ultimately simulate trace W impurity transport. Towards this goal, the fast quasilinear gyrokinetic code QuaLiKiz \cite{Bourdelle16, Citrin17}\footnote{QuaLiKiz is open source. See \href{http://www.qualikiz.com}{qualikiz.com} for details.} is coupled to the 1.5-dimensional transport code JETTO \cite{Cenacchi88, Romanelli14} and used for the first time for integrated modelling of an AUG discharge. Similar work has recently been carried out on JET, where fully self-consistent simulations by NEO-QuaLiKiz could reproduce W accumulation observed experimentally.\cite{Breton18} QuaLiKiz calculates turbulent heat, particle and momentum fluxes driven by ion temperature gradient (ITG), trapped electron (TEM) and electron temperature gradient (ETG) modes. The computed quasilinear fluxes have been validated against nonlinear simulations (see Ref. \citen{Bourdelle16} and references therein) and tested for predicting temperatures, densities and toroidal velocities in H mode pulses \cite{Citrin17, Breton17}. Thanks to recent numerical improvements \cite{Citrin17}, QuaLiKiz can now be used routinely for time evolving predictions, modelling 1\,s of plasma evolution parallelised in $\sim\,100\,$CPUh on multiple cores.

In this work, W impurity transport simulations are performed for AUG H-mode discharge \#31115 with primary NBI heating, where a saturated $(1,1)$ MHD mode is observed in the presence of central, localized ECRH. As a prerequisite, predictive heat and particle simulations in the presence of C impurities are performed in the plasma core to validate the main ion transport mechanisms calculated against experimentally obtained temperature and density profiles for AUG discharge \#31115. Since turbulent and neoclassical contributions to central transport inside the $q=1$ surface are insufficient to describe experimental fluxes, the difference observed is prescribed additionally and attributed to MHD driven transport. Applying the steady-state profiles calculated in the presence or absence of central MHD transport, W impurity transport simulations are carried out to assess the importance of neoclassical and MHD driven transport in achieving a hollow W profile in AUG discharge \#31115, as typically observed under application of central ECRH. Tungsten is treated in the trace limit for simplicity, thus assuming no impact on main ion profiles in the simulations performed. Still, impurity radiation from experimental measurements is taken into account evolving the latter profiles.

The rest of this paper is organized as follows. AUG discharge \#31115 is described in Sec.\,\ref{sec:2_Description_of_AUG_31115}. The setup of the predictive heat and particle transport simulations performed is summarized in Sec.\,\ref{sec:3_Setup_of_transport_simulations}, the results are presented in Sec.\,\ref{sec:4_Predictive_transport_simulations}. The main sensitivities, e.g. to the boundary conditions of the density and temperature profiles calculated, are discussed in Sec.\,\ref{sec:5_Simulation_sensitivity}. Simulations of trace W impurity transport are shown in Sec.\,\ref{sec:6_Predictive_W_transport_simulations}. A conclusion and an outlook are given in Sec.\,\ref{sec:7_Conclusion_and_outlook}. An overview of the JETTO-QuaLiKiz versions used for the simulations is provided in Appendix\,\ref{sec:B_Overview_of_JETTO-QuaLiKiz_simulations}.

\newpage
\section{Description of ASDEX Upgrade discharge \#31115}
\label{sec:2_Description_of_AUG_31115}
For the first time, integrated modelling of an ASDEX Upgrade plasma with the quasilinear gyrokinetic code QuaLiKiz is carried out. Discharge \#31115 is chosen, since this particular shot is part of a dedicated set of experiments on W transport in the presence of MHD instabilities (see Ref.~\citen{Sertoli15_2} for discharge \#31114, part of this experiment). These discharges aimed at better understanding why central W accumulation can usually be avoided by small amounts of ECRH deposited inside the $q=1$ surface\cite{Neu02,Dux03}and why this typically occurs in the presence of long lasting $(m,n) = (1,1)$ MHD activity\cite{Dux99,Gunter99,Nave03,Stober07}. For this purpose, typical AUG lower-single-null H-mode discharges were utilized\cite{Sertoli15_2}, with the plasma shape slightly modified to enhance measurements of impurity transport and core MHD. Additionally, small amounts of co-ECCD have been applied to ensure long sawtooth cycles throughout the discharge. Following this approach, W accumulation in the plasma centre is avoided, whereas in typical NBI-only heated AUG discharges, W accumulation is observed.

\begin{figure}[b!]
	\centering
	\includegraphics[width=\linewidth]{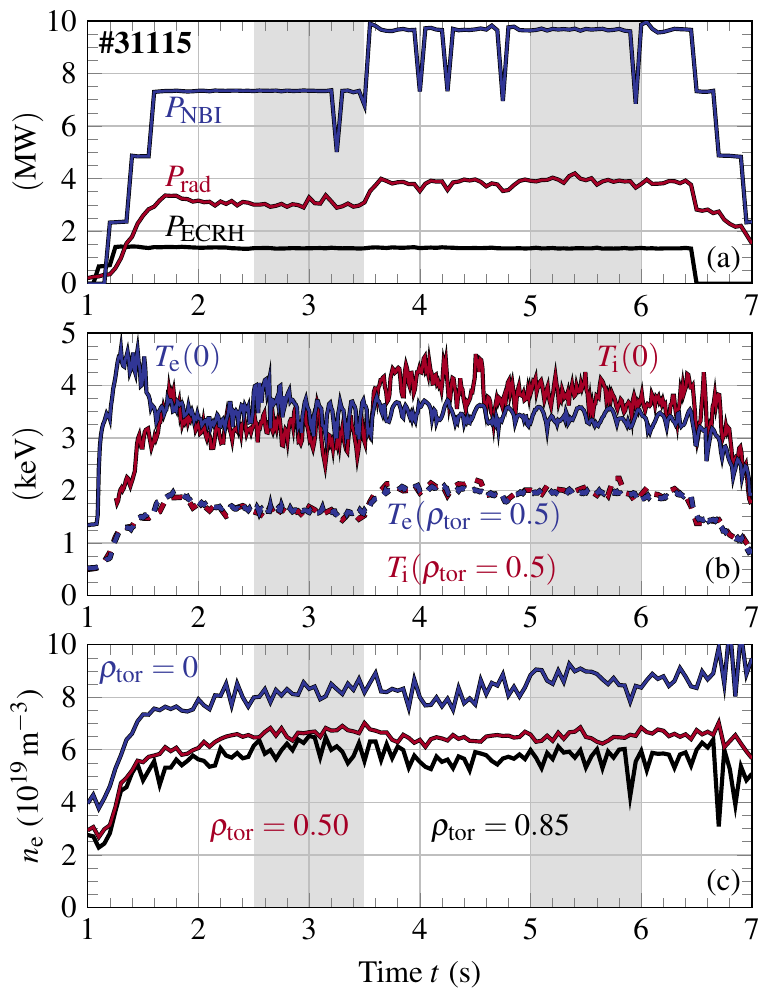}
	\stdcap{}{\label{fig:1_Time_traces_of_AUG_31115}%
		Time traces of (a) the NBI, ECRH and radiated powers, (b) the species' temperatures $T_\mathrm{s}$ on axis and at mid-radius, and (c) the electron density $n_\mathrm{e}$ on axis, at mid-radius and on top of the pedestal for AUG discharge \#31115. Fluctuations of central species' temperatures are due to MHD activity. Shaded regions indicate the time slices used for predictive heat and particle transport simulations with JETTO-QuaLiKiz in this work.
		}
\end{figure}

In discharge \#31115 chosen for this study, H-mode confinement is achieved during the current flat top phase of 1\,MA with an applied magnetic field of 2.5\,T ($q_{95} = 4.0$). The plasma is heated primarily by NBI with a base power of 7.3\,MW (see time traces in Fig.\,\ref{fig:1_Time_traces_of_AUG_31115}(a)), delivered by two 93\,kV beams and one 60\,kV beam of equal power. An additional 60\,kV beam is employed after around 2\,s of H-mode at $t = 3.5\,$s, increasing the total NBI power to 9.7\,MW. Simultaneously, the flow of NBI injected deuterium neutrals is increased from $8.5\times 10^{20}\,\mathrm{s}^{-1}$ to $12.0\times 10^{20}\,\mathrm{s}^{-1}$. Further heating of constant 1.4\,MW is provided by electron cyclotron resonance heating (ECRH)\cite{Wagner11} throughout the duration of the discharge with a narrow deposition profile close to the magnetic axis.

\begin{figure}[b!]
	\centering
	\includegraphics[width=\linewidth]{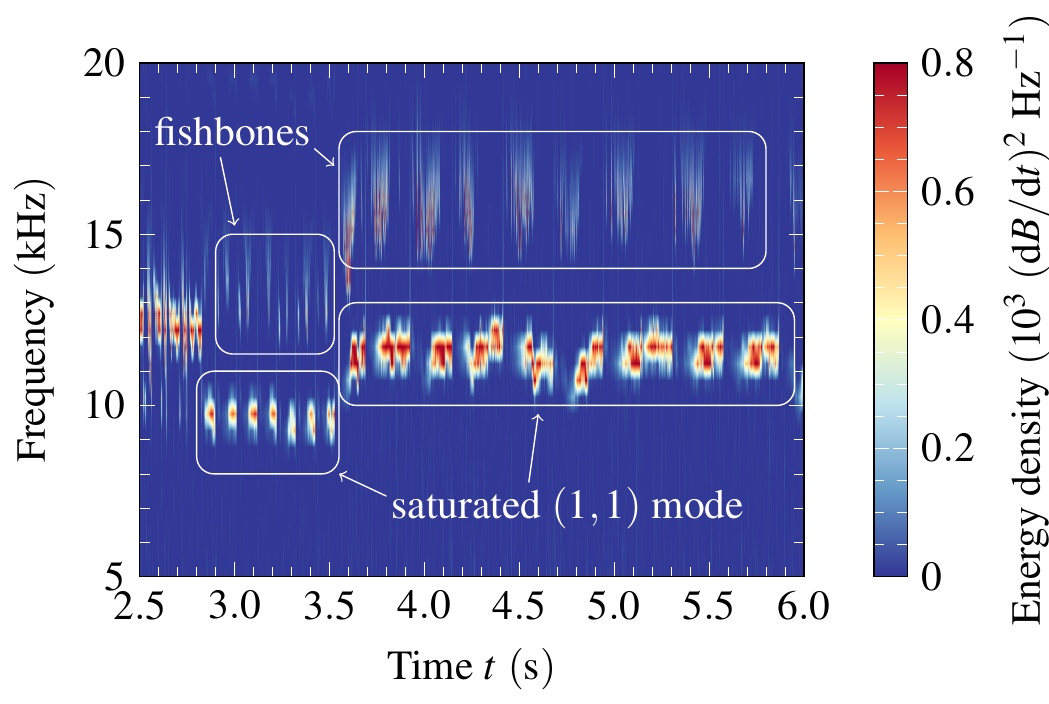}
	\stdcap{}{\label{fig:2_Spectrogram_of_MHD_modes}%
		Spectrogram of MHD modes detected by magnetic measurements throughout AUG discharge \#31115. A strong, saturated $(1,1)$ mode, as well as fishbones are periodically observed. Note that the plasma rotation velocity is increased as a fourth NBI source is coupled to the plasma from $t=3.5~$s onwards.
		}
\end{figure}

\subsection{MHD activity}
\begin{figure*}
	\centering
	\includegraphics[width=0.75\linewidth]{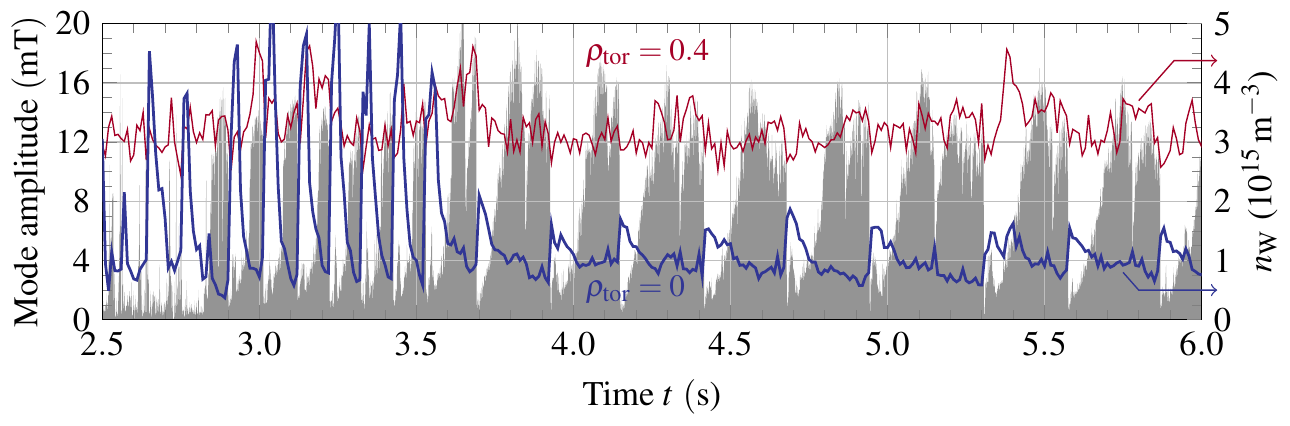}
	\stdcap{}{\label{fig:3_Amplitude_of_MHD_mode}%
		Amplitude of the saturated $(1,1)$ MHD mode (filled grey, left ordinate) and the W density $n_\mathrm{W}$ on-axis and at $\rho_\text{tor} = 0.4$ (coloured, right ordinate). Crashes of the mode amplitude are accompanied by a sudden increase of the central W density. 
		}
\end{figure*}

Throughout the duration of AUG discharge \#31115, a saturated $(m,n) = (1,1)$ MHD mode with a frequency of 10$-$11~kHz is present (see Fig.~\ref{fig:2_Spectrogram_of_MHD_modes}), as captured by magnetic measurement and soft X-ray spectroscopy (SXR). Sawtoothing is periodically observed, however the mode is strongly saturated prior to sawtooth crashes at an amplitude consistently measured throughout the discharge (see Fig.~\ref{fig:3_Amplitude_of_MHD_mode}). Additionally, fishbones with increased frequency in the range of 12$-$18~kHz arise in the course of a sawtooth cycle during growth of the $(1,1)$ mode but fade away as the latter reaches saturation.

With the increase of injected NBI power at $t=3.5~$s, the sawtooth period is extended noticeably from around 0.10~s to around 0.25~s. Simultaneously, strong mode saturation is sustained over a longer fraction of the sawtooth cycle, covering the last 60\% of the cycle instead of the last third (see Fig.~\ref{fig:3_Amplitude_of_MHD_mode}). However as mode amplitude and inversion radii are similar in both phases of the discharge, significant differences in the size of the saturated island are not expected. Through the remainder of this publication, the saturated $(1,1)$ MHD mode in either phase of the discharge will be referred to as low saturation fraction and high saturation fraction mode.

\newpage

\subsection{W density response in different phases of MHD activity}
\label{sec:2.2_W_density_response}
In the discharge analysed, W density profiles are derived from soft X-ray (SXR) Abel-inversion\cite{Sertoli15_2} and by a grazing incidence spectrometer (GIW)\cite{Sertoli15_2}. In the case of SXR, the local impurity density is obtained from Abel-inversion of the SXR emissivity from multiple line-of-sights, considering photons with energies exceeding 1\,keV. Using the GIW diagnostic, the W density is calculated from measurements of light with wavelengths around 5\,nm from two different groups of W ionization stages utilizing a single line-of-sight. Considering the fractional abundance profiles of the ionization stages involved, as well as electron temperature and density profiles, the local W density in combination of shape and average position of each of the two emissivity measurements can be determined. An in depth description is given in Ref.\,\citen{Sertoli15_2}, whereas the steps taken to obtain W density profiles from raw measurements for this discharge are discussed in Ref.\,\citen{Sertoli15}.

\begin{figure}[bt!]
	\centering
	\includegraphics[width=\linewidth]{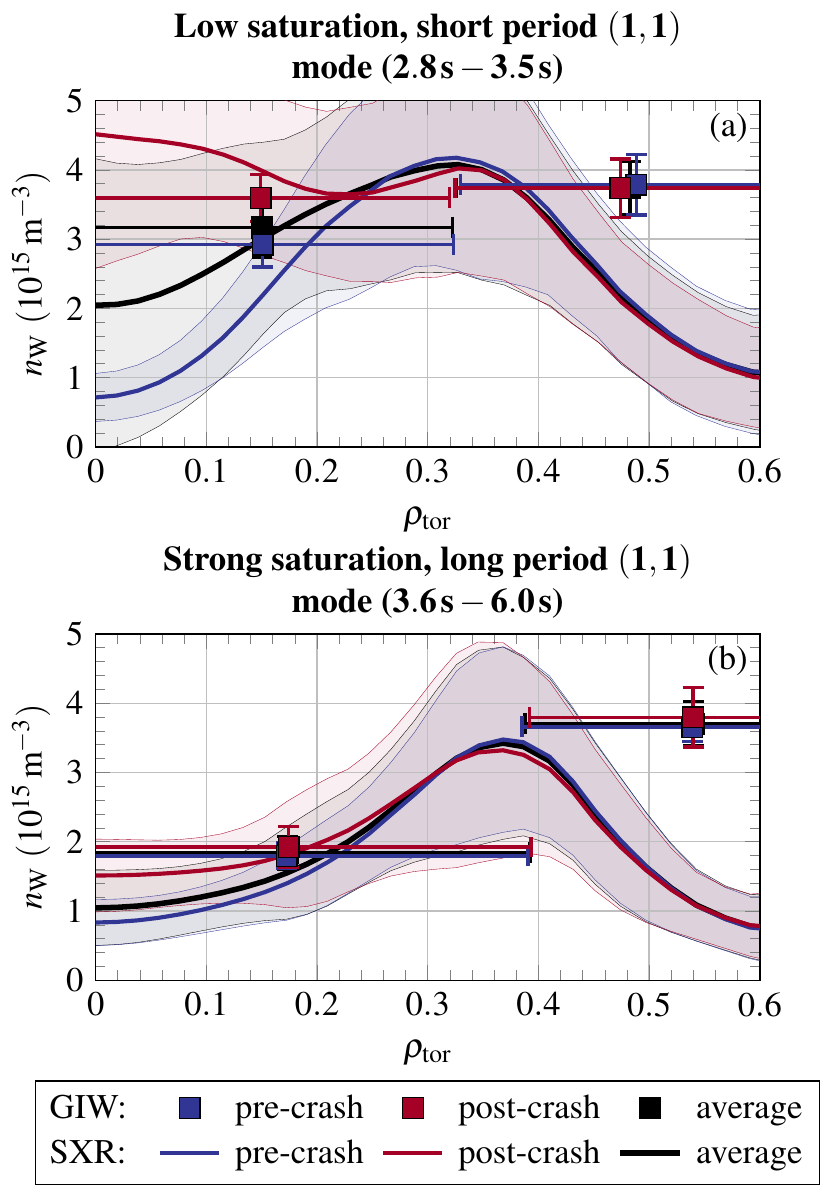}
	\stdcap{}{\label{fig:4_W_density}%
		Time-averaged W density $n_\mathrm{W}$ in the presence of a $(m,n) = (1,1)$ MHD mode of (a) short period for $t \in \left[ 2.8\,\mathrm{s}, 3.5\,\mathrm{s}\right]$ and of (b) long period for $t \in \left[ 3.6\,\mathrm{s}, 6.0\,\mathrm{s}\right]$, as measured by a grazing incidence spectrometer (GIW) and soft X-ray Abel-inversion (SXR). For each diagnostic, averages are taken prior to mode crashing (blue), after mode crashing (red), as well as over the entire cycle (black).
		}
\end{figure}

In both phases of different saturation fractions of the $(1,1)$ MHD mode, hollow W profiles, with average on-axis densities of $0.7$ and $0.8\times 10^{15}~\text{m}^{-3}$ ($n_\text{W}/n_\text{e} \sim 10^{-5}$), are observed (see Figs.\,\ref{fig:3_Amplitude_of_MHD_mode}, \ref{fig:4_W_density}), as is typically the case for ASDEX Upgrade discharges heated by central ECRH \cite{Angioni17, Sertoli17}. Following a sawtooth crash, the central W density is increased, thus exhibiting a response inverted with respect to a typical crash. However, the increase of central W density post-crash is in agreement with expected sawteething behaviour for a different initial condition than usually observed, in this case a hollow profile pre-crash, mixing the W content up to the $q=1$-surface, located at $\rho_{\mathrm{tor}} \simeq 0.4$. Flattening of hollow W profiles due to sawteething has been reported in Ref.\,\citen{Sertoli15} for the preceding discharge \#31114 and in Refs.\,\citen{Angioni17,Gude10} for similar discharges.

\begin{figure*}
	\centering
	\includegraphics[width=\linewidth]{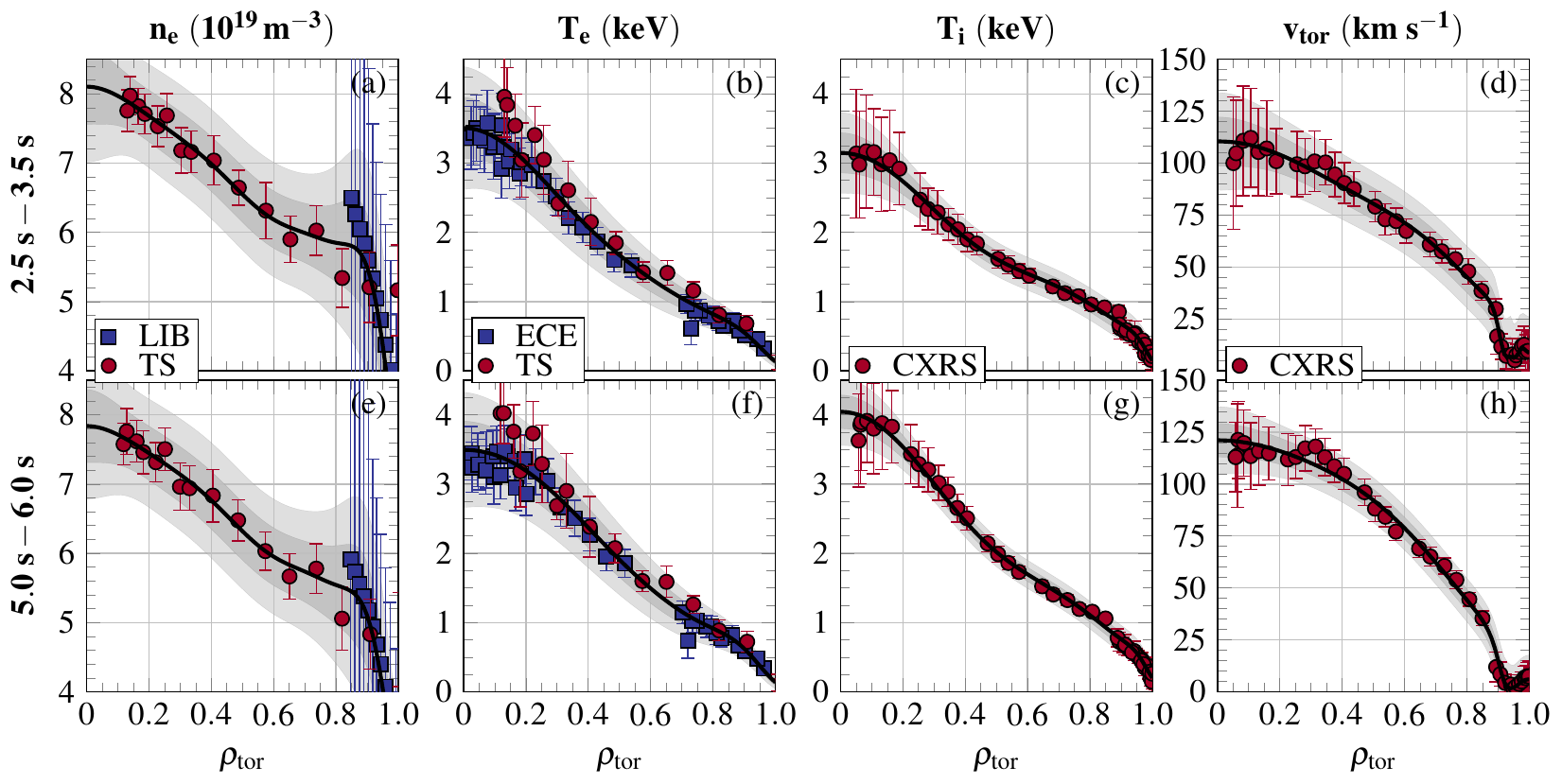}
	\stdcap{}{\label{fig:5_Application_of_GPR}%
		Application of Gaussian process regression techniques to experimental data averaged in the two time slices of interest, being $t_1 \in \left[ 2.5~\text{s}, 3.5~\text{s}\right]$ (top row) and $t_2 \in \left[ 5.0~\text{s}, 6.0~\text{s}\right]$ (bottom row), provides averaged plasma profiles (black) and associated uncertainties of $1~\sigma$ and $2~\sigma$ (filled grey). Electron density profiles (a,e) are obtained from Thomson scattering spectroscopy (TS) and lithium beam spectroscopy (LIB), electron temperature profiles (b,f) from radiometry of electron cyclotron emission (ECE) and TS. Ion temperature (c,g) and toroidal rotation velocity profiles (d,h) are both obtained from charge exchange recombination spectroscopy (CXRS).
		}
\end{figure*}

The response of the W density to a sawtooth crash is qualitatively different in each phase, increasing on-axis to on average $4.5$ and $1.5\times 10^{15}\,\mathrm{m}^{-3}$ in either phase following a crash. In the process, the W density profile in the low saturation fraction case becomes slightly peaked for about 20~ms, followed by an exponential decay on similar temporal scales, thus re-establishing a hollow W profile. In the \Snd phase, the deeply hollow W profile is preserved throughout a sawtooth cycle as the central W density increases only slightly. Additionally, the W content is noticeably reduced throughout the entire phase. The maximum averaged pre-crash W density of $4.2\times 10^{15}~\text{m}^{-3}$ at $\rho_\text{tor} = 0.33$ is both reduced down to $3.5\times 10^{15}~\text{m}^{-3}$ and shifted outwards to $\rho_\text{tor} = 0.37$ in the \Snd phase. Noticeably, W density profiles in both phases of the discharge agree within errorbars outside $\rho_\text{tor} \simeq 0.4$. Throughout the discharge, the $q=1$ surface is calculated to be located similarly at $\rho_\text{tor} \simeq 0.4$, supporting the hypothesis that 	in the present case the saturated $(1,1)$ mode facilitates outward W transport. This hypothesis is strengthened by analysis of the neoclassical contributions to W transport in the presence of poloidal asymmetries\cite{Casson15,Angioni15,Breton18_2}. Assuming kinetic profiles derived from experimental measurements (cf. Sec.~\ref{sec:2.3_Average_plasma_profile_reconstruction}), net inward neoclassical W transport is expected, as inward convective transport is likely to dominate over the effects of temperature screening, the latter being reduced due to poloidal asymmetries.

\subsection{Average plasma profile reconstruction}
\label{sec:2.3_Average_plasma_profile_reconstruction}
For quantitative comparison with predictive simulations, as well as for prescribing appropriate boundary and initial conditions to the simulations, fits of the measured kinetic quantities are performed in both intervals simulated, being $t_1 \in [2.5\,\mathrm{s}, 3.5\,\mathrm{s}]$ and $t_2 \in [5.0\,\mathrm{s}, 6.0\,\mathrm{s}]$ (shaded regions in Fig.\,\ref{fig:1_Time_traces_of_AUG_31115}). The two intervals, covering around $18$ confinement times $\tau_{\mathrm{E}}$, are chosen to represent periods of comparatively stable plasma profiles within each phase. Merely one abnormal event in the evolution of the W density is observed in the second interval at $t = 5.39\,$s, returning to usual levels within around 50\,ms (see Fig.~\ref{fig:3_Amplitude_of_MHD_mode}).

For both time slices, averaged plasma profiles are constructed from raw experimental data by applying Gaussian process regression techniques. A tool set created by A. Ho\cite{Ho18} was utilized, whose development was inspired by previous work carried out by M.A. Chilenski\cite{Chilenski15}. The Gaussian process regression techniques applied utilize Bayesian probability theory under the assumption of normally distributed weights for profile reconstruction from covariance functions, thus ensuring a robust estimation of profiles, gradients and associated uncertainties. An overview of data used and profiles obtained is illustrated in Fig.\,\ref{fig:5_Application_of_GPR}, showcasing the capabilities of Gaussian process regression techniques to produce both profiles and associated uncertainties in agreement with observations without specification of a model describing expected behaviour.

Electron density profiles are derived from combined core density measurements by Thomson scattering spectroscopy (TS) and edge density analysis by lithium beam spectroscopy (LIB). A relative radial inward shift of the TS-profiles of the order of $5\,$mm at the outer mid-plane is taken into account to compensate for estimated inaccuracies of the equilibrium and related mapping. In the case of the electron temperature, averaged profiles are calculated from radiometry of electron cyclotron emission (ECE) and TS data. Analysing charge exchange recombination spectroscopy (CXRS) measurements, both ion temperature and toroidal rotation velocity profiles are obtained.

\section{Setup of predictive transport simulations}
\label{sec:3_Setup_of_transport_simulations}
Predictive heat and particle transport simulations of ASDEX Upgrade shot \#31115 are performed by JETTO in both phases of the discharge, evolving electron and ion temperature profiles, as well as main ion and impurity density profiles. Plasma profiles obtained by application of Gaussian process regression techniques to raw experimental data (see Sec.\,\ref{sec:2.3_Average_plasma_profile_reconstruction}) are used as initial conditions and boundary conditions for the simulations. The current density response is calculated self-consistently by JETTO, considering bootstrap currents and external current drive by ECCD and NBI. The impact of (saturated) MHD activity on current diffusion is treated only through prescription of averaged profiles as initial conditions, whereas a possible impact through 3D effects (see Refs.~\citen{Jardin15,Krebs17}) is not studied in this work. Toroidal plasma rotation profiles available are taken into account for heat and particle transport predictions, yet are treated interpretively throughout the simulations. Particle deposition, auxiliary power deposition and radiative power loss profiles obtained from previous analysis with the TRANSP code\cite{Hawryluk80} are used as sources and sinks in the respective particle and energy balance equations solved by JETTO. Within TRANSP, NBI power and particle deposition profiles are calculated by the Monte Carlo fast ion module NUBEAM\cite{Pankin04}. Fluxes due to neoclassical phenomena are calculated by the code NCLASS\cite{Houlberg97}. {The 2D magnetic equilibrium and related geometric coefficients necessary for the 1D transport equations are obtained by the code ESCO\cite{Cenacchi88} from current density and pressure profiles provided by JETTO.

To ensure obtaining quasi-steady state results in this framework, the evolution of aforementioned profiles is modelled over 3~s ($\sim 50~\tau_\mathrm{E}$). Despite of a simulation duration in the order of the current diffusion time scale ($\sim 2~\mathrm{s}$), modifications of the $q$-profile over the $3~\mathrm{s}$ interval in the simulation domain are well within reasonable error bars of $q$-profile reconstruction (average/maximum change in $t_1$: $0.059$/$0.123$; $t_2$: $0.111$/$0.154$), justifying performing simulations over this duration.

\subsection{Application of QuaLiKiz}
Turbulent fluxes due to ITG, TEM and ETG modes by QuaLiKiz, taking collisions and rotational effects ($E\times B$ shearing) into account. Application of QuaLiKiz is restricted to the turbulence dominated region up to the pedestal top, i.e. $\rho_{\mathrm{tor}} \in \left[ 0.20, 0.85 \right]$. At smaller radial positions, turbulent contributions are negligible since turbulence is often found stable up to $\rho_{\mathrm{tor}} \sim 0.25$ in these simulations due to small normalized gradients. The presence of both steep gradients and additional MHD phenomena, in particular of ELMs, at locations exceeding $\rho_{\mathrm{tor}} = 0.85$ necessitates application of additional ELM and edge transport barrier models (see e.g. Ref.\,\citen{Koechl17}). However, as this study ultimately aims at core W impurity transport modelling, density and temperature profile are treated interpretively in this region of the plasma. 

\subsection{Additional transport coefficients}
\label{sec:3.2_Additional_transport_coefficients}
As the MHD activity observed is expected to set central profiles, the influence on central transport is mimicked by prescribing additional transport coefficients to the simulation inside the $q=1$ surface to reproduce experimental gradients. The corresponding transport coefficients are calculated iteratively by updating the initially vanishing coefficients by the effective transport coefficients (obtained from transport analysis of experimental profiles), weighted by the normalized difference between predicted and experimental gradients,
	\begin{align}
	\notag
		\Delta \chi_\mathrm{e/i,add} &= \left( 1 - \frac{\nabla T_\mathrm{e/i}|_\mathrm{pred}}{\nabla T_\mathrm{e/i}|_\mathrm{exp}} \right) \chi_{\mathrm{e/i,eff}}\bigl|_\mathrm{exp} ~, & \\
		\Delta D_{\mathrm{add}} &= \left( 1 - \frac{\nabla n_\mathrm{e}|_\mathrm{pred}}{\nabla n_\mathrm{e}|_\mathrm{exp}} \right) D_{\mathrm{eff}}\bigl|_\mathrm{exp} ~,
	\end{align}	 
until predicted gradients are in agreement with experimental observations. The additional transport necessary is assumed to be due to the effect of MHD activity not accounted for in the transport simulations carried out and is prescribed in following simulations. This approach ensures accurate description of central gradients in the presence of phenomena not described by neoclassical or gyrokinetic theory, yielding an estimate for the contribution of what is assumed to be MHD activity.

\subsection{Treatment of impurities}
\label{sec:3.3_Treatment_of_impurities}
Plasma impurities are treated inside the impurity transport code SANCO\cite{Lauro-Taroni94} under consideration of an average effective charge $\left\langle Z_{\mathrm{eff}}\right\rangle = 1.25 \pm 0.09$ in the \Fst time slice and $\left\langle Z_{\mathrm{eff}}\right\rangle = 1.22 \pm 0.10$ in the \Snd one. Radial profiles are calculated from measurements of B and N impurity densities by CXRS, as well as from W densities obtained by SXR. The content of additional light impurities is taken as $n_\text{He} = 0.01~n_\text{e}$ and $n_\text{C} = 2~n_\text{B}$, based on worst-case estimates from typical ASDEX Upgrade discharges. Under the assumption of a reasonably decreased content of these impurities, setting $n_\text{He} = n_\text{C} = n_\text{B}$, the average effective charge is reduced to $\left\langle Z_\text{eff} \right\rangle = 1.17 \pm 0.09$ and $1.15 \pm 0.10$ in each time slice. Since analysis of Bremsstrahlung measurements provides average values of $\left\langle Z_\text{eff} \right\rangle = 1.38 \pm 0.21$ and $1.36 \pm 0.22$ inside $\rho_\text{tor} = 0.85$, estimates from both diagnostics agree within errorbars. Noticeably, the average effective charge in the \Snd time slice is consistently found lower as compared to the \Fst phase due to a reduction of light impurity and central W content (see Sec.~\ref{sec:2.2_W_density_response} for the latter). Predictive heat and particle transport simulations are carried out in the presence of C impurities only as simulation results are found insensitive to the exact composition of impurities. Moreover, predicted plasma profiles are found to respond only weakly to a change in $\left\langle Z_\text{eff} \right\rangle$ within aforementioned boundaries of CXRS estimates.

\subsection{Performance}
Performing predictive heat and particle transport simulations as described using 51 grid points, the temporal evolution of 1\,s of ASDEX Upgrade plasma is calculated by JETTO-SANCO and QuaLiKiz within 6 - 11\,h on 16\,CPUs. As the simulation of a single time step with a maximum step size of 1\,ms takes on average around 17\,s, total computation time is determined primarily by the number of time steps needed.

\section{Predictive heat and particle transport simulations}
\label{sec:4_Predictive_transport_simulations}

\begin{figure*}
	\centering
	\includegraphics[width=\linewidth]{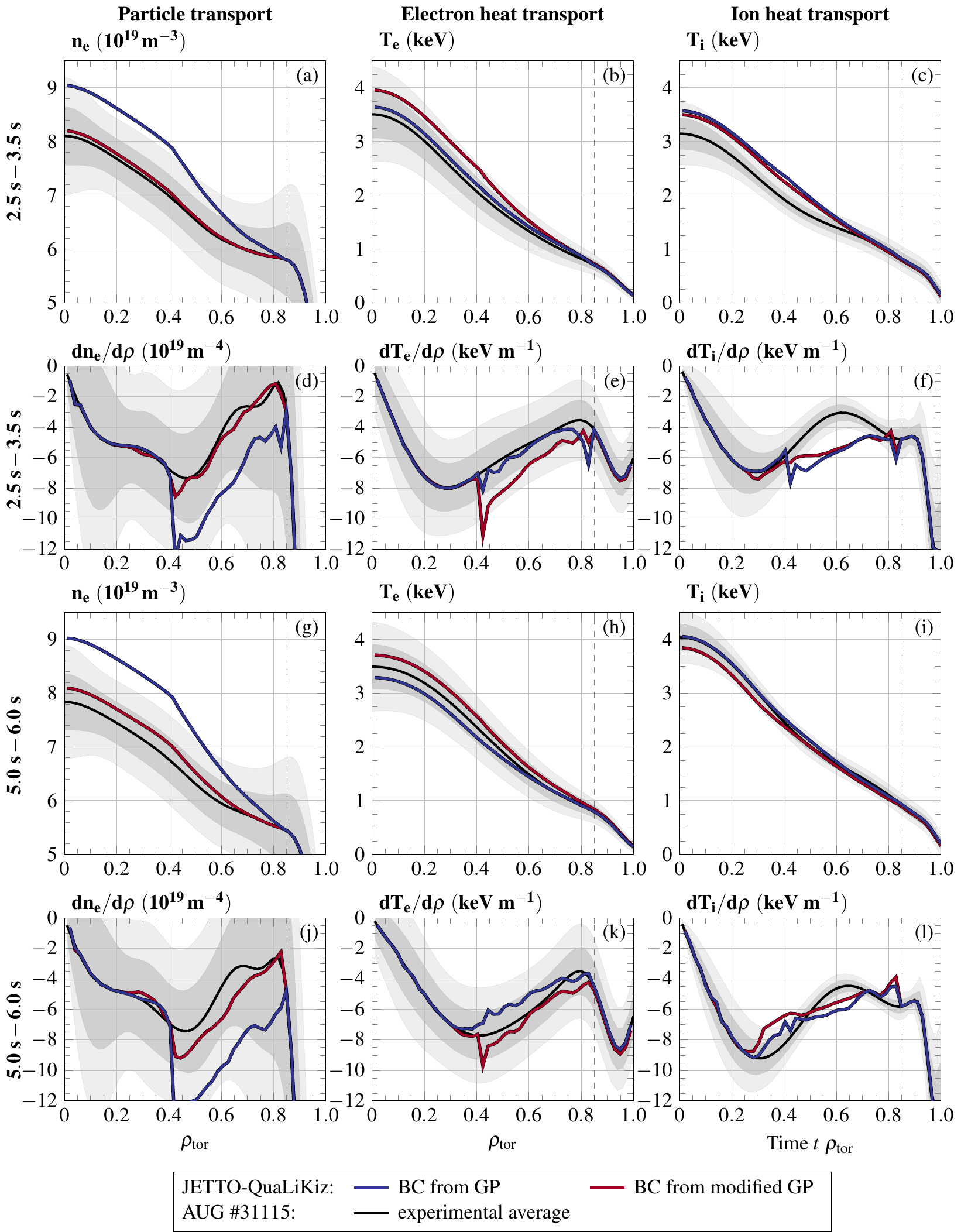}
	\stdcap{}{\label{fig:6_Predictive_transport_simulations}%
		Predictive particle and heat transport simulations of AUG discharge \#31115 performed by JETTO-QuaLiKiz (blue) for both time slices $t_1 \in \left[ 2.5\,\mathrm{s}, 3.5\,\mathrm{s}\right]$ and $t_2\in\left[ 5.0\,\mathrm{s}, 6.0\,\mathrm{s}\right]$ analysed, prescribing kinetic profiles from Gaussian process regression (GP) as boundary condition (BC) for $\rho_\mathrm{tor} \geq 0.85$ (dashed vertical line). In additional simulations (red), boundary conditions are varied within regression uncertainties. The time-independent simulations shown are carried out over 3~s to ensure reaching steady-state conditions and are averaged over a duration of $\tau_E$. Simulations are compared against the experimental average (black) with confidence intervals of $1~\sigma$ and $2~\sigma$ (grey). Turbulent fluxes are calculated by QuaLiKiz in the region $\rho_\mathrm{tor} \in \left[ 0.20, 0.85\right]$. However, turbulent transport is found relevant only outside $\rho_\text{tor} \sim 0.3$.
		}
\end{figure*}

\subsection{Modelling by JETTO-QuaLiKiz}
\label{sec:4.1_Modelling_by_JETTO-QuaLiKiz}
Flux-driven heat and particle transport over multiple confinement times of ASDEX Upgrade discharge \#31115 is predicted successfully by JETTO-QuaLiKiz, demonstrating decent agreement with the experimental average inside the boundary condition at $\rho_{\mathrm{tor}} = 0.85$ for both time slices analysed (see Fig.~\ref{fig:6_Predictive_transport_simulations}). For further comparison between simulation and experiment in the region where QuaLiKiz is applied, a simple standard deviation figure of merit is applied (see Ref.~\citen{Citrin17} and references therein):

	\begin{align}
	\label{eq:2_Figure_of_merit}
		\Delta f &= \sqrt{\int_{0.2}^{\mathrm{bc}} \left( f_\mathrm{sim} - f_\mathrm{exp} \right)^2 ~\mathrm{d}\rho} \left/ \sqrt{\int_{0.2}^{\mathrm{bc}} f_\mathrm{exp}^2 ~\mathrm{d}\rho} \right.
	\end{align}

Plasma profiles obtained by JETTO-QuaLiKiz agree within 18\% ($2.4~\sigma$) with the experimental average. Yet, agreement between simulation and experiment among different transport channels varies noticeably. Electron heat transport is well captured, as average deviations of the electron temperature gradient are within $13\%$ ($0.5~\sigma$). Inside the $q=1$ surface at $\rho_{\mathrm{tor}} = 0.4$, gradients of all kinetic profiles are typically described accurately as a result of the additional transport coefficients prescribed. Experimental ion heat transport is reproduced with less success. Whereas measurements by CXRS find the temperature gradient to change non-monotonically in the turbulence dominated region of the plasma, the reduced transport model QuaLiKiz predicts temperature profiles of approximately constant curvature (this holds for all transport channels). Average agreement of ion temperature gradients is thus within $31\%$ ($1.8~\sigma$).

In the case of particle transport, noticeably less turbulent contributions are predicted by QuaLiKiz than implied by experimental observations. Here, average deviations of electron density gradients are up to 73\% ($1.1~\sigma$). Consequently, density peaking is predicted by JETTO-QuaLiKiz in contrast to measurements (on-axis densities increased by up to 15\%, being $2.3~\sigma$). However, the curvature of experimental profiles in the turbulence dominated region is reproduced by QuaLiKiz. This suggests, that turbulence is not only stabilised at the simulation boundary of QuaLiKiz at $\rho_\mathrm{tor} = 0.85$ (due to the boundary condition), but also in the simulation domain of QuaLiKiz through a cascade of local effects.

\subsection{Modified boundary conditions}
\label{sec:4.2_Modified_boundary_conditions}
The heat and particle transport simulations of ASDEX Upgrade discharge \#31115 by JETTO-QuaLiKiz are carried out using plasma profiles obtained from Gaussian process regression of experimental measurements as boundary values. Under these conditions, turbulent particle transport is reduced with respect to experimental observations, as ITG dominated modes are found significantly stabilized. Yet, in AUG discharges heated by NBI and by additional ECRH, ion-scale modes are usually unstable, according to turbulence diagnostics\cite{Conway08, Happel15} and non-linear gyrokinetic simulations\cite{Happel15}. As the choice of plasma boundary conditions determines not only stabilisation of turbulence at the simulation boundary of QuaLiKiz, but also influences turbulence further inside the plasma core through a cascade of local effects, the boundary conditions prescribed are thus modified under consideration of uncertainties provided by Gaussian process regression such that turbulence is increased at the simulation boundary.

An increase of turbulent transport at the simulation boundary is achieved by reducing $E\times B$ shearing and the ion to electron temperature ratio $T_\mathrm{i}/T_\mathrm{e}|_\mathrm{bc}$ (see Sec.~\ref{sec:5.1_Influence_of_TiTe_BC_on_core_transport} for details). A decrease in $E\times B$ shearing is obtained by reducing the toroidal velocity gradient in the vicinity of the simulation boundary by $1~\sigma$, while allowing for a maximum change of the toroidal rotation profile of $\pm 1 ~\sigma$ throughout the plasma volume. The temperature ratio boundary condition prescribed is reduced from $T_\mathrm{i}/T_\mathrm{e}|_\mathrm{bc} = 1.16$ to $1.09$ in time slice $t_1$ and down to $1.06$ in the \Snd one. For this purpose, the ion temperature at the simulation boundary is reduced by $0.25~\sigma_{T_\mathrm{i}}$ and $0.40~\sigma_{T_\mathrm{i}}$ in each time slice, whereas the electron temperature is increased by corresponding amounts of $\sigma_{T_\mathrm{e}}$. Since under typical experimental conditions, temperature ratios of $T_\mathrm{i}/T_\mathrm{e} \sim 1.0$ are usually observed in the pedestal region across various machines\cite{Maggi14,Meyer14}, minor modifications of the temperature ratio boundary condition $T_\mathrm{i}/T_\mathrm{e}|_\mathrm{bc}$ prescribed in the simulations towards aforementioned value are reasonable.

\subsubsection{Improved agreement between simulations and experiment}
In simulations with the boundary conditions adjusted as discussed, particle transport is described in much better agreement with experimental observations (see Fig.~\ref{fig:6_Predictive_transport_simulations}). The electron density gradients calculated agree on average within 22\% ($0.3~\sigma$) with the experimental average. Correspondingly, the electron density calculated by JETTO-QuaLiKiz describes experimental observations well (average deviations within $3\%$, being $0.5~\sigma$). Ion heat transport predicted is increased only marginally by a change in boundary conditions. Hence, ion temperatures are slightly decreased throughout the simulation domain, but also as a result of the reduced boundary condition applied. In contrast, electron temperature profiles are predicted with reduced confidence, as turbulent electron heat transport is found stabilized under these conditions (see Sec.~\ref{sec:5.1_Influence_of_TiTe_BC_on_core_transport}).

\begin{table*}[bt!]
	\centering
	\begin{small}
	\stdcap{Agreement of predictive heat and particle transport simulations with experimental average}{Deviation to the experimental average of density and temperature profiles, as well as of associated gradients, calculated in predictive particle and heat transport simulations by JETTO-QuaLiKiz using modified boundary conditions in both time slices $t_1 \in \left[ 2.5\,\mathrm{s}, 3.5\,\mathrm{s}\right]$ and $t_2 \in \left[ 5.0\,\mathrm{s}, 6.0\,\mathrm{s}\right]$. Average (see Eq.~\eqref{eq:2_Figure_of_merit}), maximum and on-axis deviation are expressed with respect to both the absolute value (\%) and the standard deviation ($\sigma$) of the experimental average. Average deviations are taken inside the region $\rho_{\mathrm{tor}} \in \left[ 0.20, 0.85 \right]$ where QuaLiKiz is applied.
	}
	\label{tbl:simulation_agreement}
	\begin{tabular}{llD{.}{.}{-1}D{.}{.}{-1}|D{.}{.}{-1}D{.}{.}{-1}|D{.}{.}{-1}D{.}{.}{-1}}
	\toprule
		&&\multicolumn{2}{c|}{Average deviation}
		& \multicolumn{2}{c|}{Maximum deviation}
		& \multicolumn{2}{c}{On-axis deviation} 
	\\
		&&\multicolumn{1}{c}{(\%)} & \multicolumn{1}{c|}{($\sigma$)}
		& \multicolumn{1}{c}{(\%)} & \multicolumn{1}{c|}{($\sigma$)}
		& \multicolumn{1}{c}{(\%)} & \multicolumn{1}{c}{($\sigma$)}
	\\
	\midrule 	
		
		Electron density $n_{\mathrm{e}}$:
		& $t_1$ &  0.9 & 0.14 &  1.2 & 0.25 & 1.2 & 0.18
	\\	
		& $t_2$ &  3.1 & 0.46 &  4.4 & 0.71 & 3.3 & 0.49
	\\	
		Gradient $\mathrm{d}n_{\mathrm{e}}/\mathrm{d}\rho$:
		& $t_1$ &  9.6 & 0.13 & 34.0 & 0.48 & &
	\\	
		& $t_2$ & 21.4 & 0.31 & 57.1 & 0.78 & &
	\\	
	\midrule
		
		Electron temperature $T_{\mathrm{e}}$:
		& $t_1$ &  17.6 & 1.38 & 22.4 & 1.71 & 13.0 & 1.03
	\\	
		& $t_2$ &   8.0 & 0.65 &  10.0 & 0.73 & 6.3 & 0.53
	\\
		
		Gradient $\mathrm{d}T_{\mathrm{e}}/\mathrm{d}\rho$:
		& $t_1$ & 24.0 & 1.00 & 55.5 & 2.44 & &
	\\  
		& $t_2$ &  9.7 & 0.40 & 38.0 & 0.98 & &
	\\	
	\midrule
		
		Ion temperature $T_{\mathrm{i}}$:
		& $t_1$ & 14.2 & 1.90 &  17.8 & 2.59 & 11.2 & 1.21
	\\	
		& $t_2$ &  5.0 & 0.98 &  7.8 & 1.47 & 4.8 & 0.80
	\\
		
		Gradient $\mathrm{d}T_{\mathrm{i}}/\mathrm{d}\rho$:
		& $t_1$ & 26.0 & 1.50 & 75.7 & 4.26 & &
	\\	
		& $t_2$ & 15.6 & 1.50 & 34.1 & 4.94 & &
	\\	
	\bottomrule
	\end{tabular}
	\end{small}
\end{table*}

Taking all three transport channels considered into account, overall agreement between simulations and experiment is increased using the adjusted boundary conditions. A detailed overview of average, maximum and on-axis deviation for each of the three profiles predicted as well as for associated gradients is presented in Table~\ref{tbl:simulation_agreement}. Especially particle transport is captured more successfully under application of adjusted boundary conditions. Since this study ultimately aims at core W transport studies, following simulations are carried out with adjusted boundary conditions prescribed. To verify the simulation results presented, a successful benchmark of QuaLiKiz used within the transport code ASTRA was carried out (see Appendix~\ref{sec:A_Comparison_ASTRA_and_JETTO} for details).

Comparing the simulations performed for both sets of plasma parameters, the transport predicted is strongly sensitive to modifications of the boundary conditions and to a lesser extent to modifications of the $E\times B$ shearing prescribed (as observed in separate simulations), even when varied well within $1~\sigma$ of Gaussian process regression results. A rigorous propagation of input uncertainties through the integrated modelling process is thus necessary for proper assessment of simulation results.

\subsubsection{Influence of changing operational conditions on simulations}
\label{sec:4.2.2_Influence_of_changing_operational_conditions}
Comparing simulations carried out for both time slices with each other, the response of experimental kinetic profiles to changing operational and plasma conditions (such as NBI heating, fuelling, impurity composition, radiative power loss, plasma rotation) is not captured accurately by the simulations. In the case of particle transport, measurements find the core electron density to be slightly decreased in the \Snd phase of the discharge and mid-radius gradients to be approximately unaffected by a change in operational conditions. In contrast, simulations by JETTO-QuaLiKiz predict an increase of density gradients. Analysis of the turbulence present with stand-alone QuaLiKiz suggests the overprediction of gradients being due to an over proportional increase of inward pinches (see Sec.~\ref{sec:5.3_Core_density_dependence_on_NBI_fuelling}). Consequently, agreement between simulation and experiment deteriorates slightly in the \Snd phase of the discharge.

The difference in heat transport observed experimentally in both phases of the discharge is similarly not accurately reproduced by JETTO-QuaLiKiz simulations. Contrary to measurements, temperature gradients calculated in the turbulence dominated region are approximately constant throughout both time slices despite coupling an additional NBI source of $2.4~$MW to the plasma. Experimentally, most power is found to heat the ion population, thus significantly increasing ion temperature and gradients. Through net thermal energy exchange and additional NBI electron heating, the electron temperature is also increased to some extent (despite larger radiative power loss). In the simulations carried out, turbulent heat transport coefficients are found to be increased instead, thus ensuring energy balance. However, predicted temperature profiles are in better agreement with the experimental profiles of the \Snd phase.

It should be emphasized, that the experimental averages, the profiles obtained by JETTO-QuaLiKiz are compared against, are the result of Gaussian process regression and are thus subjected to uncertainties. Consequently, assessing the capability of JETTO-QuaLiKiz to reproduce ASDEX Upgrade discharges, both uncertainties obtained from regression as well as from raw experimental data directly are to be taken into account. As simulation results are found to agree within regression uncertainties, even for the cases of least agreement, and confidently represent raw experimental data, JETTO-QuaLiKiz is found capable of simulating ASDEX Upgrade discharges within errorbars.

\subsection{Influence of MHD induced transport on central profile agreement}
\label{sec:4.3_Influence_of_MHD_transport_on_central_profile_agreement}
Agreement in particle and heat transport inside the $q=1$ surface at $\rho_{\mathrm{tor}} \sim 0.4$ is obtained only when prescribing additional transport coefficients to mimic the effect of central MHD activity. Recall, that the transport coefficients required are calculated iteratively from the effective transport coefficients, weighted by the difference of predicted and experimental normalized gradients (see Sec.~\ref{sec:3.2_Additional_transport_coefficients}). When omitting this contribution, central transport is found significantly reduced, requiring predictive simulations over a prolonged duration, as steady state is not reached after evolving the plasma profiles over 3\,s. However, as the simulations omitting additional transport are qualitative only, profiles obtained after 3\,s of temporal evolution are showcased. To demonstrate the absence of central turbulence transport, QuaLiKiz is applied down to the magnetic axis, as opposed to the simulations including additional transport coefficients, where turbulent contributions are calculated down to only $\rho_{\mathrm{tor}} = 0.2$. The influence of MHD induced transport on central profile agreement is discussed for the set of plasma parameters presented in Sec.~\ref{sec:4.2_Modified_boundary_conditions} (qualitatively valid also for the set of parameters of Sec.~\ref{sec:4.1_Modelling_by_JETTO-QuaLiKiz}).

\begin{figure*}
	\centering
	\includegraphics[width=\linewidth]{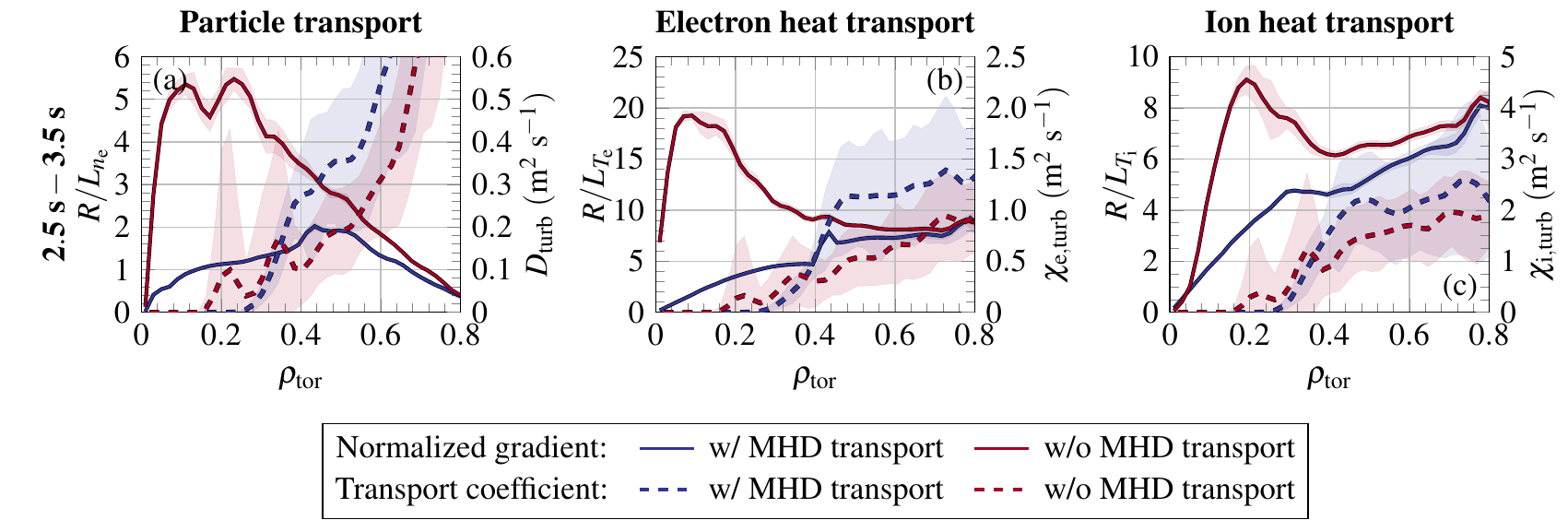}
	\stdcap{}{\label{fig:7_Simulations_without_additional_transport}%
		(a) Particle, (b) electron heat and (c) ion heat transport from predictive simulations with JETTO-QuaLiKiz in the presence (blue) or in the absence (red) of additional transport coefficients inside $\rho_{\mathrm{tor}} = 0.4$, mimicking the effect of MHD activity. Associated normalized gradients (solid, left ordinate), as well as the turbulent transport coefficients $D_{\mathrm{turb}}$ and $\chi_{\mathrm{turb}}$ calculated by QuaLiKiz (dashed, right ordinate) are averaged over the last 0.5\,s of the simulations of the \Fst time slice. Note, that the transport coefficients shown do not include contributions due to MHD activity or neoclassical phenomena. The shaded regions represent the standard deviations of the quantities presented, demonstrating the non-stationary behaviour of the normalized plasma gradients in the last 0.5\,s of plasma evolution when omitting additional transport mimicking the impact of MHD activity.
		}
\end{figure*}

\subsubsection{Transport simulations excluding additional central transport}
Omitting additional transport coefficients mimicking MHD activity, transport inside the $q=1$ surface is severely underestimated, associated central plasma profiles severely overestimated. In the case of particle transport, simulations predict on-axis densities increased by as much as 78\% compared to simulations including additional contributions to central transport in both time slices analysed ($t_1$: $+6.4\times 10^{19}\,\mathrm{m}^{-3}$, $t_2$: $+5.3\times 10^{19}\,\mathrm{m}^{-3}$). Similarly, heat transport predictions put the on-axis electron temperature at around 21\,keV, corresponding to an increase of around 17\,keV, and find the central ion temperature to be increased by $2.5~$keV and $1.9$~keV in both time slices. A comparison of normalized gradients for both cases is illustrated in Fig.\,\ref{fig:7_Simulations_without_additional_transport} for the \Fst time slice analysed only, as the situation is similar in both phases of the discharge. As additional transport is included to obtain central transport agreement, normalized gradients in this case are close to the experimental profiles. Hence, normalized gradients are severely overestimated in the absence of enhanced central transport coefficients. Note, that turbulence spreading is yet to be investigated in QuaLiKiz \cite{Bourdelle16}.

All gradients shown in Fig.\,\ref{fig:7_Simulations_without_additional_transport} are obtained by averaging the last 0.5\,s of the simulations. The finite standard deviation of gradients inside the $q=1$ surface in the absence of additional central transport illustrates that simulated plasma profiles are still in the process of converging in the last $0.5~$s of the simulation, whereas a steady state central solution is obtained when including enhanced transport coefficients, demonstrated by the vanishing standard deviation in this case.

In the case of particle transport, omitting additional contributions, normalized gradients inside the $q=1$ surface are as high as $R/L_{n_{\mathrm{e}}} = 5.5$, as compared to values ranging between $R/L_{n_{\mathrm{e}}} \sim 1 -2$ in the turbulence dominated region. Similarly, central normalized electron temperature gradients are found up to $R/L_{T_{\mathrm{e}}} = 20$, constituting a significant increase from the average value of $R/L_{T_{\mathrm{e}}} \sim 7$ present in the turbulence dominated region. The same holds for central normalized ion temperature gradients, where an increase of $R/L_{T_\mathrm{i}}$ from $5-7$ in the turbulence dominated region up to to $R/L_{T_\mathrm{i}} \sim 9$ in the vicinity of the magnetic axis is observed. In the presence of additional central transport, normalized gradients flatten monotonously between $q=1$ surface and magnetic axis. On a side note, as normalized density gradients observed in this discharge are comparatively low (see Fig.\,\ref{fig:7_Simulations_without_additional_transport}(a)), ion scale turbulence is found to be determined by ITG dominated modes. Hence, critical density gradients for TEM destabilization are not reached\cite{Garbet04,Romanelli04}.

Including additional central transport, turbulence is stabilized inside $\rho_{\mathrm{tor}} \sim 0.25$. Here, no unstable modes are present, as indicated by turbulent particle and heat transport coefficients, since normalized gradients are decreased below their critical values by enhanced central transport. Turbulent transport coefficients presented in Fig.\,\ref{fig:7_Simulations_without_additional_transport} for the \Fst time slice analysed are averaged over the last 0.5\,s of plasma evolution, highlighting the reduced turbulence levels inside $\rho_{\mathrm{tor}} \sim 0.4$, as standard deviations of the averaged transport coefficients are small. Neglecting additional central transport, neoclassical transport is insufficient to maintain central plasma profiles, even though neoclassical transport increases with respect to case considering additional transport coefficients for all channels. Gradients increase until turbulence is driven sufficiently unstable to balance transport, thus setting central profiles. Correspondingly, finite turbulent transport coefficients are observed down to $\rho_{\mathrm{tor}} \sim 0.15$, although decreased as compared to the amplitude found in the turbulence dominated region, as steep normalized gradients are present in the plasma centre. Hence, additional transport coefficients mimicking the influence of central MHD activity are vital to obtain agreement in central transport.

\begin{figure*}
	\centering
	\includegraphics[width=\linewidth]{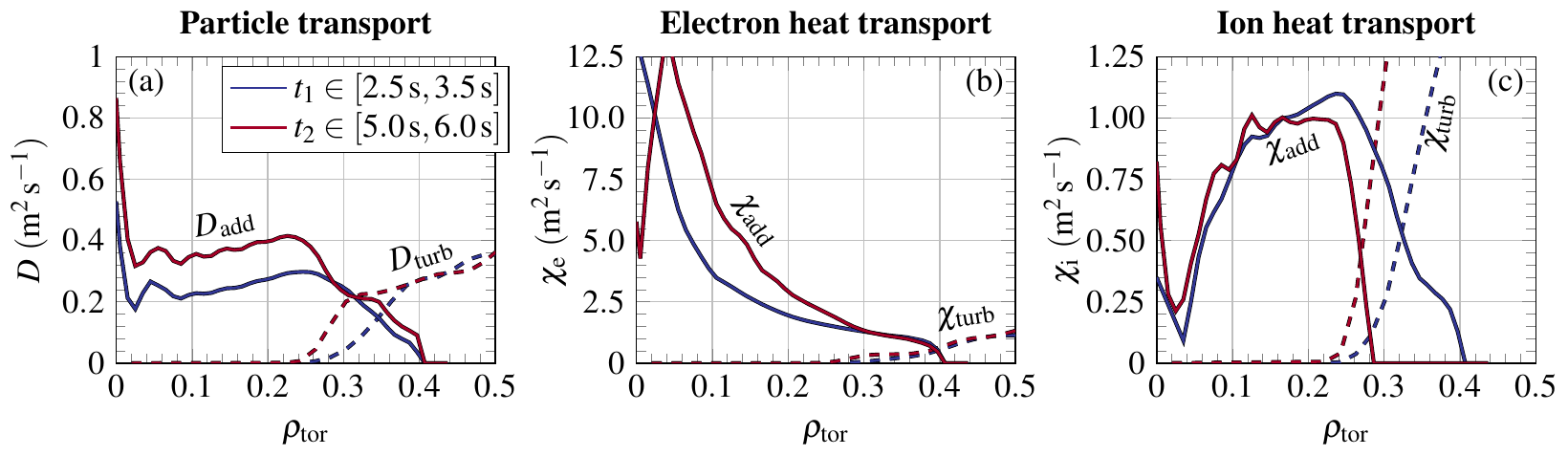}
	\stdcap{}{\label{fig:8_Additional_transport_coefficients}%
		Particle (a), electron heat (b) and ion heat (c) transport coefficients in the inner-half radius. Additional transport coefficients (solid) prescribed in predictive transport simulations to mimic the effect of a saturated $(1,1)$ MHD mode with a low saturation fraction, short period in time slice $t_1 \in \left[ 2.5\,\mathrm{s}, 3.5\,\mathrm{s}\right]$ (blue) and with a high saturation fraction, long period in time slice $t_2 \in \left[ 5.0\,\mathrm{s}, 6.0\,\mathrm{s}\right]$ (red) inside the $q=1$ surface are compared to turbulent contributions (dashed) calculated by QuaLiKiz.
		}
\end{figure*}

\subsubsection{Magnitude of MHD induced central transport}
Additional transport coefficients prescribed to mimic the effect of MHD activity are of comparable magnitude as turbulent contributions in the vicinity of the $q=1$ surface in both phases of the discharge (see Fig.~\ref{fig:8_Additional_transport_coefficients}). Here, particle diffusivity is around $0.2$ to $0.3~\mathrm{m}^2~\mathrm{s}^{-1}$, whereas heat transport is roughly $1\,\mathrm{m}^{2}\,\mathrm{s}^{-1}$ for both turbulent and additional transport. At plasma radii below $\rho_\mathrm{tor} \sim 0.3$, additional transport necessary exceeds turbulent contributions for all transport channels, as turbulence is found increasingly stabilized\footnote{Note, that turbulent ion heat transport is increased in the 2$^{\text{nd}}$ phase of the discharge (see Sec.~\ref{sec:4.2.2_Influence_of_changing_operational_conditions}). Additional transport is thus necessary only inside $\rho_\mathrm{tor} = 0.3$.}. Approaching the magnetic axis, the additional particle transport necessary is approximately radially constant. In the case of additional heat transport however, contributions to electron heat transport increase significantly, while contributions to ion heat transport decrease noticeably.

In the presence of the high saturation fraction $(1,1)$ MHD mode, additional particle diffusivity necessary to obtain transport agreement inside the $q=1$ surface is noticeably larger than in the phase of the low saturation fraction mode, especially inside $\rho_\mathrm{tor} \sim 0.25$ where the additional diffusivity is found to be increased by on average $0.12~\mathrm{m}^2~\mathrm{s}^{-1}$ (neglecting the magnetic axis). Similarly, additional electron heat transport is also found to be increased in the \Snd phase of the discharge. In the case of additional ion heat transport however, a pronounced increase is not necessary to reproduce central experimental gradients. Nevertheless, the additional transport coefficients prescribed demonstrate increased outward transport in the presence of the high saturation fraction, long period saturated $(1,1)$ MHD mode. 

It should be noted, that the additional transport coefficients prescribed are prone to errors due to uncertainties of the input quantities of the respective balance equations. Additionally, small gradients in the vicinity of the magnetic axis demand comparatively large transport coefficients to match heat and particle fluxes deposited. As the relative uncertainties in the gradients obtained are quite large (see Fig.\,\ref{fig:6_Predictive_transport_simulations}), confidence in the calculated effective transport coefficients close to the magnetic axis is reduced. In the case of electron heat transport, the footprint of the ECRH power deposition profile is imprinted in the effective heat diffusivity, as the entire $1.4\,$MW of auxiliary power are deposited within $\rho_{\mathrm{tor}} = 0.2$, demanding a large heat conductivity to ensure energy balance. Since turbulent and neoclassical contributions to both ion and electron heat transport are negligible inside $\rho_{\mathrm{tor}} = 0.2$, additional heat transport coefficients are determined primarily by the effective transport coefficients and associated uncertainties. In the case of particle transport, considerable contributions from inward pinches and neoclassical phenomena reduce the influence of the effective particle diffusivity in calculating additional transport coefficients necessary, yet introduce further uncertainties.

Considering the effect of the additional particle diffusivity prescribed on central W transport under the assumption of identical transport coefficients for electrons and W, the increase in diffusivity observed between both phases of the discharge is in alignment with the experimental observation, that the central W content is noticeably reduced with the onset of the long period, saturated $(1,1)$ mode. This suggests that the mechanism present in the \Snd phase of the discharge is more efficient. Since a difference between both phases is the longer period and higher saturation fraction of the $(1,1)$ MHD mode present, this saturated MHD mode might perhaps indeed facilitate outward W transport. The importance of the different transport channels for W impurity transport is investigated in dedicated simulations presented in Sec.~\ref{sec:6_Predictive_W_transport_simulations}.

\section{Simulation sensitivity on plasma parameters}
\label{sec:5_Simulation_sensitivity}

\begin{figure*}
	\centering
	\includegraphics[width=\linewidth]{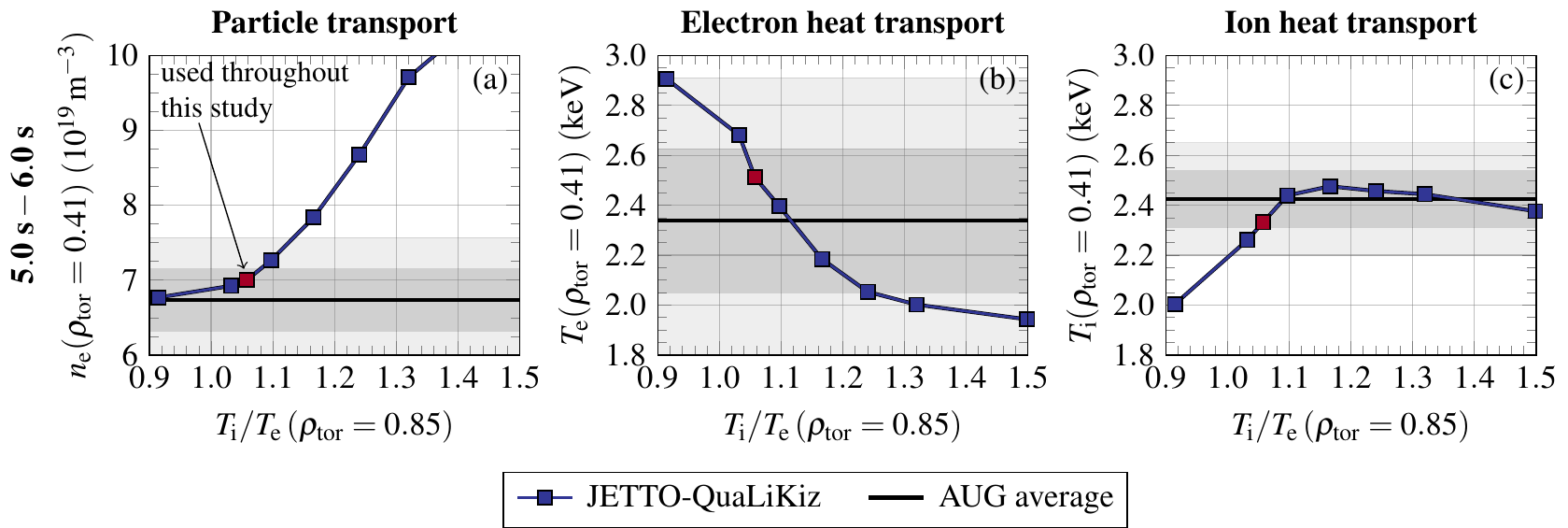}
	\stdcap{}{\label{fig:9_Plasma_reponse_to_TiTe_BC}%
		Plasma response at $\rho_\mathrm{tor} = 0.41$ to the $\left.T_\mathrm{i}/T_\mathrm{e}\right|_\mathrm{bc}$ boundary condition for the time slice $t_2 \in [5.0\,\mathrm{s}, 6.0\,\mathrm{s}]$. From the temperature profiles obtained by Gaussian process regression ($\left.T_\mathrm{i}/T_\mathrm{e}\right|_\mathrm{bc} = 1.17$), different ratios $T_\mathrm{i}/T_\mathrm{e}|_{\mathrm{bc}}$ are obtained by asymmetrically changing the species' temperatures $T_\mathrm{s}$ by multiples of the respective standard deviations, in the case presented by $\pm 1.00~\sigma_{T_\mathrm{s}}$, $\pm 0.50~\sigma_{T_\mathrm{s}}$ and $\pm 0.25~\sigma_{T_\mathrm{s}}$. The value $T_\mathrm{i}/T_\mathrm{e}|_{\mathrm{bc}}= 1.06$ used for predictive heat and particle transport simulations throughout the rest of this study is emphasized by a red square. Predicted (a) electron density, (b) electron temperature and (c) ion temperature (solid blue) compared to the averaged plasma profile (dashed black) with confidence intervals of $1\,\sigma$ and $2\,\sigma$ (grey). The behaviour illustrated is qualitatively observed at all radial positions.
		}
\end{figure*}

Throughout the preparation of this study, plasma profiles calculated in predictive heat and particle transport simulations by JETTO-QuaLiKiz were found sensitive to a reasonable variation of various plasma parameters. The dependence of density and temperature profile predictions on the following quantities is discussed in the upcoming sections: dependence on the ion to electron temperature ratio \TiTe prescribed at the simulation boundary (Sec.\,\ref{sec:5.1_Influence_of_TiTe_BC_on_core_transport}), influence of the average effective charge $\left\langle Z_\mathrm{eff} \right\rangle$ (Sec.\,\ref{sec:5.2_ETG_stabilization_by_impurities}), effect of the NBI particle source employed (Sec.\,\ref{sec:5.3_Core_density_dependence_on_NBI_fuelling}), impact of reduced collisionality (Sec.\,\ref{sec:5.4_Particle_transprot_reduction_with_reduced_collisionality}).

Note, that the following simulations presented are evaluated primarily in the turbulence dominated region between the $q=1$ surface at $\rho_{\mathrm{tor}} \sim 0.4$ and the simulation boundary at $\rho_{\mathrm{tor}} = 0.85$ to eliminate the possibility of over-/underestimation of additional central transport coefficients inside $\rho_{\mathrm{tor}}=0.4$ due to changing plasma parameters and thus of influencing the sensitivity analyses carried out.

In the simulations carried out within this work, sources and sinks are prescribed from previous analysis with the TRANSP code (see Sec.~\ref{sec:3_Setup_of_transport_simulations}). Simulation results are expected to be affected only slightly by a variation of sources and sinks within uncertainties - those being $\sim 15\%$ for ECRH, 4-6\% for NBI and 5\% for the radiated power - as their magnitude is relatively small (compare with variation of the particle source in Sec.~\ref{sec:5.3_Core_density_dependence_on_NBI_fuelling}) and steady-state density profiles are observed to be transport-dominated instead of source-dominated (see Sec.~\ref{sec:5.3_Core_density_dependence_on_NBI_fuelling}).

\subsection{Influence of the ion to electron temperature boundary condition on core transport}
\label{sec:5.1_Influence_of_TiTe_BC_on_core_transport}
The predictive heat and particle transport simulations performed are found sensitive to the imposed ion to electron temperature $T_{\mathrm{i}}/T_{\mathrm{e}}|_{\mathrm{bc}}$ boundary condition at $\rho_{\mathrm{tor}} = 0.85$. For values of $T_{\mathrm{i}}/T_{\mathrm{e}}|_{\mathrm{bc}}$ exceeding $1.2$, severe density peaking is observed throughout the core plasma whereas heat transport is less affected by a variation in $T_\mathrm{i}/T_{\mathrm{e}}|_{\mathrm{bc}}$ (see Fig.\,\ref{fig:9_Plasma_reponse_to_TiTe_BC}). A detailed description of the effects of an elevated temperature ratio boundary condition on the simulations is given in Sec.\,\ref{sec:5.1.1_Observations_from_transport_simulations}. A physical interpretation of the phenomena observed is presented in Sec.\,\ref{sec:5.1.2_Observations_from_gyrokinetic_calculations}. Even though this effect is present in both time slices considered, the discussion will be limited to time slice $t_2 \in \left[ 5.0\,\mathrm{s}, 6.0\,\mathrm{s}\right]$. Finally, consequences for simulation setup and validation are drawn in Sec.~\ref{sec:5.1.3_Implications_for_transport_simulations}.

\subsubsection{Observations from predictive transport simulations}
\label{sec:5.1.1_Observations_from_transport_simulations}

\newcommand{\footnoteFiveOneOne}{%
	Note, that the plasma parameters illustrated in Fig.\,\ref{fig:9_Plasma_reponse_to_TiTe_BC} do not represent the on-axis values of the respective profiles. Instead, the plasma response to a change in the $T_\text{i}/T_\text{e}|_\text{bc}$ boundary condition is evaluated at $\rho_{\mathrm{tor}} = 0.41$, corresponding to the innermost radial position where no additional transport is prescribed. However, the plasma response observed at $\rho_\mathrm{tor} = 0.41$ is found qualitatively at all radial positions. Additionally, as the uncertainty of the regression results is increased on the magnetic axis (see Fig.\,\ref{fig:6_Predictive_transport_simulations}), a comparison between the plasma response and the experimental average in terms of the respective standard deviation in the vicinity of the magnetic axis would demonstrate a higher degree of agreement than suggested by evaluation at mid-radius.
	}

Performing simulations evolving the initial profiles obtained with Gaussian process regression from experimental measurements, the ion to electron temperature boundary condition is prescribed as $T_{\mathrm{i}}/T_{\mathrm{e}}|_{\mathrm{bc}} = 1.17$. Under these conditions, the electron density is overestimated on-axis by $1.2\times 10^{19}\,\mathrm{m}^{-3}$, corresponding to 15\,\% of the central experimental average (cf. Sec.~\ref{sec:4.1_Modelling_by_JETTO-QuaLiKiz}). However since plasma profiles are treated interpretively beyond $\rho_{\mathrm{tor}} = 0.85$, electron density profile evolution is modelled only on top of the H-mode pedestal. Expressed in terms of the increase between pedestal shoulder and magnetic axis, the central density is overpredicted by more than 49\%. Approaching a temperature ratio of unity at the simulation boundary, agreement of the predicted on-axis electron density with the experimental average is improved significantly to within 2\,\% (cf. Fig.\,\ref{fig:9_Plasma_reponse_to_TiTe_BC}(a)\footnote{\footnoteFiveOneOne}). Starting from the temperature profiles obtained through regression, different temperature ratio boundary conditions are constructed by asymmetrically changing the species' temperatures in the respective domain of the plasma by up to one standard deviation for this analysis.

\begin{figure*}
	\centering
	\includegraphics[width=\linewidth]{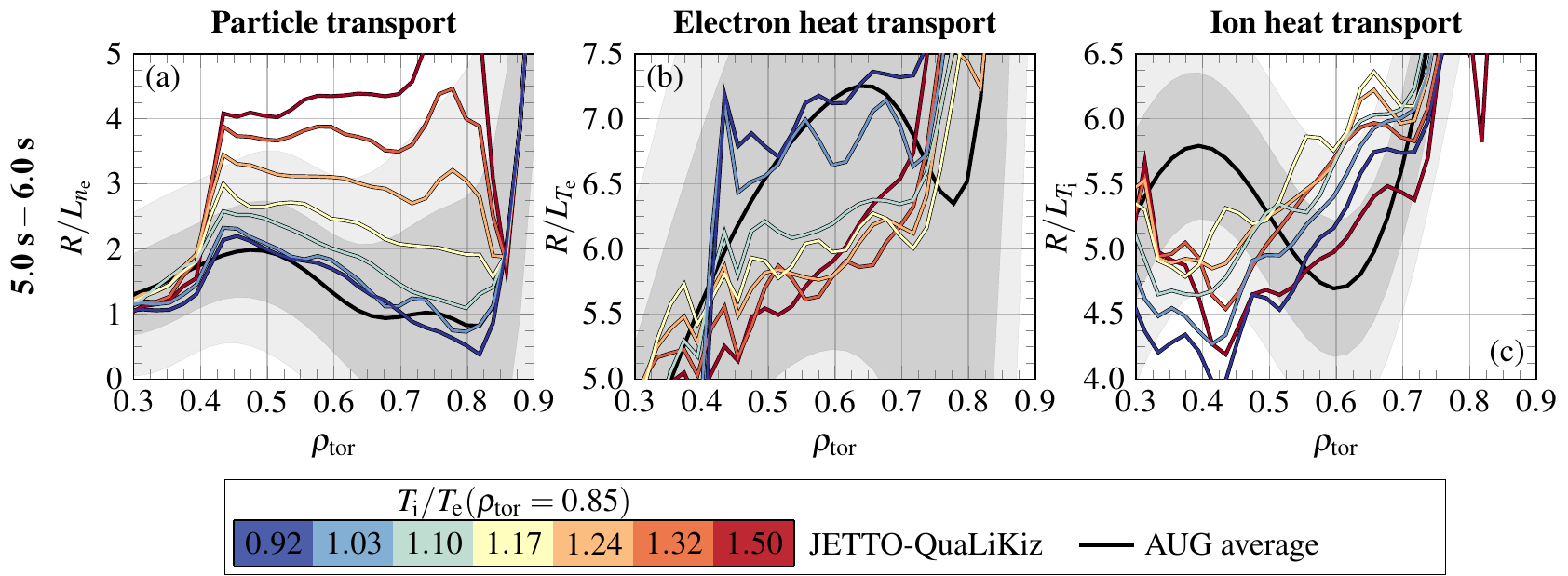}
	\stdcap{}{\label{fig:10_Gradients_for_various_TiTe_BC}%
		Normalized plasma profile gradients for different values of the $\left.T_\mathrm{i}/T_\mathrm{e}\right|_\mathrm{bc}$ boundary condition, gradually increasing from $\left.T_\mathrm{i}/T_\mathrm{e}\right|_\mathrm{bc} = 0.92$ (blue) to $1.50$ (red), for the time slice $t_2 \in [5.0\,\mathrm{s}, 6.0\,\mathrm{s}]$. From the temperature profiles obtained by Gaussian process regression ($\left.T_\mathrm{i}/T_\mathrm{e}\right|_\mathrm{bc} = 1.17$), different ratios $T_\mathrm{i}/T_\mathrm{e}|_{\mathrm{bc}}$ are obtained by asymmetrically changing the species' temperatures $T_\mathrm{s}$ by multiples of the respective standard deviations. Predicted normalized (a) electron density, (b) electron temperature and (c) ion temperature gradients compared to the experimental average (black) with confidence intervals of $1\,\sigma$ and $2\,\sigma$ (grey).
		}
\end{figure*}

Regarding electron heat transport, on-axis temperatures are underestimated by 6\,\% using the Gaussian process regression results of $T_{\mathrm{i}}/T_{\mathrm{e}}|_{\mathrm{bc}} = 1.17$, thus being less affected than particle transport calculations by a temperature ratio boundary condition noticeably exceeding unity. The electron temperature response to a reduction of $T_{\mathrm{i}}/T_\mathrm{e}|_\mathrm{bc}$ is qualitatively different, exhibiting increased temperature peaking as the temperature ratio boundary condition is reduced to unity (see Fig.\,\ref{fig:9_Plasma_reponse_to_TiTe_BC}(b)). Still, in the predictive heat transport simulation performed, central electron temperatures agree within $2\,\sigma$ with the experimental average.

A flattening of the ion temperature profile is generally observed when reducing the temperature ratio boundary condition from $T_\mathrm{i}/T_\mathrm{e}|_\mathrm{bc} = 1.17$ to unity (see Fig.\,\ref{fig:9_Plasma_reponse_to_TiTe_BC}(c)). Albeit comparatively similar in the turbulence dominated region ($\Delta R/L_{T_\mathrm{i}} \approx 1$), a noticeable flatting of mid-radius gradients is observed decreasing the boundary condition from $T_{\mathrm{i}}/T_\mathrm{e}|_\mathrm{bc} = 1.17$ down to unity (cf. Fig.\,\ref{fig:10_Gradients_for_various_TiTe_BC}(c)). This effect is pronounced for temperature ratio boundary conditions close to unity, i.e. $T_\mathrm{i}/T_\mathrm{e}|_\mathrm{bc} \lesssim 1.1$. The corresponding reduction of on-axis ion temperatures is hence only partially attributed to a decrease of the boundary ion temperature. Noticeably, as the temperature ratio boundary condition is increased from $T_{\mathrm{i}}/T_\mathrm{e}|_\mathrm{bc} = 1.17$, flattening of ion temperature gradients and a reduction of on-axis ion temperatures is observed as well.

In agreement with the observed density peaking as the temperature ratio boundary condition deviates significantly from unity, normalized density gradients are severely overestimated, exceeding experimental gradients by a factor of up to 4 (cf. Fig.\,\ref{fig:10_Gradients_for_various_TiTe_BC}(a)). Unlike normalized temperature gradients, normalized density gradients are highly sensitive to a reduction of the temperature ratio boundary condition, exhibiting noticeable flattening for each step in $T_\mathrm{i}/T_\mathrm{e}|_\mathrm{bc}$ performed towards unity. Furthermore, overestimation of density gradients is observed throughout the entire turbulence dominated region.

In the predictive heat and particle transport simulations carried out, plasma profiles are treated interpretively beyond $\rho_\mathrm{tor} = 0.85$. As the plasma profiles evolved change gradually, the ion to electron temperature ratio is clamped in the vicinity of the simulation boundary, decreasing by typically $\Delta T_\mathrm{i}/T_\mathrm{e} < 0.2$ down to the $q=1$ surface. The decrease of the temperature ratio across the turbulence dominated region is particularly large for values of the boundary condition around or below unity. Under these conditions, a decrease of $\Delta T_\mathrm{i}/T_\mathrm{e} \gtrsim 0.2$ is observed as opposed to a decrease of $\Delta T_\mathrm{i}/T_\mathrm{e} \lesssim 0.1$ in the case of ion temperatures noticeably exceeding electron temperatures. Consequently, the effect of an elevated temperature ratio boundary condition propagates inwards particularly well, thus reinforcing the decrease of particle transport in the turbulence dominated region.

As presented, increased density peaking is predicted by JETTO-QuaLiKiz for an ion to electron temperature ratio boundary condition significantly exceeding unity. A physical interpretation of this phenomenon is given in the next section.

\begin{figure*}[h!]
	\centering
	\includegraphics[width=.75\linewidth]{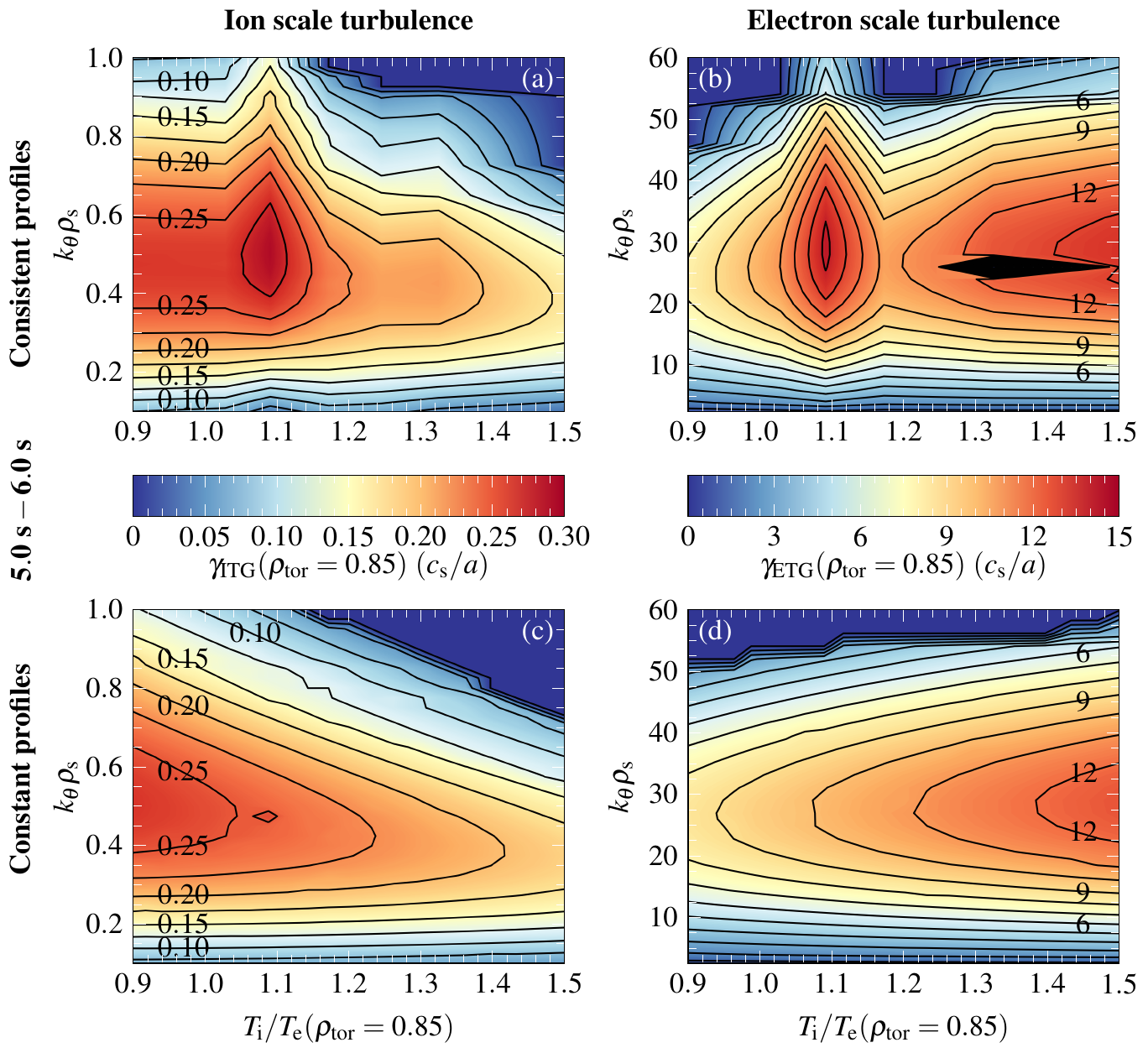}
	\stdcap{}{\label{fig:11_Growth_rates_for_various_TiTe_BC}%
		Growth rates $\gamma$ at the simulation boundary at $\rho_{\mathrm{tor}} = 0.85$ for (a,c) ion and (b,d) electron scale turbulence calculated by stand-alone QuaLiKiz for varying values of the $T_{\mathrm{i}}/T_{\mathrm{e}}$ boundary condition using plasma parameters of the \Snd time slice analysed. In (a,b), converged plasma profiles obtained with JETTO-QuaLiKiz for different values of $T_{\mathrm{i}}/T_{\mathrm{e}}|_{\mathrm{bc}}$ are used to calculate turbulence growth rates. In (c,d), plasma parameters from simulation results for $T_{\mathrm{i}}/T_{\mathrm{e}} = 1.17$ are used for the same calculations with $T_{\mathrm{i}}/T_{\mathrm{e}}$ set independently.
		}
\end{figure*}

\subsubsection{Observations from gyrokinetic calculations}
\label{sec:5.1.2_Observations_from_gyrokinetic_calculations}

The impact of the ion to electron temperature boundary condition on both ion and electron scale turbulence present is analysed by stand-alone QuaLiKiz utilizing the converged plasma profiles calculated by JETTO-QuaLiKiz for different ratios $T_\mathrm{i}/T_\mathrm{e}|_\mathrm{bc}$. Under these conditions, ion scale turbulence is stabilized qualitatively as the temperature ratio boundary condition is increased from just below unity up to the value $T_\mathrm{i}/T_\mathrm{e}|_\mathrm{bc} = 1.50$, corresponding to the upper bound obtained from Gaussian process regression (see Fig.\,\ref{fig:11_Growth_rates_for_various_TiTe_BC}(a)). In the latter case, the maximum instability growth rate is noticeably reduced. Additionally, turbulence is driven unstable over fewer wavenumbers. The opposite effect is observed for electron scale turbulence. As the ion temperature boundary condition exceeds progressively the electron temperature, instability growth rates generally increase for all wavenumbers (see Fig.\,\ref{fig:11_Growth_rates_for_various_TiTe_BC}(b)), corresponding to an increase of electron scale turbulence.

Note, that converged self-consistent profiles of predictive heat and particle transport simulations are used as input for stand-alone QuaLiKiz. As a result, simulation parameters for different values of the temperature ratio boundary condition differ slightly in additional plasma parameters, most importantly in the normalized gradients prescribed. As the turbulence sensitivity analysis performed consequently does not solely depend on $T_\mathrm{i}/T_\mathrm{e}|_\mathrm{bc}$, the growth rates calculated for both ion and electron scale turbulence do not change monotonously for all wavenumbers considered, an exception being in particular the case $T_\mathrm{i}/T_\mathrm{e}|_\mathrm{bc} = 1.10$.

To emphasize the effect of a variation in the temperature ratio boundary condition, additional calculations are performed with QuaLiKiz, utilizing the converged profiles from predictive transport simulations for $T_\mathrm{i}/T_\mathrm{e}|_\mathrm{bc} = 1.17$ and manually adjusting boundary electron and ion temperature only. Consequently, all remaining plasma parameters are identically constant throughout this analysis, including the normalized gradients used. Following this approach, ion scale modes are again found to be stabilized monotonously, electron scale modes destabilized monotonously as the temperature ratio increases (see Figs.\,\ref{fig:11_Growth_rates_for_various_TiTe_BC}(c,d)). Sensitivity of the growth rates on the ion to electron temperature ratio is expected as this dimensionless parameter is introduced in the gyrokinetic dispersion relation solved by QuaLiKiz through summation of each particle species' Vlasov equation in the formulation of quasineutrality (cf. e.g. Eq.\,(4) in Ref.\,\citen{Bourdelle16}). A reduction of ITG dominated modes with an increase in $T_\mathrm{i}/T_\mathrm{e}$ is expected from analytical considerations, as the linear threshold for the emergence of these modes is shifted to larger values\cite{Guo93,Casati08}. Similarly, the threshold of ETG modes is reduced under these conditions, resulting in a destabilization of electron scale modes as $T_\mathrm{i}/T_\mathrm{e}$ is increased\cite{Jenko01}.

Compared to the calculations performed using input parameters obtained from converged plasma profiles for all values of \TiTeE , growth rates of both methods are in qualitative agreement, especially when neglecting the case with $T_\mathrm{i}/T_\mathrm{e}|_\mathrm{bc} = 1.10$. Consequently, the observed stabilization of ion scale modes and destabilization of electron scale modes when using plasma parameters from transport simulations can indeed be attributed to a change in \TiTe instead of an unfavourable change of the gradients prescribed.

Following the stabilization of ITG dominated modes as the temperature ratio boundary increases progressively from unity, a reduction in the particle diffusivity and the outward particle flux is observed for both cases, where either self-consistent or constant plasma profiles are used as input for stand-alone QuaLiKiz. Even though ETG modes are destabilized in the process, no increase in particle fluxes occurs as short wavelength turbulence does not drive particle transport. Consequently for temperature ratio boundary conditions significantly exceeding unity, turbulent particle transport is noticeably underestimated, resulting in the formation of peaked density profiles.

\begin{figure*}
	\centering
	\includegraphics[width=\linewidth]{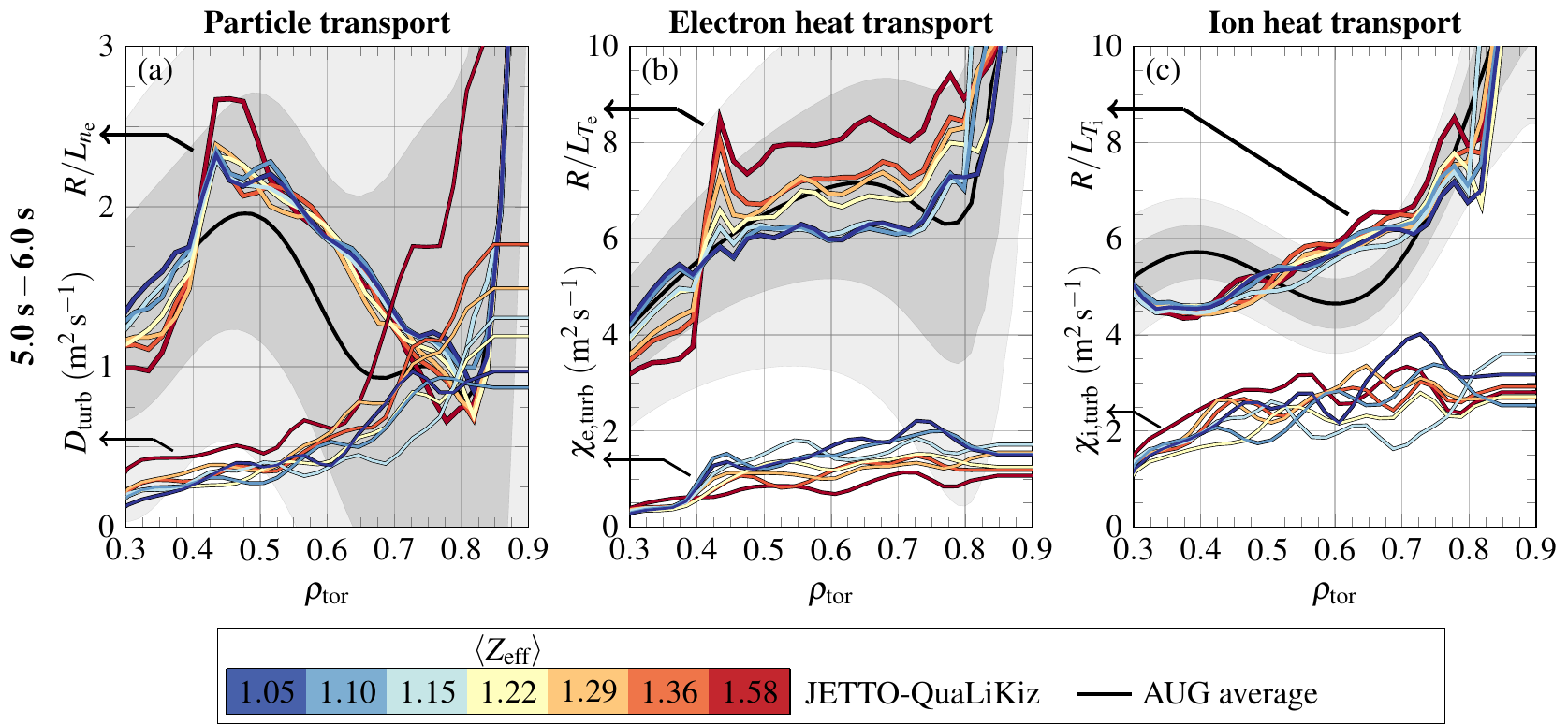}
	\stdcap{}{\label{fig:12_Simulations_for_various_Zeff}%
		Predictive particle and heat transport simulations for different values of the average effective charge $\left\langle Z_\mathrm{eff} \right\rangle$, gradually increasing from $\left\langle Z_\mathrm{eff}\right\rangle = 1.05$ (blue) to $1.58$ (red), corresponding to experimental range of uncertainty. Normalized gradients (coloured, thick) of the (a) electron density, (b) electron temperature and (c) ion temperature are compared to the experimental average (black) with confidence intervals of $1\,\sigma$ and $2\,\sigma$ (grey). Corresponding transport coefficients (coloured, thin) in (a-c) are averaged over the last 0.25\,s of steady-state plasma evolution to account for fluctuations in these parameters. 
		}
\end{figure*}

The electron temperature response to a variation of the temperature ratio boundary condition is determined by the collective effects of stabilization of ITG dominated modes and ETG destabilization. As the increase in electron heat transport driven by destabilized short wavelength turbulence outweighs the reduced drive due to ion scale turbulence stabilization significantly, electron heat transport is noticeably enhanced for values of \TiTe exceeding unity. Consequently, a reduction of the central electron temperature is observed.

In the case of ion heat transport, the stabilization of ITG dominated modes with increasing temperature ratio boundary condition reduces turbulent contributions. As a result, ion temperature peaking is encountered. However in contrast to the steady increase in core density as \TiTe progressively exceeds unity, increasing net electron heat exchange counters further peaking of the ion temperature profiles, thus stabilizing ion temperature gradients for $T_\mathrm{i}/T_\mathrm{e}|_\mathrm{bc} \gtrsim 1.1$ and noticeably reducing ion temperature peaking for $T_\mathrm{i}/T_\mathrm{e}|_\mathrm{bc} \gtrsim 1.2$. Further ion temperature peaking due to stabilization of ITG dominated modes and further electron temperature flattening due to ETG destabilization is thus countered by net ion to electron heat transfer. Therefore, heat transport predictions are less sensitive to temperature ratio boundary conditions noticeably exceeding unity. In the case of particle transport however, enhanced transport due to stabilization of ion scale turbulence is not compensated by another mechanism, resulting in the severe density peaking observed.

\subsubsection{Implications for predictive transport simulations}
\label{sec:5.1.3_Implications_for_transport_simulations}
The simulations discussed in Sec.~\ref{sec:5.1.1_Observations_from_transport_simulations} demonstrate a strong sensitivity of particle transport on the $T_\mathrm{i}/T_\mathrm{e}$ boundary condition, advocating thorough analysis of experimental data to prescribe reasonable values. Yet, even for variations of $T_\mathrm{i}/T_\mathrm{e}|_\mathrm{bc}$ within experimental uncertainties of temperature measurements, significant changes in particle transport are encountered as ion-scale modes are observed to be stabilized with an increase of the temperature ratio boundary condition, particularly for $T_\mathrm{i}/T_\mathrm{e}|_\mathrm{bc} > 1.2$. Simultaneously, electron-scale modes are destabilized. In future work, the sensitivity of simulation results to a variation of the $T_\mathrm{i}/T_\mathrm{e}$ boundary condition within experimental uncertainties could thus be improved by imposing further experimental constraints from turbulence fluctuation measurements at low and high $k$ in the validation (see e.g. Ref.~\citen{Rhodes11} for a validation at DIII-D), using e.g. the Correlation Electron Cyclotron Emission diagnostic at AUG (see Ref.~\citen{Freethy18}).

\begin{figure*}[h!]
	\centering
	\includegraphics[width=0.85\linewidth]{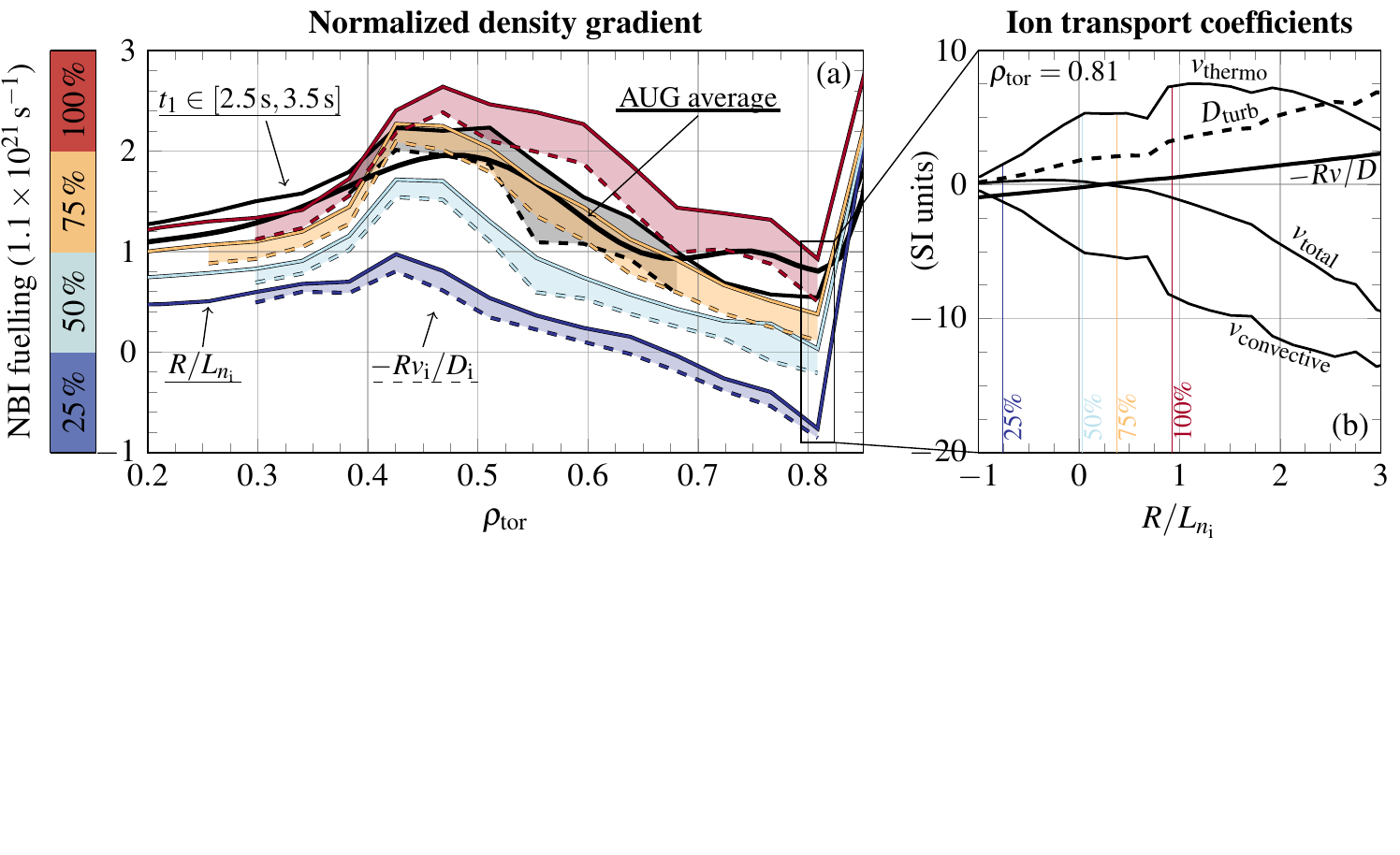}
	\vspace*{-3.2cm}
	\stdcap{}{\label{fig:13_Simulations_for_varying_NBI_fuelling}%
		Predictions by JETTO-QuaLiKiz in the time slice $t_2 \in \left[ 5.0\,\mathrm{s}, 6.0\,\mathrm{s} \right]$ for varying amounts of NBI fuelling (between one and four averaged NBI injectors; coloured blue to red) of (a) the core main ion density gradient $R/L_{n_\mathrm{i}}$ (solid) compared to the experimental average (thick black) and to the ratio of turbulent convective to turbulent diffusive main ion transport $-Rv_\mathrm{i}/D_\mathrm{i}$ (dashed) in each case. The difference (shaded) between predicted $R/L_{n_\mathrm{i}}$ and $-Rv_\mathrm{i}/D_\mathrm{i}$ denotes the source term (cf. Eq.\,\eqref{eq:3_Steady_state_particle_balance}). The same is additionally shown for time slice $t_1 \in \left[ 2.5\,\mathrm{s}, 3.5\,\mathrm{s}\right]$ utilizing three NBI injectors (thin black). (b) Main ion transport coefficients calculated by stand-alone QuaLiKiz at $\rho_\mathrm{tor} = 0.81$ for varying $R/L_{n_\mathrm{i}}$: turbulent diffusivity $D_\mathrm{turb}$, thermodiffusive pinch $v_\mathrm{thermo}$, pure convective term $v_\mathrm{convective}$, total pinch $v_\mathrm{total}$, as well as the ratio of turbulent convective to turbulent diffusive transport $-Rv_\mathrm{i}/D_\mathrm{i}$. The values of $R/L_{n_\mathrm{i}}(\rho_\mathrm{tor}=0.81)$ of the JETTO-QuaLiKiz simulations in Fig.\,\ref{fig:13_Simulations_for_varying_NBI_fuelling}(a) are highlighted (vertical, coloured).
		}
\end{figure*}

\subsection{ETG stabilization by impurities}
\label{sec:5.2_ETG_stabilization_by_impurities}

\newcommand{\footnoteFiveTwo}{%
	Note, that the density profile of C impurities present is evolved by SANCO throughout the entire plasma from a radially constant profile, reaching steady state within 0.2\,s of plasma evolution.
	}

Electron heat transport simulations are found moderately sensitive to the average effective charge prescribed for this discharge\footnote{\footnoteFiveTwo}. To investigate the underlying dependencies in simulations of the \Snd time slice, the value of the effective charge applied is varied within experimental uncertainties of impurity density analysis and Bremsstrahlung estimates, ranging from $Z_\mathrm{eff} = 1.05$ to $1.58$ (see Sec.~\ref{sec:3.3_Treatment_of_impurities}).

With an increase in impurity content, electron temperature peaking is observed, as illustrated by the corresponding normalized gradients in Fig.\,\ref{fig:12_Simulations_for_various_Zeff}. A noticeable reduction in turbulent heat transport is achieved with the introduction of additional C impurities, resulting in monotonously increasing gradients in the turbulence dominated region for all values of $\left\langle Z_\mathrm{eff} \right\rangle$ considered (see Fig.\,\ref{fig:12_Simulations_for_various_Zeff}(b)). With respect to the simulations performed for the cleanest plasma with $Z_\mathrm{eff} = 1.05$, on-axis electron temperature is increased by 0.7\,keV ($+20\,\%$) when applying an effective charge of $1.58$. Both particle and ion heat transport are less affected, as on-axis values of the corresponding profiles are found slightly decreased ($-0.2\times 10^{-19}~\mathrm{m}^{-3}$, being 2.4\%) or increased ($+0.2~$keV, being 4.8\%), respectively, under these conditions. Correspondingly, variations in normalized gradients predicted are less distinct.
	
Analysing the influence of microturbulence on the profile peaking observed, turbulent electron heat transport is significantly reduced when the impurity content increases progressively, as demonstrated by the change in turbulent heat diffusivity (see Fig.\,\ref{fig:12_Simulations_for_various_Zeff}(b)). In the case of the highest average effective charge of $\left\langle Z_\mathrm{eff} \right\rangle = 1.58$ considered, heat diffusivity is reduced by $0.5 - 1.0\,\mathrm{m}^{2}\,\mathrm{s}^{-1}$ throughout the turbulence dominated region as compared to the situation of a clean deuterium plasma ($Z_\mathrm{eff} = 1.05$). Simultaneously, no clear change in neither particle nor ion heat turbulent transport coefficients is generally observed for an increase in the average effective charge applied (see Fig.\,\ref{fig:12_Simulations_for_various_Zeff}(a,c)).

As only turbulent electron heat transport is noticeably affected by an increase in impurity content, ETG modes are found stabilized under these conditions. Consequently, steep electron temperature gradients are required to drive electron-scale turbulence unstable, in agreement with expectations\cite{Jenko01}, resulting in a peaked electron temperature profile. In the process, the net heat exchange from ions to electrons in the case of a plasma with $Z_\mathrm{eff} = 1.05$ is steadily reduced, reversing direction to net ion heating for $\left\langle Z_\mathrm{eff} \right\rangle \geq 1.15$. The resulting peaking in ion temperature is sufficient to drive ITG dominated modes further unstable, countering the stabilization of ITG dominated modes by an increase in  $\left\langle Z_\mathrm{eff} \right\rangle$, thus leaving these modes roughly unaffected. These observations are confirmed by stand-alone QuaLiKiz using the converged profiles of predictive simulations with JETTO-QuaLiKiz. The density peaking observed is driven by a significant increase in the inward neoclassical pinch, ranging from around $3\,\mathrm{cm}\,\mathrm{s}^{-1}$ for a the cleanest plasma considered to up to $8\,\mathrm{cm}\,\mathrm{s}^{-1}$ when applying $\left\langle Z_\mathrm{eff} \right\rangle = 1.58$.

Applying an average effective charge within errorbars of the impurity density analysis estimates by the CXRS diagnostic to the simulations, i.e. $\left\langle Z_\text{eff} \right\rangle = 1.25 \pm 0.09$ and $1.22 \pm 0.10$ in each time slice, good agreement between predictive heat and particle transport simulations and the experimental average is achieved. Only under application of the upper bounds of Bremsstrahlung estimates of $Z_\mathrm{eff}$, noticeable disagreement between simulations and experiment is observed.

\subsection{Core density dependence on NBI fuelling}
\label{sec:5.3_Core_density_dependence_on_NBI_fuelling}
At $t=3.5\,$s of the discharge, an additional 60\,kV NBI injector is coupled to the plasma, supplementing the three NBI sources already employed. In the process, plasma fuelling and heating are increased by $3.1\times 10^{20}\,\mathrm{s}^{-1}$ and $2.4\,\mathrm{MW}$, respectively. Simultaneously, period and saturation fraction of the central, saturated $(1,1)$ MHD mode is increased. Under these conditions, the electron density is slightly reduced by $(0.25 \pm 0.03)\times 10^{19}\,\mathrm{m}^{-3}$ at all radial positions of the core plasma ($- 3.3\,\%$ on-axis), as captured by measurements from TS (see Fig.\,\ref{fig:5_Application_of_GPR}(a,e)). Correspondingly, gradients change only to a minor extent between both phases of the discharge, being virtually unaltered in the turbulence dominated region. A slight decrease in core electron density is regularly observed in AUG H-mode discharges when auxiliary heating is provided to a larger fraction by NBI as compared to the contributions of ECRH to total heating\cite{Sommer12, Sommer15}.

\begin{figure*}
	\centering
	\includegraphics[width=1.3\linewidth]{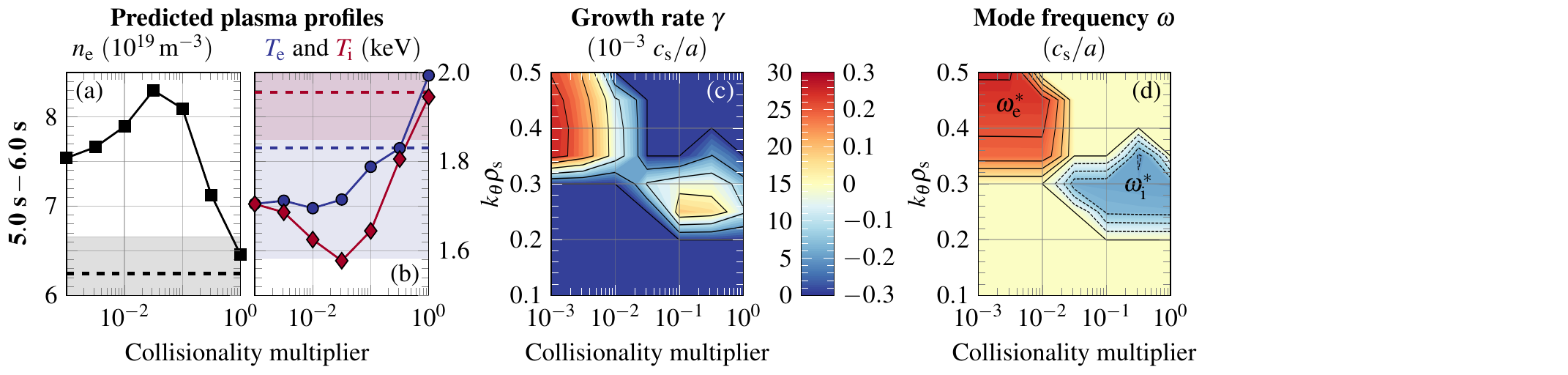}
	\stdcap{}{\label{fig:14_Simulations_for_varying_collisionality}%
		Predictive heat and particle transport simulations performed by JETTO-QuaLiKiz by progressively reducing a multiplier in the calculation of the collisionality used in QuaLiKiz at $\rho_\mathrm{tor} = 0.52$ in the \Snd time slice $t_2 \in \left[ 5.0\,\mathrm{s}, 6.0\,\mathrm{s} \right]$. Calculated (a) electron density (black squares), (b) electron temperature (blue circles), and ion temperature (red diamonds) compared to the experimental averages (dashed, same colour) and associated uncertainties of $1\,\sigma$. Instability growth rate (c) and mode frequency (d) calculated by stand-alone QuaLiKiz,  utilizing converged plasma profiles of JETTO-QuaLiKiz simulations with a collisionality multiplier of unity. Note, that the ion diamagnetic drift frequency $\omega_\mathrm{i}^{*}$ is defined negative, whereas the electron diamagnetic drift frequency $\omega_{\mathrm{e}}^{*}$ is defined positive in QuaLiKiz.
		}
\end{figure*}

The minor reduction in plasma density observed experimentally is not described by the predictive heat and particle transport simulations performed. Instead, an increase in density peaking is predicted in the \Snd time slice (see Fig.\,\ref{fig:13_Simulations_for_varying_NBI_fuelling}(a)). Analysing the relevant contributions of the steady state particle balance
	\begin{align}
		\label{eq:3_Steady_state_particle_balance}
		R/L_n &= \frac{R\,S_n}{n D} - \frac{Rv}{D} 
	\end{align}
in the presence of a source $S_n$, the ratio of convective to diffusive main ion transport in the turbulence dominated region increases in the \Snd phase of the discharge (see Fig.\,\ref{fig:13_Simulations_for_varying_NBI_fuelling}(a)) since the increase in the inward pinch outweighs the general increase of the diffusivity. Simultaneously, the stronger source $S_n$ is countered by the increased diffusivity and density. This suggests that the increase in $R/L_n$ in the \Snd phase of the discharge is primarily driven by an over proportional increase of the turbulent inward pinch. Note that only turbulent contributions to the total inward pinch and diffusivity are considered as neoclassical contributions are negligible ($v_\mathrm{nc} \sim - 0.05\,\mathrm{m}\,\mathrm{s}^{-1}$, $D_\mathrm{nc} \sim 0.02\,\mathrm{m}^2\,\mathrm{s}^{-1}$).  

Performing simulations of the \Snd phase of the discharge with a particle source artificially reduced to 75\,\%, mimicking the source of time slice $t_1$, good agreement in particle transport is achieved in the turbulence dominated region as compared to the experimental average (see Fig.\,\ref{fig:13_Simulations_for_varying_NBI_fuelling}(a)). Note that transport agreement inside the $q=1$ surface is not obtained, since additional transport coefficients prescribed are tailored for simulations incorporating the full particle source. Utilizing only one or two average NBI injectors in the presence of the full 10\,MW of NBI heating, a further flattening of the electron density profile is predicted, highlighting the sensitivity of the simulations on the particle source employed. With a progressive reduction of the particle source throughout this analysis, an increase in both electron and ion temperature occurs.

Reducing only the particle source in the \Snd phase of the discharge, the ratio of turbulent convective to turbulent diffusive main ion transport decreases significantly, yet still constitutes the dominant contribution to the predicted normalized density gradient (see Fig.\,\ref{fig:13_Simulations_for_varying_NBI_fuelling}(a)). Simulations by stand-alone QuaLiKiz of the steady-state solution obtained with JETTO-QuaLiKiz identify a strong sensitivity of both the thermodiffusive pinch and the pure convective term on the particle source $S_n$ employed, or rather on the normalized density gradient $R/L_n$ as observed in additional simulations with stand-alone QuaLiKiz where only $R/L_n$ was modified (see Fig.\,\ref{fig:13_Simulations_for_varying_NBI_fuelling}(b)). The latter simulations suggest an increased sensitivity of the density predictions on the source term through modification of the initial $R/L_{n}$ (including the electron density gradient) and thus of the aforementioned pinches. This occurs due to modifications of the electron source. In multiple ion simulations with fixed electron source and electron density gradient, where only the ion sources are modified, ion density peaking in the ITG regime is found to be transport-dominated independent of the ion source\cite{Bourdelle18}.

\subsection{Particle transport reduction with reduced collisionality}
\label{sec:5.4_Particle_transprot_reduction_with_reduced_collisionality}
In this study, the effect of a reduction of the collisionality utilized in QuaLiKiz on core profile predictions was additionally analysed to evaluate the applicability of the Krook collision operator employed in QuaLiKiz\cite{Romanelli07} to AUG conditions. For this purpose, a free parameter is introduced in the calculation of the collisionality. Consequently, consistent evaluation of the collisionality in accordance with plasma parameters is ensured, yet allowing for an arbitrary modification of collisionality.

Reducing the multiplier from unity down to $10^{-3}$, an increase in electron density well above experimental levels is observed in the turbulence dominated region (see Fig.\,\ref{fig:14_Simulations_for_varying_collisionality}(a)).  Simultaneously, a decrease in both electron and ion temperature below the experimental average occurs in the same region (see Fig.\,\ref{fig:14_Simulations_for_varying_collisionality}(b)). Noticeably, deviations of the plasma profiles predicted are greatest for values of the collisionality multiplier between $10^{-2}$ and $10^{-1}$. Note, that the plasma response presented in Fig.~\ref{fig:14_Simulations_for_varying_collisionality} at $\rho_\mathrm{tor} = 0.52$ is representative of the behaviour observed throughout the simulation domain.

Analysing turbulence with stand-alone QuaLiKiz\footnote{Stand-alone QuaLiKiz simulations of varying collisionality were carried out utilizing converged profiles from JETTO-QuaLiKiz simulations with default collisionality to isolate the impact of changing collisionality.}, ion scale modes are stabilized as the collisionality multiplier is reduced from unity to values between $10^{-2}$ and $10^{-1}$ (see Fig.~\ref{fig:14_Simulations_for_varying_collisionality}(c)), thus reducing particle transport. For even smaller values, ion scale modes are destabilized again. However as collisionality is low, de-trapping of electrons in banana orbits is less efficient. Hence, particle transport is still found reduced as compared to the default situation. With a reduction of the collisionality multiplier, the emergence of TEM dominated modes is predicted by QuaLiKiz whereas ITG dominated modes are found stable (see Fig.\,\ref{fig:14_Simulations_for_varying_collisionality}(d)) in the presence of a larger population of trapped electrons. Reproduction of increased density peaking with reduced collisionality as the mode frequency becomes closest to zero in absolute terms is in agreement with other works\cite{Fable10}, thus increasing the confidence in the applicability of the reduced gyrokinetic code QuaLiKiz. 

Even though the heat and particle transport predictions performed are found sensitive to the value of the collisionality multiplier used, the good agreement between simulations and experiment discussed in Sec.\,\ref{sec:4.1_Modelling_by_JETTO-QuaLiKiz} is obtained only when imposing a collisionaliy multiplier close to its default value of unity. Since the quasilinear fluxes calculated by QuaLiKiz were validated using JET parameters, applicability of the Krook collision operator used in QuaLiKiz to conditions of AUG discharges is an additional success for the reduced code QuaLiKiz.

\begin{figure*}[h!]
	\centering
	\includegraphics[width=.85\linewidth]{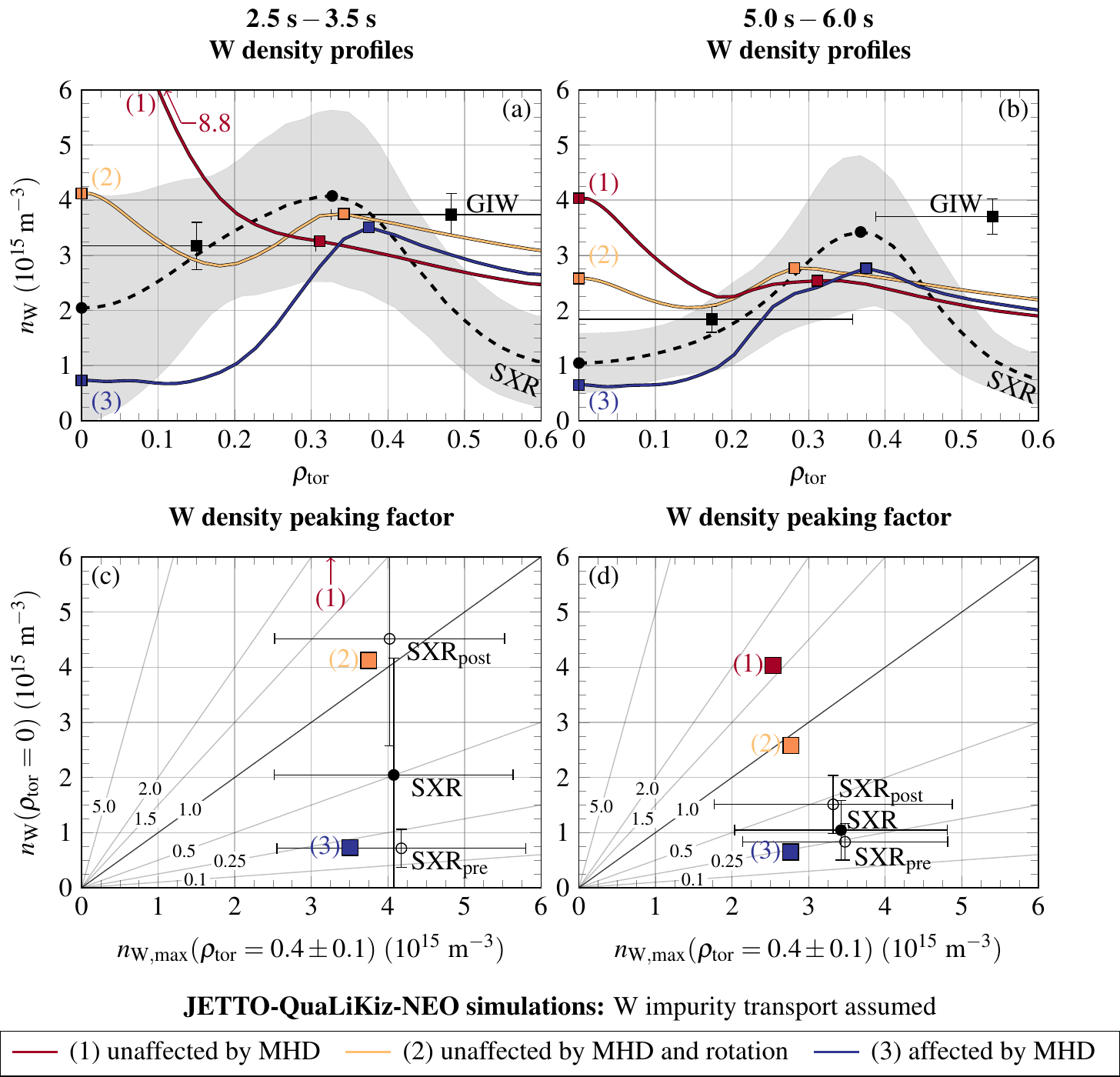}
	\stdcap{}{\label{fig:15_W_transport_simulations}%
		Simulations of trace W impurity transport by JETTO-QuaLiKiz-NEO compared to the experimental average derived from grazing incidence spetrometry (GIW) and soft X-ray Abel-inversion (SXR). Simulations are carried out assuming W impurities to be 1) unaffected by MHD activity (red), 2) unaffected by MHD activity and rotation (yellow) or 3) affected by MHD activity (blue) through prescription of additional central transport. W density profiles predicted in the inner half radius for (a) time slice $t_1 \in \left[ 2.5~\mathrm{s}, 3.5~\mathrm{s} \right]$ and (b) $t_2 \in \left[ 5.0~\mathrm{s}, 6.0~\mathrm{s}\right]$. Note that case 2) is a physically inconsistent sensitivity analysis of case 1), neglecting the impact of poloidal asymmetries. Corresponding W density peaking factors in (c,d), defined as the ratio of on-axis W density $n_\mathrm{W}(\rho_\mathrm{tor} = 0)$ to the maximum W density $n_\mathrm{W,max}(\rho_\mathrm{tor} = 0.4 \pm 0.1)$ in the vicinity of the $q=1$-surface. Additionally, pre- and post-crash averages of SXR measurements are shown.
		}
\end{figure*}

\section{Predictive trace W impurity transport simulations}
\label{sec:6_Predictive_W_transport_simulations}
Transport of W impurities in typical AUG or JET discharges is determined by both neoclassical and turbulent effects. Since W transport is thus influenced by main ion density and temperature profiles, successful modelling of the corresponding transport channels is a vital prerequisite for accurate W transport simulations. As demonstrated in the first part of this study (see Sec.\,\ref{sec:4.1_Modelling_by_JETTO-QuaLiKiz}), JETTO-QuaLiKiz is found capable of predicting main plasma profiles of AUG discharge \#31115 in the turbulence dominated region, thus paving the way for W impurity transport simulations.

\subsection{Setup}
Predictive W transport simulations are performed with the impurity transport code SANCO within JETTO. The temporal evolution of the W density is simulated over 3\,s to ensure reaching steady-state conditions. Turbulent contributions to W transport are calculated by QuaLiKiz, taking poloidal asymmetries of  heavy impurities into account. To account for neoclassical transport in the presence of rotation induced poloidal asymmetries of the W density distribution, the code NEO\cite{Belli08, Belli12} is used. To assess the significance of neoclassical, turbulent and MHD driven transport in the presence of poloidal asymmetries for the avoidance of central W accumulation, the following three cases are modelled (see Table~\ref{tbl:2_W_cases}).

\begin{table}[t!]
	\centering
	\stdcap{}{%
		\label{tbl:2_W_cases}In the simulations carried out, the importance of different mechanisms on W impurity transport is investigated. 
		}
	\begin{tabular}{ccc}
	\toprule
	\multicolumn{3}{l}{W impurity transport is assumed affected by} \\
		& MHD activity & rotation \\
	\midrule	
		1) & {\color{Cmap100}no} & {\color{Cmap0}yes} \\
		2) & {\color{Cmap100}no} & {\color{Cmap100}no} \\
		3) & {\color{Cmap0}yes} & {\color{Cmap0}yes} \\
	\bottomrule
	\end{tabular}
\end{table}

The first case describes W impurity transport incorporating validated theory only, considering both neoclassical and turbulent effects in the presence of poloidal asymmetries. Case 2) represents a sensitivity study of case 1), where rotation and consequently poloidal asymmetries are neglected, thus being inconsistent with known theory. With the final case 3), the impact of additional transport on the W impurity population is explored. 
	
Remaining simulation parameters are kept identical throughout this analysis. The influence of central MHD activity on W impurity transport is assumed identical as for deuterium, i.e. $D_\mathrm{add,W} = D_\mathrm{add,D}$, such that the additional core particle transport coefficients obtained in the previous part of the study can be applied. The additional diffusivity $D_\mathrm{add}$ attributed to MHD activity was prescribed in predictive main ion heat and particle transport simulations inside $\rho_\mathrm{tor} = 0.4$ to obtain agreement between simulated and effective main ion particle transport. In W impurity transport simulations, the additional transport is expressed as a convective velocity through $v_\mathrm{add} = - D_\mathrm{add} \nabla_\rho n_\mathrm{e}/n_\mathrm{e}$. For the calculation of the MHD attributed outward velocity, the steady-state density profiles predicted by JETTO-QuaLiKiz are used.  The impact of MHD activity on different transport channels is taken as either convective (W impurities) or diffusive (main ion heat and particles) based on technical reasons only. Expressing the effect of central MHD activity in terms of a convective velocity $v_\mathrm{add}$ instead of a diffusivity $D_\mathrm{add}$, consistent results for the main ion heat and particle transport simulations are obtained. 

For simplicity, the W population is treated in the trace limit, thus assuming no influence of W impurities on the evolution of main ion profiles. Hence, the impurity radiation calculated by SANCO is not used to solve the energy balance equations, the impact of plasma dilution and collisional effects are neglected. Instead, converged plasma profiles obtained through predictive modelling with JETTO-QuaLiKiz under consideration of experimental measurements of impurity radiation and effective charge (both including contributions of W impurities) can be used interpretively. 

The W impurity population prescribed in the transport simulations carried out is obtained from Abel-inversion of SXR-emissivity considering photons with energies exceeding 1~keV\cite{Sertoli15_2}. Thus, the W impurity density in the outer region of the plasma is challenging to estimate. The same holds for the total W impurity content. Hence, simulations carried out are not intended to quantitatively describe central measurements, but aim at qualitatively reproducing observations.

\begin{figure*}[h!]
	\centering
	\includegraphics[width=.85\linewidth]{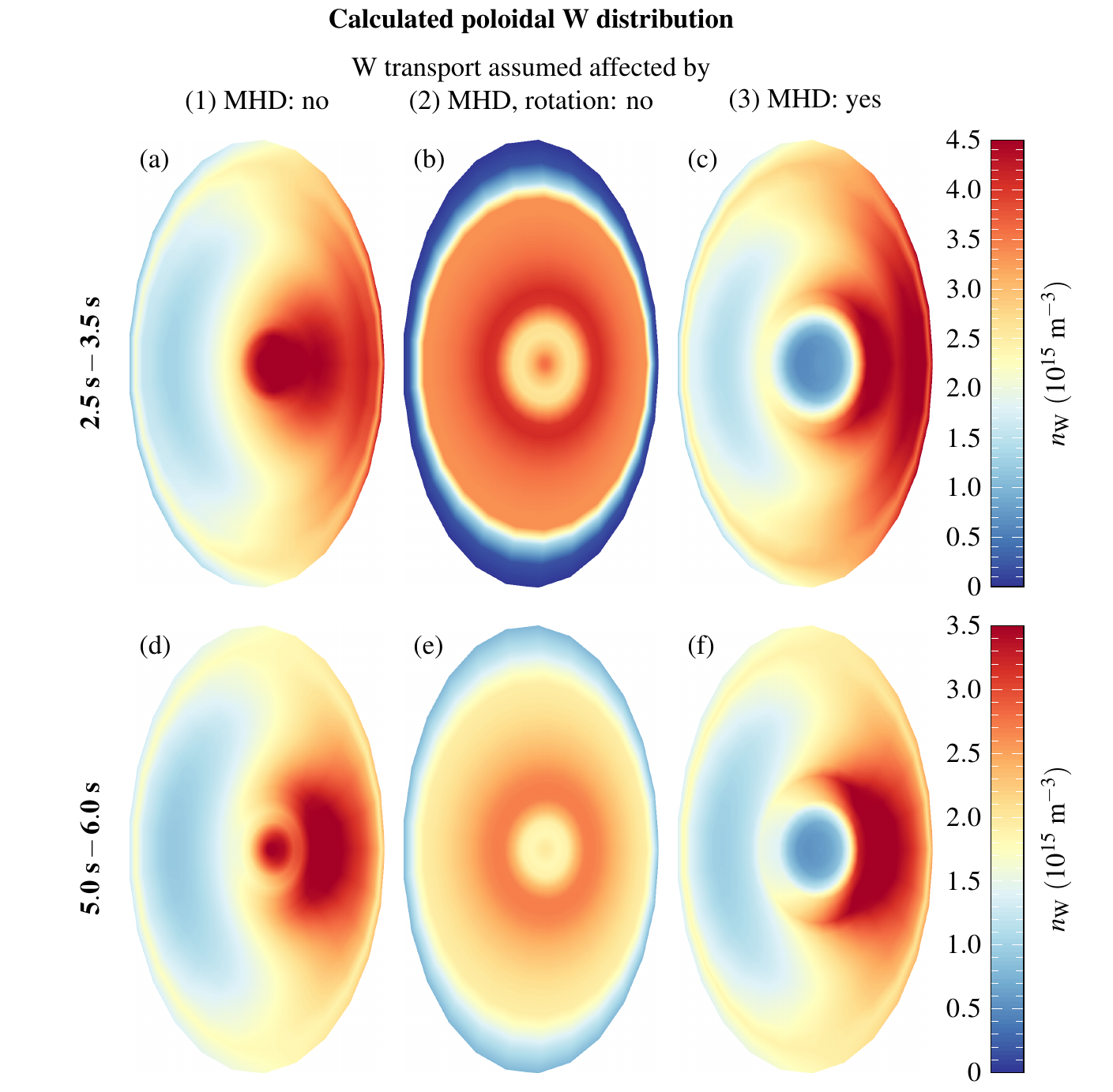}
	\stdcap{}{\label{fig:16_Poloidal_W_distribution}%
		Poloidal distribution of W impurities in AUG \#31115 calculated by JETTO-QuaLiKiz-NEO in both time slices $t_1 \in \left[ 2.5~\mathrm{s}, 3.5~\mathrm{s} \right]$ (top row) and $t_2 \in \left[ 5.0~\mathrm{s}, 6.0~\mathrm{s}\right]$ (bottom row). Simulations are carried out assuming W impurities to be 1) unaffected by MHD activity (left column), 2) unaffected by MHD activity and rotation (centre column) or 3) affected by MHD activity (right column) through prescription of additional central transport. Note that case 2) is a physically inconsistent sensitivity analysis of case 1), neglecting the impact of poloidal asymmetries.
		}
\end{figure*}

\subsection{Results}
Trace W impurity transport simulations carried out by JETTO-QuaLiKiz-NEO are found capable of qualitatively reproducing experimental observations under the assumption of a direct impact of the effects of a saturated $(m,n) = (1,1)$ MHD mode on W impurity transport (see. Fig.~\ref{fig:15_W_transport_simulations}) through prescription of additional central transport coefficients. In both phases of the discharge, where either a short or a long period mode is present, deeply hollow W density profiles are predicted to occur by the set of tools used (see. Figs.~\ref{fig:15_W_transport_simulations}, \ref{fig:16_Poloidal_W_distribution}).

Considering only neoclassical and turbulent W impurity transport in the simulations carried out, corresponding to case 1), significant W density peaking is predicted by JETTO-QuaLiKiz-NEO in contrast to experimental observations of deeply hollow W density profiles. Peaking of W density profiles is especially pronounced in time slice $t_1$, where a peaking factor of 2.7 is predicted\footnote{The W density peaking factor is defined as the ratio of on-axis to maximum W density in the vicinity of the $q=1$ surface, i.e. $n_\mathrm{W}(\rho_\mathrm{tor} = 0)/n_\mathrm{W,max}(\rho_\mathrm{tor} = 0.4 \pm 0.1)$.} (peaking factor of 1.6 in $t_2$). As a result of strong centrifugal forces, W impurities are calculated to be primarily localised on the high-field side (see Figs.~\ref{fig:16_Poloidal_W_distribution}(a,d)).
 
Omitting plasma rotation in the second case considered, a poloidally uniform W impurity distribution is obtained. Central profiles predicted are comparatively flat, with densities on-axis and in the vicinity of the $q=1$ surface being in qualitative agreement. These results are in contrast to both experimental observations but also to simulations incorporating full neoclassical and turbulent effects (first case). Consequently, the impact of rotation induced poloidal asymmetries cannot be neglected in the treatment of W impurity transport.

Analysing the impact of neoclassical effects on the W population in case 2), neoclassical transport is found to reverse direction from net inward transport close to the magnetic axis to net outward transport for $\rho_{\mathrm{tor}} \gtrsim 0.1 - 0.2$. In contrast, experimental measurements imply net outward transport inside $\rho_\mathrm{tor} \sim 0.4$. Still, simulations for the \Snd phase of the discharge predict less inward transport as compared to the \Fst phase. With increased NBI fuelling and heating, electron density flattening and ion temperature peaking are observed experimentally. As a result, the impact of the neoclassical pinch is reduced, while temperature screening becomes more efficient (see e.g. Ref.~\citen{Angioni15}).

Treating neoclassical phenomena consistently in case 1), i.e. taking rotation induced poloidal asymmetries into account, the effect of temperature screening is noticeably reduced in both time slices. Finite neoclassical outward transport at a small magnitude is predicted only for the \Snd phase of the discharge. Thus in the presence of poloidal asymmetries, central W transport is determined by the dominant contribution of the neoclassical pinch. Consideration of poloidal asymmetries for heavy impurity transport is thus mandatory to accurately describe the effect of temperature screening.

Qualitative agreement between simulations and experiment is obtained only when additionally assuming a direct impact of MHD activity on W impurity transport through prescription of additional central transport coefficients, corresponding to case 3). Under these conditions, the strong neoclassical pinch in the presence of rotation induced poloidal asymmetries is countered by additional outward transport, prescribed to mimic the effect of the saturated $(1,1)$ MHD mode. Consequently, deeply hollow W density profiles are calculated by JETTO-QuaLiKiz-NEO in both phases of the discharge. Whereas W density profiles predicted in the \Snd phase of the discharge are in quantitative agreement with average estimates by SXR inside the $q=1$-surface, simulations of the \Fst phase find the W density to be noticeably lower then measurements of the W density averaged between $2.5~$s and $3.5~$s of the discharge (see Figs.~\ref{fig:15_W_transport_simulations}(a,b)). However, averaging only the W density profiles measured prior to sawtooth crashes, agreement between the experimental average constructed and simulations is increased, as illustrated by the corresponding W density peaking factor (see Fig.~\ref{fig:15_W_transport_simulations}(c)).

\subsection{Discussion}
Comparing the different cases of W impurity transport simulated, only under consideration of an additional transport mechanism beside neoclassical phenomena, central W impurity transport can be described in qualitative agreement with observations. The indirect effect of MHD activity on heavy impurity transport through modifications of kinetic profiles and thus of neoclassical transport is insufficient in establishing hollow W density profiles in the simulations carried out. Similar as in the main ion heat and particle transport simulations discussed in Sec.~\ref{sec:4.1_Modelling_by_JETTO-QuaLiKiz}, turbulent heavy impurity transport under consideration of poloidal asymmetries contributes only for $\rho_\mathrm{tor} \gtrsim 0.25$. As both neoclassical and turbulent phenomena predict W density profiles inconsistent with experimental observations, an additional mechanism for central W impurity transport is suspected to be present.

In this work, the additional transport required is attributed to the emergence of a saturated $(m,n) = (1,1)$ MHD mode, as suggested by the temporal correlation of mode activity and W behaviour. The additional transport coefficients prescribed to mimic the impact of this mode on W transport are here assumed identical for deuterium and W. However as shown in both this section and Sec.~\ref{sec:4.1_Modelling_by_JETTO-QuaLiKiz}, experimental central main ion and heavy impurity transport are described in qualitative agreement with experimental observations following this approach. Exploring the possibility of a modification of the additional transport coefficients by either the particle mass or charge, prescribed convective transport would either completely dominate overall central transport or be completely irrelevant. Considering the uncertainties of the simulation results, a dependence on the charge to mass ratio cannot be discarded due to a differences of only a factor 2 comparing both ion species under the present conditions. However, the tools used in this study are not suitable to accurately determine the effect of MHD activity on W impurity transport. Further investigation of the interplay of MHD activity with particle and impurity transport in an integrated framework is thus necessary (for work on non-integrated modelling see e.g. Refs.\,\citen{Garbet16, Ahn16, Nicolas12, Nicolas14}).

Under consideration of additional transport coefficients in simulations of the \Fst phase of the discharge, central W density profiles are slightly more hollow (up to $\rho_\mathrm{tor} \sim 0.2$) than suggested by experimental measurements. As sawtooth crashes are observed in this phase, describing the effect of a sawtooth cycle by a constant, averaged transport coefficient introduces inaccuracies. Experimentally, flattening of both peaked main ion and hollow W density profiles is observed during a sawtooth crash, pointing to a strong diffusive rather than a convective effect of sawteeth\cite{Sertoli15_2}. Given the rapid redistribution of heat and particles, neoclassical W impurity transport is expected to change drastically throughout a sawtooth cycle. Hence, simulations of plasma conditions averaged over several sawtooth cycles do not capture the averaged effect of sawteeth accurately. The W density response throughout a sawtooth cycle will be examined in future work (see e.g. Ref.~\citen{Koechl17}).

Finally, the simulations performed can neither rule out that the additional drive of central outward W impurity transport necessary to explain experimental observations is caused by a mechanism not considered in this study, such as e.g. direct drive by ECRH, nor determine the significance of possible additional mechanisms for outward W impurity transport.

\section{Conclusion and outlook}
\label{sec:7_Conclusion_and_outlook}
In this study, integrated modelling of main ion heat and particle transport in AUG discharge \#31115 was carried out by JETTO-QuaLiKiz for the first time. Good agreement between simulations and experiment is found for all transport channels in the turbulence dominated region, with average deviations of plasma profiles and associated gradients being in the order of $1-18\%$ and $10-26\%$ respectively. While the quasilinear turbulent fluxes calculated by QuaLiKiz have been validated against JET discharges\cite{Bourdelle16}, correct reproduction of experimental conditions in the mid-sized tokamak ASDEX Upgrade demonstrates the capabilities of the reduced model QuaLiKiz. The temperature and density profiles calculated by JETTO-QuaLiKiz are also obtained utilizing ASTRA-QuaLiKiz. For practical purposes, this benchmark between both implementations of QuaLiKiz is excellent. Consequently, the very fast quasilinear gyrokinetic code QuaLiKiz can be utilized for the calculation of turbulent fluxes in integrated transport simulations with even more increased confidence between the $q=1$ surface and the pedestal top.

\begin{figure*}[h!]
	\centering
	\includegraphics[width=.85\linewidth]{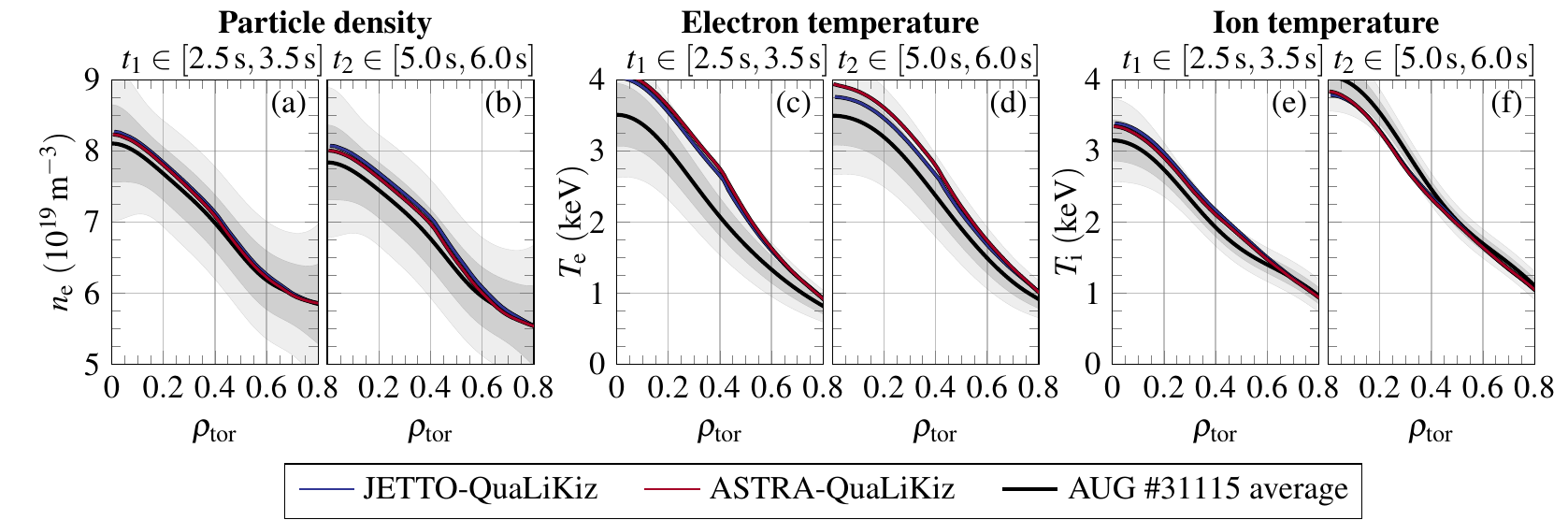}
	\stdcap{}{\label{fig:17_Comparison_ASTRA_and_JETTO}%
		Predictive heat and particle transport simulations performed by JETTO-QuaLiKiz (blue) and ASTRA-QuaLiKiz (red) showing (a,b) electron density, (c,d) electron temperature and (e,f) ion temperature profiles compared to the experimental average of AUG \#31115 (black) with confidence intervals of $1\,\sigma$ and $2\,\sigma$ (grey) in both time slices $t_1\in \left[ 2.5\,\mathrm{s}, 3.5\,\mathrm{s} \right]$ and $t_2 \in \left[ 5.0\,\mathrm{s}, 6.0\,\mathrm{s} \right]$. Note, that the benchmark was carried out with a modified version of JETTO-QuaLiKiz.
		}
\end{figure*}

Thorough analysis and careful interpretation of experimental measurements proved crucial for successful prediction of transport in the turbulence dominated region between the $q=1$ surface at $\rho_\mathrm{tor} \approx 0.4$ and the fixed boundary condition at $\rho_\mathrm{tor} = 0.85$. Utilizing ion and electron temperature measurements without consideration of uncertainties, an averaged ion to electron temperature ratio noticeably exceeding unity is prescribed at the boundary condition. Under these conditions, ITG dominated modes are stabilized and particle transport severely underpredicted as a result. A comparable underprediction of ion heat transport is not observed due to increased net ion to electron heat transfer, as ETG modes are additionally destabilized. Applying Gaussian process regression techniques, reliable estimates of plasma profiles and corresponding uncertainties are obtained from experimental data. Through prescription of modified boundary conditions well within experimental uncertainties, good agreement between simulations and experiment is obtained. As the validation and interpretation of experimental measurements is hence vital for simulation success, following this approach is strongly encouraged as good practice.

Central transport inside the $q=1$ surface is impacted by MHD activity, being a saturated $(m,n) = (1,1)$ mode with a short period of low saturation fraction in the first phase of the discharge and a long period of high saturation fraction in the second one. Neglecting this transport channel leads to severe density and temperature peaking overpredictions. Hence to achieve central transport agreement, additional transport coefficients derived from balance equations are prescribed in the $q=1$ surface to the simulations performed. 

Simulations of trace W impurity transport based on the successful modelling of main ion heat and particle transport are capable of qualitatively reproducing central hollow W density profiles measured in the presence of the saturated $(1,1)$ MHD mode. Neglecting the impact of MHD activity on heavy impurity transport, peaked central W density profiles are predicted by neoclassical heavy impurity transport under consideration of poloidal asymmetries. The impact of MHD on only the main ion profiles is insufficient to create conditions of reversed net neoclassical heavy impurity transport in AUG discharge \#31115. Instead, qualitative agreement with experimental measurements is obtained only when assuming trace W impurities to be impacted by the saturated $(1,1)$ MHD mode through prescription of additional central transport. 

The impact of MHD activity on central transport is derived from balance equations only in this study. However, to advance predictive integrated modelling of discharges such as AUG \#31115, an accurate first-principles based reduced model of MHD driven transport is necessary for all transport channels. Future work should also fully exploit the capabilities of the JINTRAC integrated modelling suite\cite{Romanelli14}, providing not only all codes presented in this study, but also tools for the self-consistent simulation of auxiliary heating and particle sources, of neutrals and of certain MHD phenomena (sawteeth, edge localised modes). Fully self-consistent modelling will additionally have to incorporate the calculation of plasma rotation, as well as the interplay between impurities and main plasma ions through radiation. Regarding the influence of the $(1,1)$ MHD mode on heavy impurity transport, several questions need further attention:
\begin{itemize}
	\item Is the outward trace W impurity transport due to the direct effect of the saturated $(1,1)$ MHD mode or due to the direct effect of ECRH?
	\item Is the outward convective velocity assumed for the MHD related transport identical for both W trace impurities and main ions (as assumed in this study)?
	\item Are further contributions to central W trace impurity transport necessary to match experimental measurements?
\end{itemize}

\section*{Acknowledgements}
The authors are deeply grateful to R.M. McDermott and R. Fischer for assistance with the analysis of experimental data.
DIFFER is a partner in the Trilateral Euregio Cluster TEC. 
This work has been carried out within the framework of the EUROfusion Consortium and has received funding from the Euratom research and training programme 2014-2018 under grant agreement No 633053. The views and opinions expressed herein do not necessarily reflect those of the European Commission. 
	
\appendix
\begin{table*}[h!]
	\centering
	\begin{small}
	\stdcap{}{The simulations presented throughout this work are accessible through the local catalogue manager or the PPF system on the Freia cluster of the UKAEA. Versions of JETTO-QuaLiKiz used are defined below by the first 10 digits of the corresponding git SHA1-key.
		}
	\label{tbl:3_JETTO-QuaLiKiz_versions}
	\begin{tabular}{lllll}
		\toprule 
		Section & Catalogue entry & PPF seq num & Time & Description\\
				& \textit{\small{olinder/jetto/aug/31115/}} & \textit{\small{shot = 31115}} & slice\\
				& & \textit{\small{userid = olinder}} \\
		\midrule
		\multicolumn{5}{l}{\textbf{Simulations with 2f3626e1e2}} \\
		\midrule
		\ref{sec:4.1_Modelling_by_JETTO-QuaLiKiz}
			& apr2618/seq\#1 & \#122 & $t_1$ & original regression results\\
			& apr2718/seq\#1 & \#123 & $t_2$ & original regression results\\
		\midrule
		\ref{sec:4.2_Modified_boundary_conditions}
			& apr2618/seq\#2 & \#124 & $t_1$ & modified boundary conditions\\
			& apr2718/seq\#2 & \#125 & $t_2$ & modified boundary conditions\\
		\midrule
		\ref{sec:4.3_Influence_of_MHD_transport_on_central_profile_agreement}
			& apr0618/seq\#1 & \#126 & $t_1$ & w/o MHD transport\\
			& may1818/seq\#1 & \#127 & $t_2$ & w/o MHD transport\\
		\midrule
		\ref{sec:5.1_Influence_of_TiTe_BC_on_core_transport} 
			& may1818/seq\#2 & \#130 & $t_2$ & $T_\mathrm{i}/T_\mathrm{e}|_\mathrm{bc} = 0.92$\\
			& may1818/seq\#3 & \#131 & $t_2$ & $T_\mathrm{i}/T_\mathrm{e}|_\mathrm{bc} = 1.03$\\
			& may1818/seq\#4 & \#132 & $t_2$ & $T_\mathrm{i}/T_\mathrm{e}|_\mathrm{bc} = 1.10$\\
			& may1818/seq\#5 & \#133 & $t_2$ & $T_\mathrm{i}/T_\mathrm{e}|_\mathrm{bc} = 1.17$\\
			& may1818/seq\#6 & \#134 & $t_2$ & $T_\mathrm{i}/T_\mathrm{e}|_\mathrm{bc} = 1.24$\\
			& may1818/seq\#7 & \#135 & $t_2$ & $T_\mathrm{i}/T_\mathrm{e}|_\mathrm{bc} = 1.32$\\
			& may1818/seq\#8 & \#136 & $t_2$ & $T_\mathrm{i}/T_\mathrm{e}|_\mathrm{bc} = 1.50$\\
		\midrule
		\ref{sec:5.2_ETG_stabilization_by_impurities} 
			& may1818/seq\#9 & \#137 & $t_2$ & $Z_\mathrm{eff} = 1.05$\\
			& may1818/seq\#10& \#138 & $t_2$ & $Z_\mathrm{eff} = 1.10$\\
			& may1818/seq\#11& \#139 & $t_2$ & $Z_\mathrm{eff} = 1.15$\\
			& may1818/seq\#12& \#140 & $t_2$ & $Z_\mathrm{eff} = 1.29$\\
			& may1818/seq\#13& \#141 & $t_2$ & $Z_\mathrm{eff} = 1.36$\\
			& may1818/seq\#14& \#142 & $t_2$ & $Z_\mathrm{eff} = 1.58$\\
		\midrule
		\ref{sec:5.3_Core_density_dependence_on_NBI_fuelling}
			& may1818/seq\#15& \#143 & $t_2$ & NBI fuelling: 75\%\\
			& may1818/seq\#16& \#144 & $t_2$ & NBI fuelling: 50\%\\
			& may1818/seq\#17& \#145 & $t_2$ & NBI fuelling: 25\%\\
		\midrule
		\ref{sec:5.4_Particle_transprot_reduction_with_reduced_collisionality} 
			& may2318/seq\#3 & \#146 & $t_2$ & collisionality: $3.16\times 10^{-1}$\\
			& may2318/seq\#4 & \#147 & $t_2$ & collisionality: $1.00\times 10^{-1}$\\
			& may2318/seq\#5 & \#148 & $t_2$ & collisionality: $3.16\times 10^{-2}$\\
			& may2318/seq\#6 & \#149 & $t_2$ & collisionality: $1.00\times 10^{-2}$\\
			& may2318/seq\#7 & \#150 & $t_2$ & collisionality: $3.16\times 10^{-3}$\\
			& may2318/seq\#8 & \#151 & $t_2$ & collisionality: $1.00\times 10^{-3}$\\
		\midrule
		\ref{sec:6_Predictive_W_transport_simulations}
			& jun0818/seq\#1 & \#152 & $t_1$ & W transp. w/o MHD, rotation\\
			& jun0818/seq\#2 & \#153 & $t_2$ & W transp. w/o MHD, rotation\\
			& jun0818/seq\#3 & \#154 & $t_1$ & W transp. w/o MHD\\
			& jun0818/seq\#4 & \#155 & $t_2$ & W transp. w/o MHD\\
			& jun0818/seq\#5 & \#156 & $t_1$ & W transp. w/ MHD\\
			& jun0818/seq\#6 & \#157 & $t_2$ & W transp. w/ MHD\\
		\midrule
		\multicolumn{5}{l}{\textbf{Simulations with 26c9b722e6}} \\
		\midrule
		\ref{sec:A_Comparison_ASTRA_and_JETTO}
			& may2318/seq\#1 & \#128 & $t_1$ & ASTRA-JETTO comparison\\
			& may2318/seq\#2 & \#129 & $t_2$ & ASTRA-JETTO comparison \\
		\bottomrule
	\end{tabular}
	\end{small}
\end{table*}

\section{Comparison of transport simulations between ASTRA and JETTO}
\label{sec:A_Comparison_ASTRA_and_JETTO}
To ensure a consistent implementation of QuaLiKiz in standard transport codes, the predictive heat and particle transport simulations performed by JETTO-QuaLiKiz for AUG \#31115 discussed in Sec.~\ref{sec:4.2_Modified_boundary_conditions} are repeated exchanging the transport code JETTO for the Automated System for TRansport Analysis (ASTRA)\cite{Pereverzev02, Fable13}. For this purpose, a modified version of JETTO-QuaLiKiz is used to ensure identical treatment of the radial electric field in both ASTRA and JETTO.\footnote{In default JETTO-QuaLiKiz, the radial electric field $E_r$ is calculated from $E_r =  v_\varphi B_\theta  - v_\theta B_\varphi + \frac{\mathrm{d}}{\mathrm{d} r} p_\mathrm{i}/Z_\mathrm{i} e n_\mathrm{i}$, whereas in ASTRA-QuaLiKiz only the term $v_\varphi B_\theta$ is taken into account.} Simulations carried out with this implementation of JETTO-QuaLiKiz are found to be in close agreement with simulations based on the full expression of the radial electric field. The following calculations with ASTRA-QuaLiKiz are carried out for both time slices evolving the steady-state solutions obtained with JETTO-QuaLiKiz until plasma profiles are converged. Since steady-state profiles are calculated usually independently of the initial conditions prescribed, this approach is expected to yield almost identical steady-state solutions as simulations performed utilizing the experimental averaged plasma profiles initially.

The steady-state density and temperature profiles obtained by ASTRA-QuaLiKiz are in good agreement with predictions by JETTO-QuaLiKiz in both time slices analysed (see Fig.\,\ref{fig:17_Comparison_ASTRA_and_JETTO}). Both on-axis deviations and average deviations in the turbulence dominated region between plasma profiles predicted by both implementations are within 4\,\%. Similarly, good agreement is observed in the corresponding transport coefficients calculated by QuaLiKiz inside both transport codes. Consequently, deviations in predicted gradients are small as well, resulting in the good agreement of calculated plasma profiles. Even though agreement is not strictly exact for any of the transport channels, the minor differences observed may not necessarily point towards a difference in implementation of QuaLiKiz between both transport codes. Small differences in predicted plasma profiles may result from a different treatment of the magnetic geometry, from a difference in numerical schemes utilized or from applying different smoothing methods to the transport coefficients obtained.

\section{Overview of JETTO-QuaLiKiz simulations and versions used}
\label{sec:B_Overview_of_JETTO-QuaLiKiz_simulations}
The simulations presented throughout this work were carried out on the Freia cluster of the UK Atomic Energy Authority (UKAEA) and are accessible through the local catalogue manager and the Processed Pulse File (PPF) system. Catalogue entries and PPF sequence numbers are provided in Table\,\ref{tbl:3_JETTO-QuaLiKiz_versions} for each simulation discussed. The overview of simulations performed is given separated by JETTO version used, which is defined by the first 10 digits of the corresponding git SHA1-key.
	
	\footnotesize{\bibliography{references}}

\begin{thebibliography}{10}

\bibitem{Bolt04}
H.~Bolt, V.~Barabash, W.~Krauss, J.~Linke, R.~Neu, S.~Suzuki, N.~Yoshida, and
  ASDEX~Upgrade Team.
\newblock Materials for the plasma-facing components of fusion reactors.
\newblock \href{http://dx.doi.org/10.1016/j.jnucmat.2004.04.005}{{\em J. Nucl.
  Mater.}} \textbf{329-333}, 66 (2004).

\bibitem{Causey02}
R.A. Causey.
\newblock Hydrogen isotope retention and recycling in fusion reactor
  plasma-facing components.
\newblock \href{http://dx.doi.org/10.1016/S0022-3115(01)00732-2}{{\em J. Nucl.
  Mater.}} \textbf{300}, 91 (2002).

\bibitem{Philipps11}
V.~Philipps.
\newblock Tungsten as material for plasma-facing components in fusion devices.
\newblock \href{http://dx.doi.org/doi.org/10.1016/j.jnucmat.2011.01.110}{{\em
  J. Nucl. Mater.}} \textbf{415}, S2 (2011).

\bibitem{Neu07}
R.~Neu, M.~Balden, V.~Bobkov, R.~Dux, O.~Gruber, A.~Herrmann, A.~Kallenbach,
  M.~Kaufmann, C.~F. Maggi, H.~Maier, H.W. Müller, T.~Pütterich, R.~Pugno,
  V.~Rohde, A.~C.~C. Sips, J.~Stober, W.~Suttrop, C.~Angioni, C.~V. Atanasiu,
  W.~Becker, K.~Behler, K.~Behringer, A.~Bergmann, T.~Bertoncelli, R.~Bilato,
  A.~Bottino, M.~Brambilla, F.~Braun, A.~Buhler, A.~Chankin, G.~Conway, D.~P.
  Coster, P.~de. Marné, S.~Dietrich, K.~Dimova, R.~Drube, T.~Eich,
  K.~Engelhardt, H.-U. Fahrbach, U.~Fantz, L.~Fattorini, J.~Fink, R.~Fischer,
  A.~Flaws, P.~Franzen, J.~C. Fuchs, K.~Gál, M.~García. Muñoz,
  M.~Gemisic-Adamov, L.~Giannone, S.~Gori, S.~da. Graca, H.~Greuner, A.~Gude,
  S.~Günter, G.~Haas, J.~Harhausen, B.~Heinemann, N.~Hicks, J.~Hobirk,
  D.~Holtum, C.~Hopf, L.~Horton, M.~Huart, V.~Igochine, S.~Kálvin, O.~Kardaun,
  M.~Kick, G.~Kocsis, H.~Kollotzek, C.~Konz, K.~Krieger, T.~Kurki-Suonio,
  B.~Kurzan, K.~Lackner, P.~T. Lang, P.~Lauber, M.~Laux, J.~Likonen, L.~Liu,
  A.~Lohs, K.~Mank, A.~Manini, M.-E. Manso, M.~Maraschek, P.~Martin, Y.~Martin,
  M.~Mayer, P.~McCarthy, K.~McCormick, H.~Meister, F.~Meo, P.~Merkel,
  R.~Merkel, V.~Mertens, F.~Merz, H.~Meyer, M.~Mlynek, F.~Monaco, H.~Murmann,
  G.~Neu, J.~Neuhauser, B.~Nold, J.-M. Noterdaeme, G.~Pautasso, G.~Pereverzev,
  E.~Poli, M.~Püschel, G.~Raupp, M.~Reich, B.~Reiter, T.~Ribeiro, R.~Riedl,
  J.~Roth, M.~Rott, F.~Ryter, W.~Sandmann, J.~Santos, K.~Sassenberg,
  A.~Scarabosio, G.~Schall, J.~Schirmer, A.~Schmid, W.~Schneider, G.~Schramm,
  R.~Schrittwieser, W.~Schustereder, J.~Schweinzer, S.~Schweizer, B.~Scott,
  U.~Seidel, F.~Serra, M.~Sertoli, A.~Sigalov, A.~Silva, E.~Speth, A.~Stäbler,
  K.-H. Steuer, E.~Strumberger, G.~Tardini, C.~Tichmann, W.~Treutterer,
  C.~Tröster, L.~Urso, E.~Vainonen-Ahlgren, P.~Varela, L.~Vermare, D.~Wagner,
  M.~Wischmeier, E.~Wolfrum, E.~W\"ursching, D.~Yadikin, Q.~Yu, D.~Zasche,
  T.~Zehetbauer, M.~Zilker, and H.~Zohm.
\newblock Plasma wall interaction and its implication in an all tungsten
  divertor tokamak.
\newblock \href{http://dx.doi.org/10.1088/0741-3335/49/12B/S04}{{\em Plasma
  Phys. Control. Fusion}} \textbf{49}, B59 (2007).

\bibitem{Matthews10}
G.F. Matthews, M.~Beurskens, S.~Brezinsek, M.~Groth, E.~Joffrin, A.~Loving,
  M.~Kear, M.-L. Mayoral, R.~Neu, P.~Prior, V.~Riccardo, F.~Rimini, M.~Rubel,
  G.~Sips, E.~Villedieu, P.~de~Vries, M.~L. Watkins, and EFDA-JET contributors.
\newblock \text{JET} \text{ITER}-like wall—overview and experimental
  programme.
\newblock \href{http://dx.doi.org/10.1088/0031-8949/2011/T145/014001}{{\em
  Phys. Scr.}} \textbf{T145}, 014001 (2011).

\bibitem{Bucalossi14}
J.~Bucalossi, M.~Missirlian, P.~Moreau, F.~Samaille, E.~Tsitrone, D.~van
  Houtte, T.~Batal, C.~Bourdelle, M.~Chantant, Y.~Corre, X.~Courtois,
  L.~Delpech, L.~Doceul, D.~Douai, H.~Dougnac, F.~Fa\"isse, C.~Fenzi,
  F.~Ferlay, M.~Firdaouss, L.~Gargiulo, P.~Garin, C.~Gil, A.~Grosman,
  D.~Guilhem, J.~Gunn, C.~Hernandez, D.~Keller, S.~Larroque, F.~Leroux,
  M.~Lipa, P.~Lotte, A.~Martinez, O.~Meyer, F.~Micolon, P.~Mollard, E.~Nardon,
  R.~Nouailletas, A.~Pilia, M.~Richou, S.~Salasca, and J.-M. Trav\'ere.
\newblock The \text{WEST} project: \text{T}esting \text{ITER} divertor high
  heat flux component technology in a steady state tokamak environment.
\newblock \href{http://dx.doi.org/10.1016/j.fusengdes.2014.01.062}{{\em Fusion
  Eng. Des.}} \textbf{89}, 907 (2014).

\bibitem{Post77}
D.E. Post and R.V. Jensen.
\newblock Steady-state radiative cooling rates for low-density,
  high-temperature plasmas.
\newblock \href{http://dx.doi.org/10.1016/0092-640X(77)90026-2}{{\em At. Data
  Nucl. Data Tables}} \textbf{20}, 397 (1977).

\bibitem{Neu97}
R.~Neu, K.~Asmussen, S~Deschka, A.~Thoma, M.~Bessenrodt-Weberpals, R.~Dux,
  W.~Engelhardt, J.C. Fuchs, J.~Gaffert, C.~Garci\'ia-Rosales, A.~Herrmann,
  K.~Krieger, F.~Mast, J.~Roth, V.~Rohde, M.~Weinlich, U.~Wenzel, ASDEX~Upgrade
  Team, and ASDEX NI-Team.
\newblock The tungsten experiment in \text{ASDEX} \text{U}pgrade.
\newblock \href{http://dx.doi.org/10.1016/S0022-3115(97)80121-3}{{\em J. Nucl.
  Mater.}} \textbf{241-243}, 678 (1997).

\bibitem{Neu02}
R.~Neu, R.~Dux, A.~Geier, A.~Kallenbach, R.~Pugno, V.~Rohde, D.~Bolshukhin,
  J.C. Fuchs, O.~Gehre, O.~Gruber, J.~Hobirk, M.~Kaufmann, K.~Krieger, M.~Laux,
  C.~Maggi, H.~Murmann, J.~Neuhauser, F.~Ryter, A.C.C. Sips, A.~St\"abler,
  J.~Stober, W.~Suttrop, H.~Zohm, and ASDEX~Upgrade Team.
\newblock Impurity behaviour in the \text{ASDEX} \text{U}pgrade divertor
  tokamak with large area tungsten walls.
\newblock \href{http://dx.doi.org/10.1088/0741-3335/44/6/313}{{\em Plasma Phys.
  Control. Fusion}} \textbf{44}, 811 (2002).

\bibitem{Dux03}
R.~Dux, C.~Giroud, R.~Neu, A.G. Peeters, J.~Stober, K.-D. Zastrow, Contributors
  to~the EFDA-JET~Workprogramme, and ASDEX~Upgrade Team.
\newblock Accumulation of impurities in advanced scenarios.
\newblock \href{http://dx.doi.org/10.1016/S0022-3115(02)01508-8}{{\em J. Nucl.
  mater.}} \textbf{313-316}, 1150 (2003).

\bibitem{Putterich13}
T.~P\"utterich, R.~Dux, R.~Neu, M.~Bernert, M.N.A. Beurskens, V.~Bobkov,
  S.~Brezinsek, C.~Challis, J.W. Coenen, I.~Coffey, A.~Czarnecka, C.~Giroud,
  P.~Jacquet, E.~Joffrin, A.~Kallenbach, M.~Lehnen, E.~Lerche, E.~de~la Luna,
  S.~Marsen, G.~Matthews, M.-L. Mayoral, R.M. McDermott, A.~Meigs, J.~Mlynar,
  M.~Sertoli, G.~van Rooij, the ASDEX Upgrade~Team, and JET~EFDA Contributors.
\newblock Observations on the \text{W}-transport in the core plasma of
  \text{JET} and \text{ASDEX} \text{U}pgrade.
\newblock \href{http://dx.doi.org/10.1088/0741-3335/55/12/124036}{{\em Plasma
  Phys. Control. Fusion}} \textbf{55}, 124036 (2013).

\bibitem{Angioni14}
C.~Angioni, P.~Mantica, T.~P\"utterich, M.~Valisa, M.~Baruzzo, E.A. Belli,
  P.~Belo, F.J. Casson, C.~Challis, P.~Drewelow, C.~Giroud, N.~Hawkes, T.C.
  Hender, J.~Hobirk, T.~Koskela, L.~Lauro~Taroni, C.F. Maggi, J.~Mlynar,
  T.~Odstrcil, M.L. Reinke, M.~Romanelli, and JET~EFDA Contributors.
\newblock Tungsten transport in \text{JET} \text{H}-mode plasmas in hybrid
  scenario, experimental observations and modelling.
\newblock \href{http://dx.doi.org/10.1088/0029-5515/54/8/083028}{{\em Nucl.
  Fusion}} \textbf{54}, 083028 (2014).

\bibitem{Casson15}
F.J. Casson, C.~Angioni, E.A. Belli, R.~Bilato, P.~Mantica, T.~Odstrcil,
  T.~P\"utterich, M.~Valisa, L.~Garzotti, C.~Giroud, J.~Hobirk, C.F. Maggi,
  J.~Mlynar, M.L. Reinke, JET~EFDA Contributors, and ASDEX~Upgrade Team.
\newblock Theoretical description of heavy impurity transport and its
  application to the modelling of tungsten in \text{JET} and \text{ASDEX}
  upgrade.
\newblock \href{http://dx.doi.org/10.1088/0741-3335/57/1/014031}{{\em Plasma
  Phys. Control. Fusion}} \textbf{57}, 014031 (2015).

\bibitem{Angioni15}
C.~Angioni.
\newblock The impact of poloidal asymmetries on tungsten transport in the core
  of \text{JET} \text{H}-mode plasmas.
\newblock \href{http://dx.doi.org/10.1063/1.4919036}{{\em Phys. Plasmas}}
  \textbf{22}, 055902 (2015).

\bibitem{Angioni17_2}
C.~Angioni, R.~Bilato, F.J. Casson, E.~Fable, P.~Mantica, T.~Odstrcil,
  M.~Valisa, ASDEX~Upgrade Team, and JET Contributors.
\newblock Gyrokinetic study of turbulent convection of heavy impurities in
  tokamak plasmas at comparable ion and electron heat fluxes.
\newblock \href{http://dx.doi.org/10.1088/0029-5515/57/2/022009}{{\em Nucl.
  Fusion}} \textbf{57}, 022009 (2017).

\bibitem{Reinke12}
M.L. Reinke, I.H. Hutchinson, J.E. Rice, N.T. Howard, A.~Bader, S.~Wukitch,
  Y.~Lin, D.C. Pace, A.~Hubbard, J.W. Hughes, and Y.~Podpaly.
\newblock Poloidal variation of high-$z$ impurity density due to hydrogen
  minority ion cyclotron resonance heating on \text{A}lcator
  \text{C}-\text{M}od.
\newblock \href{http://dx.doi.org/10.1088/0741-3335/54/4/045004}{{\em Plasma
  Phys. Control. Fusion}} \textbf{54}, 045004 (2012).

\bibitem{Angioni17}
C.~Angioni, M.~Sertoli, R.~Bilato, V.~Bobkov, A.~Loarte, R.~Ochoukov,
  T.~Odstrcil, T.~P\"utterich, J.~Stober, and the ASDEX Upgrade~Team.
\newblock A comparison of the impact of central \text{ECRH} and central
  \text{ICRH} on the tungsten behaviour in \text{ASDEX} \text{U}pgrade
  \text{H}-mode plasmas.
\newblock \href{http://dx.doi.org/10.1088/1741-4326/aa6453}{{\em Nucl. Fusion}}
  \textbf{57}, 056015 (2017).

\bibitem{Sertoli17}
M.~Sertoli, C.~Angioni, T.~Odstrcil, the ASDEX Upgrade~Team, and the EUROFusion
  MST1~Team.
\newblock Parametric dependencies of the experimental tungsten transport
  coefficients in \text{ICRH} and \text{ECRH} assisted \text{ASDEX}
  \text{U}pgrade \text{H}-modes.
\newblock \href{http://dx.doi.org/10.1063/1.4996412}{{\em Phys. Plasmas}}
  \textbf{24}, 112503 (2017).

\bibitem{Sertoli15}
M.~Sertoli, T.~Odstrcil, C.~Angioni, and the ASDEX Upgrade~Team.
\newblock Interplay between central \text{ECRH} and saturated $(m,n)=(1,1)$
  \text{MHD} activity in mitigating tungsten accumulation at \text{ASDEX}
  \text{U}pgrade.
\newblock \href{http://dx.doi.org/10.1088/0029-5515/55/11/113029}{{\em Nucl.
  Fusion}} \textbf{55}, 113029 (2015).

\bibitem{Sertoli15_2}
M.~Sertoli, R.~Dux, T.~P\"utterich, and the ASDEX Upgrade~Team.
\newblock Modification of impurity transport in the presence of saturated
  $(m,n)=(1,1)$ mhd activity at asdex upgrade.
\newblock \href{http://dx.doi.org/10.1088/0741-3335/57/7/075004}{{\em Plasma
  Phys. Control. Fusion}} \textbf{57}, 075004 (2015).

\bibitem{Gunter99}
S.~G\"unter, A.~Gude, K.~Lackner, M.~Maraschek, S.~Pinches, S.~Sesnic, R.~Wolf,
  and the ASDEX Upgrade~Team.
\newblock The influence of fishbones on the background plasma.
\newblock \href{http://dx.doi.org/10.1088/0029-5515/39/11/304}{{\em Nucl.
  Fusion}} \textbf{39}, 1535 (1999).

\bibitem{Nave03}
M.F.F. Nave, J.~Rapp, T.~Bolzonella, R.~Dux, M.J. Mantsinen, R.~Budny,
  P.~Dumortier, M.~von Hellermann, S.~Jachmich, H.R. Koslowski, G.~Maddison,
  A.~Messiaen, P.~Monier-Garbet, J.~Ongena, M.E. Puiatti, J.~Strachan,
  G.~Telesca, B.~Unterberg, M.~Valisa, P.~de~Vries, and contributors to~the
  JET-EFDA~Workprogramme.
\newblock Role of sawtooth in avoiding impurity accumulation and maintaining
  good confinement in \text{JET} radiative mantle discharges.
\newblock \href{http://dx.doi.org/10.1088/0029-5515/43/10/023}{{\em Nucl.
  Fusion}} \textbf{43}, 1204 (2003).

\bibitem{Hender16}
T.C. Hender, P.~Buratti, F.J. Casson, B.~Alper, Y.F. Baranov, M.~Baruzzo, C.D.
  Challis, F.~Koechl, K.D. Lawson, C.~Marchetto, M.F.F. Nave, T.~P\"utterich,
  S.~Reyes~Cortes, and JET Contributors.
\newblock The role of \text{MHD} in causing impurity peaking in \text{JET}
  hybrid plasmas.
\newblock \href{http://dx.doi.org/10.1088/0029-5515/56/6/066002}{{\em Nucl.
  Fusion}} \textbf{56}, 066002 (2016).

\bibitem{Goniche17}
M.~Goniche, R.J. Dumont, V.~Bobkov, P.~Buratti, S.~Brezinsek, C.~Challis,
  L.~Colas, A.~Czarnecka, P.~Drewelow, N.~Fedorczak, J.~Garcia, C.~Giroud,
  M.~Graham, J.P. Graves, J.~Hobirk, P.~Jacquet, E.~Lerche, P.~Mantica,
  I.~Monakhov, P.~Monier-Garbet, M.F.F. Nave, C.~Noble, I.~Nunes,
  T.~P\"utterich, F.~Rimini, M.~Sertoli, M.~Valisa, D.~Van~Eester, and JET
  Contributors.
\newblock Ion cyclotron resonance heating for tungsten control in various
  \text{JET} \text{H}-mode scenarios.
\newblock \href{http://dx.doi.org/10.1088/1361-6587/aa60d2}{{\em Plasma Phys.
  Control. Fusion}} \textbf{59}, 055001 (2017).

\bibitem{Sertoli11}
M.~Sertoli, C.~Angioni, R.~Dux, R.~Neu, T.~P\"utterich, V.~Igochine, and the
  ASDEX Upgrade~Team.
\newblock Local effects of \text{ECRH} on argon transport in \text{L}-mode
  discharges at \text{ASDEX} \text{U}pgrade.
\newblock \href{http://dx.doi.org/10.1088/0741-3335/53/3/035024}{{\em Plasma
  Phys. Control. Fusion}} \textbf{53}, 035024 (2011).

\bibitem{Bourdelle16}
C.~Bourdelle, J.~Citrin, B.~Baiocchi, A.~Casati, P.~Cottier, X.~Garbet,
  F.~Imbeaux, and JET Contributors.
\newblock Core turbulent transport in tokamak plasmas: bridging theory and
  experiment with \text{Q}ua\text{L}i\text{K}iz.
\newblock \href{http://dx.doi.org/10.1088/0741-3335/58/1/014036}{{\em Plasma
  Phys. Control. Fusion}} \textbf{58}, 014036 (2016).

\bibitem{Citrin17}
J.~Citrin, C.~Bourdelle, F.J. Casson, C.~Angioni, N.~Bananomi, Y.~Camenen,
  X.~Garbet, L.~Garzotti, T.~G\"orler, O.~G\"urcan, F.~Koechl, F.~Imbeaux,
  O.~Linder, K.~van~de Plassche, P.~Strand, G.~Szepesi, and JET Contributors.
\newblock Tractable flux-driven temperature, density, and rotation profile
  evolution with the quasilinear gyrokinetic transport model
  \text{Q}ua\text{L}i\text{K}iz.
\newblock \href{http://dx.doi.org/10.1088/1361-6587/aa8aeb}{{\em Plasma Phys.
  Control. Fusion}} \textbf{59}, 124005 (2017).

\bibitem{Cenacchi88}
G.~Cenacchi and A.~Taroni.
\newblock {\em \text{JETTO}: \text{A} \text{F}ree-\text{B}oundary \text{P}lasma
  \text{T}ransport \text{C}ode}.
\newblock Rapporto ENEA RT/TIB/88/5, 1988.

\bibitem{Romanelli14}
M.~Romanelli, G.~Corrigan, V.~Parail, S.~Wiesen, R.~Ambrosino, P.~da~Silva
  Aresta~Belo, L.~Garzotti, D.~Harting, F.~K\"ochl, T.~Koskela,
  L.~Lauro-Taroni, C.~Marchetto, M.~Mattei, E.~Militello-Asp, M.F.F. Nave,
  S.~Pamela, A.~Salmi, P.~Strand, G.~Szepesi, and EFDA-JET Contributors.
\newblock \text{JINTRAC}: \text{A} \text{S}ystem of \text{C}odes for
  \text{I}ntegrated \text{S}imulation of \text{T}okamak \text{S}cenarios.
\newblock \href{http://dx.doi.org/10.1585/pfr.9.3403023}{{\em Plasma and Fusion
  Research}} \textbf{9}, 3403023 (2014).

\bibitem{Breton18}
S.~Breton, F.J. Casson, C.~Bourdelle, J.~Citrin, Y.~Baranov, J.~Challis,
  C.~Garcia, G.~Corrigan, L.~Garzotti, S.~Henderson, F.~Koechl, M.~OMullane,
  T.~P\"utterich, M.~Sertoli, M.~Valisa, and JET contributors.
\newblock First principle integrated modeling of multi-channel tranport
  including \text{T}ungsten in \text{JET}.
\newblock \href{http://dx.doi.org/10.1088/1741-4326/aac780}{{\em Nucl. Fusion}}
  \textbf{58}, 096003 (2018).

\bibitem{Breton17}
S.~Breton, F.J. Casson, C.~Bourdelle, Y.~Camenen, J.~Citrin, Y.~Baranov,
  C.~Challis, J.~Garcia, G.~Corrigan, L.~Garzotti, S.~Henderson, F.~Koechl,
  M.~O'Mullane, T.~P\"utterich, M.~Sertoli, M.~Valisa, and JET contributors.
\newblock Integrated modelling of multi-channel transport including
  \text{T}ungsten in \text{JET}.
\newblock Paper presented at the 44th EPS Conference on Plasma Physics,
  Belfast, 26-30 June 2017, O4.124.

\bibitem{Dux99}
R.~Dux, A.G. Peeters, A.~Gude, A.~Kallenbach, and ASDEX~Upgrade Team.
\newblock $z$ dependence of the core impurity transport in \text{ASDEX}
  \text{U}pgrade \text{H} mode discharges.
\newblock \href{http://dx.doi.org/10.1088/0029-5515/39/11/302}{{\em Nucl.
  Fusion}} \textbf{39}, 1509 (1999).

\bibitem{Stober07}
J.~Stober, A.~Gude, F.~Leuterer, A.~Manini, R.~Neu, T.~P\"utterich, A.C.C.
  Sips, D.~Wagner, H.~Zohm, and the ASDEX Upgrade~team.
\newblock First experiments with the extended \text{ECRH} system on
  \text{ASDEX} \text{U}pgrade.
\newblock Paper presented at the 34th EPS Conference on Plasma Physics, Warsaw,
  2-6 July 2007, P5.138.

\bibitem{Wagner11}
D.~Wagner, J.~Stober, F.~Leuterer, F.~Monaco, M.~M\"unich, D.~Schmid-Lorch,
  H.~Sch\"utz, H.~Zohm, M.~Thumm, T.~Scherer, A.~Meier, G.~Gantenbein, Flamm.
  J., W.~Kasparek, H.~H\"ohnle, C.~Lechte, A.G. Litvak, G.G. Denisov,
  A.~Chirkov, L.G. Popov, V.O. Nichiporenko, V.E. Myasnikov, E.M. Tai, E.A.
  Solyanova, and S.A. Malygin.
\newblock Recent \text{U}pgrades and \text{E}xtensions of the \text{ASDEX}
  \text{U}pgrade \text{ECRH} \text{S}ystem.
\newblock \href{http://dx.doi.org/10.1007/s10762-010-9703-3}{{\em J. Infrared
  Millim. Terahertz Waves}} \textbf{32}, 274 (2011).

\bibitem{Gude10}
A.~Gude, M.~Maraschek, C.~Angioni, J.~Stober, and the ASDEX Upgrade~Team.
\newblock Hollow central radiation profiles and inverse sawtooth-like crashes
  in \text{ASDEX} \text{U}pgrade plasmas with central wave heating.
\newblock Paper presented at the 37th EPS Conference on Plasma Physics, Dublin,
  21-25 June 2010, P4.124.

\bibitem{Breton18_2}
S.~Breton, F.J. Casson, C.~Bourdelle, C.~Angioni, E.~Belli, Y.~Camenen,
  J.~Citrin, X.~Garbet, Y.~Sarazin, M.~Sertoli, and JET contributors.
\newblock High \text{Z} neoclassical transport: application and limitation of
  analytical formulae for modelling \text{JET} experimental parameters.
\newblock \href{http://dx.doi.org/10.1063/1.5019275}{{\em Phys. Plasmas}}
  \textbf{25}, 012303 (2018).

\bibitem{Ho18}
A.~Ho, J.~Citrin, F.~Auriemma, C.~Bourdelle, F.J. Casson, H.-T. Kim, P.~Manas,
  G.~Szepesi, H.~Weisen, and JET Contributors.
\newblock Turbulent transport model validation against a \text{JET} plasma via
  integrated modelling enhanced by \text{G}aussian process regression.
\newblock {\em Submitted to Nucl. Fusion.}

\bibitem{Chilenski15}
M.A. Chilenski, M.~Greenwald, Y.~Marzouk, N.T. Howard, A.E. White, J.E. Rice,
  and J.R. Walk.
\newblock Improved profile fitting and quantification of uncertainty in
  experimental measurements of impurity transport coefficients using
  \text{G}aussian process regression.
\newblock \href{http://dx.doi.org/10.1088/0029-5515/55/2/023012}{{\em Nucl.
  Fusion}} \textbf{55}, 023012 (2015).

\bibitem{Jardin15}
S.C. Jardin, N.~Ferraro, and I.~Krebs.
\newblock \text{S}elf-\text{O}rganized \text{S}tationary \text{S}tates of
  \text{T}okamaks.
\newblock \href{http://dx.doi.org/10.1103/PhysRevLett.115.215001}{{\em Phys.
  Rev. Lett.}} \textbf{115}, 215001 (2015).

\bibitem{Krebs17}
I.~Krebs, S.C. Jardin, S.~G\"unter, K.~Lackner, M.~Hoelzl, E.~Strumberger, and
  N~Ferraro.
\newblock Magnetic flux pumping in 3\text{D} nonlinear magnetohydrodynamic
  simulations.
\newblock \href{http://dx.doi.org/10.1063/1.4990704}{{\em Phys. Plasmas}}
  \textbf{24}, 102511 (2017).

\bibitem{Hawryluk80}
R.J. Hawryluk.
\newblock An \text{E}mpirical \text{A}pproach to \text{T}okamak
  \text{T}ransport.
\newblock In B.~Coppi, G.G. Leotta, D.~Pfirsch, R.~Pozzoli, and E.~Sindoni,
  {\em Physics of Plasmas Close to Thermonuclear Conditions}. Pergamon Press,
  1981.

\bibitem{Pankin04}
A.~Pankin, D.~McCune, R.~Andre, G.~Bateman, and A.~Kritz.
\newblock The tokamak \text{M}onte \text{C}arlo fast ion module \text{NUBEAM}
  in the \text{N}ational \text{T}ransport \text{C}ode \text{C}ollaboration
  library.
\newblock \href{http://dx.doi.org/10.1016/j.cpc.2003.11.002}{{\em Comput. Phys.
  Commun.}} \textbf{159}, 157 (2004).

\bibitem{Houlberg97}
W.A. Houlberg, K.C. Shaing, S.P. Hirshman, and M.C. Zarnstorff.
\newblock Bootstrap current and neoclassical transport in tokamaks of arbitrary
  collisionality and aspect ratio.
\newblock \href{http://dx.doi.org/10.1063/1.872465}{{\em Phys. Plasmas}}
  \textbf{4}, 3230 (1997).

\bibitem{Koechl17}
F.~Koechl, A.~Loarte, V.~Parail, P.~Belo, M.~Brix, G.~Corrigan, D.~Harting,
  T.~Koskela, A.S. Kukushkin, A.R. Polevoi, M.~Romanelli, G.~Saibene,
  R.~Sartori, T.~Eich, and JET Contributors.
\newblock Modelling of transitions between \text{L}- and \text{H}-mode in
  \text{JET} high plasma current plasmas and application to \text{ITER}
  scenarios including tungsten behaviour.
\newblock \href{http://dx.doi.org/10.1088/1741-4326/aa7539}{{\em Nucl. Fusion}}
  \textbf{57}, 086023 (2017).

\bibitem{Lauro-Taroni94}
L.~Lauro-Taroni, B.~Alper, R.~Giannella, K.~Lawson, F.~Marcus, M.~Mattioli,
  P.~Smeulders, and M.~von Hellermann.
\newblock {\em Proc. 21st EPS on Plasma Physics} (1994).

\bibitem{Conway08}
G.D. Conway.
\newblock Turbulence measurements in fusion plasmas.
\newblock \href{http://dx.doi.org/10.1088/0741-3335/50/12/124026}{{\em Plasma
  Phys. Control. Fusion}} \textbf{50}, 124026 (2008).

\bibitem{Happel15}
T.~Happel, A.~Ba\~n\'on Navarro, G.D. Conway, C.~Angioni, M.~Bernert, M.~Dunne,
  E.~Fable, B.~Geiger, T.~G\"orler, F.~Jenko, R.M. McDermott, F.~Ryter,
  U.~Stroth, and the ASDEX Upgrade~Team.
\newblock Core turbulence behavior moving from ion-temperature-gradient regime
  towards trapped-electron-mode regime in the \text{ASDEX} \text{U}pgrade
  tokamak and comparison with gyrokinetic simulation.
\newblock \href{http://dx.doi.org/10.1063/1.4914153}{{\em Phys. Plasmas}}
  \textbf{22}, 032503 (2015).

\bibitem{Maggi14}
C.F. Maggi, E.~Delabie, T.M. Biewer, M.~Groth, N.C. Hawkes, M.~Lehnen, E.~de~la
  Luna, K.~McCormick, C.~Reux, F.~Rimini, E.R. Solano, Y.~Andrew, C.~Bourdelle,
  V.~Bobkov, M.~Brix, G.~Calabro, A.~Czarnecka, J.~Flanagan, E.~Lerche,
  S.~Marsen, I.~Nunes, D.~Van~Eester, M.F. Stamp, and JET~EFDA Contributors.
\newblock L-\text{H} power threshold studies in \text{JET} with
  \text{B}e/\text{W} and \text{C} wall.
\newblock \href{http://dx.doi.org/10.1088/0029-5515/54/2/023007}{{\em Nucl.
  Fusion}} \textbf{54}, 023007 (2014).

\bibitem{Meyer14}
H.~Meyer, E.~Delabie, C.F. Maggi, C.~Bourdelle, P.~Drewelow, I.~Carvalho,
  P.~Lang, F.~Rimini, and JET~EFDA Contributors.
\newblock The role of divertor and \text{SOL} physics for access to
  \text{H}-mode on \text{JET}.
\newblock Paper presented at the 41st EPS Conference on Plasma Physics, Berlin,
  23-27 June 2014, P1.013.

\bibitem{Garbet04}
X.~Garbet, P.~Mantica, C.~Angioni, E.~Asp, Y.~Baranov, C.~Bourdelle, R.~Budny,
  F.~Crisanti, G.~Cordey, L.~Garzotti, N.~Kirneva, D.~Hogeweij, T.~Hoang,
  F.~Imbeaux, E.~Joffrin, X.~Litaudon, A.~Manini, D.C. McDonald, H.~Nordman,
  V.~Prail, A.~Peeter, F.~Ryter, C.~Sozzi, M.~Valovic, T.~Tala, A.~Thyagaraja,
  I.~Voitsekhovitch, J.~Weiland, H.~Weisen, A.~Zabolotsky, and the JET
  EFDA~Contributors.
\newblock Physics of transport in tokamaks.
\newblock \href{http://dx.doi.org/10.1088/0741-3335/46/12B/045}{{\em Plasma
  Phys. Control. Fusion}} \textbf{46}, B557 (2004).

\bibitem{Romanelli04}
M.~Romanelli, C.~Bourdelle, and W.~Dorland.
\newblock Effects of high density peaking and high collisionality on the
  stabilization of the electrostatic turbulence in the \text{F}rascati
  \text{T}okamak \text{U}pgrade.
\newblock \href{http://dx.doi.org/10.1063/1.1766031}{{\em Phys. Plasmas}}
  \textbf{11}, 3845 (2004).

\bibitem{Guo93}
S.C. Guo and F.~Romanelli.
\newblock The linear threshold of the ion-temperature-gradient-driven mode.
\newblock \href{http://dx.doi.org/10.1063/1.860537}{{\em Phys. Fluids B}}
  \textbf{5}, 520 (1993).

\bibitem{Casati08}
A.~Casati, C.~Bourdelle, X.~Garbet, and F.~Imbeaux.
\newblock Temperature ratio dependence of ion temperature gradient and trapped
  electron mode instability thresholds.
\newblock \href{http://dx.doi.org/10.1063/1.2906223}{{\em Phys. Plasmas}}
  \textbf{15}, 042310 (2008).

\bibitem{Jenko01}
F.~Jenko, W.~Dorland, and G.W. Hammett.
\newblock Critical gradient formula for toroidal electron temperature gradient
  modes.
\newblock \href{http://dx.doi.org/10.1063/1.1391261}{{\em Phys. Plasmas}}
  \textbf{8}, 4096 (2001).

\bibitem{Rhodes11}
T.L. Rhodes, C.~Holland, S.P. Smith, A.E. White, K.H. Burrell, J.~Candy, J.C.
  DeBoo, E.J. Doyle, J.C. Hillesheim, J.E. Kinsey, G.R. McKee, D.~Mikkelsen,
  W.A. Peebles, C.C. Petty, R.~Prater, S.~Parker, Y.~Chen, L.~Schmitz, G.M.
  Staebler, R.E. Waltz, G.~Wang, Z.~Yan, and L.~Zeng.
\newblock L-mode validation studies of gyrokinetic turbulence simulations via
  multiscale and mutifield turbulence measurements on the \text{DIII}-\text{D}
  tokamak.
\newblock \href{http://dx.doi.org/10.1088/0029-5515/51/6/063022}{{\em Nucl.
  Fusion}} \textbf{51}, 063022 (2011).

\bibitem{Freethy18}
S.J. Freethy, T.~G\"orler, A.J. Creely, G.D. Conway, S.S. Denk, T.~Happel,
  C.~Koenen, P.~Hennequin, A.E. White, and ASDEX~Upgrade Team.
\newblock Validation of gyrokinetic simulations with measurements of electron
  temperature fluctuations and density-temperature phase angles on \text{ASDEX}
  \text{U}pgrade.
\newblock \href{http://dx.doi.org/10.1063/1.5018930}{{\em Phys. Plasmas}}
  \textbf{25}, 055903 (2018).

\bibitem{Sommer12}
F.~Sommer, J.~Stober, C.~Angioni, M.~Bernert, A.~Burckhart, V.~Bobkov,
  R.~Fischer, C.~Fuchs, R.M. McDermott, W.~Suttrop, E.~Viezzer, and the ASDEX
  Upgrade~Team.
\newblock H-mode characterization for dominant \text{ECRH} and comparison to
  dominant \text{NBI} and \text{ICRF} heating at \text{ASDEX} \text{U}pgrade.
\newblock \href{http://dx.doi.org/10.1088/0029-5515/52/11/114018}{{\em Nucl.
  Fusion}} \textbf{52}, 114018 (2012).

\bibitem{Sommer15}
F.~Sommer, J.~Stober, C.~Angioni, E.~Fable, M.~Bernert, A.~Burckhardt,
  V.~Bobkov, R.~Fischer, C.~Fuchs, R.M. McDermott, W.~Suttrop, E.~Viezzer, and
  the ASDEX Upgrade~Team.
\newblock Transport properties of \text{H}-mode plasmas with dominant electron
  heating in comparison to dominant ion heating at \text{ASDEX} \text{U}pgrade.
\newblock \href{http://dx.doi.org/10.1088/0029-5515/55/3/033006}{{\em Nucl.
  Fusion}} \textbf{55}, 033006 (2015).

\bibitem{Bourdelle18}
C.~Bourdelle, Y.~Camenen, J.~Citrin, M.~Marin, F.J. Casson, F.~Koechl,
  M.~Maslov, and the JET~Contributors.
\newblock Fast \text{H} isotope and impurity mixing in ion-temperature-gradient
  turbulence.
\newblock \href{http://dx.doi.org/10.1088/1741-4326/aacd57}{{\em Nucl. Fusion}}
  \textbf{58}, 076028 (2018).

\bibitem{Romanelli07}
M.~Romanelli, G.~Regnoli, and C.~Bourdelle.
\newblock Numerical study of linear dissipative drift electrostatic modes in
  tokamaks.
\newblock \href{http://dx.doi.org/10.1063/1.2755981}{{\em Phys. Plasmas}}
  \textbf{14}, 082305 (2007).

\bibitem{Fable10}
E.~Fable, C.~Angioni, and O.~Sauter.
\newblock The role of ion and electron electrostatic turbulence in
  characterizing stationary particle transport in the core of tokamak plasmas.
\newblock \href{http://dx.doi.org/10.1088/0741-3335/52/1/015007}{{\em Plasma
  Phys. Control. Fusion}} \textbf{52}, 015007 (2010).

\bibitem{Belli08}
E.A. Belli and J.~Candy.
\newblock Kinetic calculation of neoclassical transport including
  self-consistent electron and impurity dynamics.
\newblock \href{http://dx.doi.org/10.1088/0741-3335/50/9/095010}{{\em Plasma
  Phys. Control. Fusion}} \textbf{50}, 095010 (2008).

\bibitem{Belli12}
E.A. Belli and J.~Candy.
\newblock Full linearized \text{F}okker–\text{P}lanck collisions in
  neoclassical transport simulations.
\newblock \href{http://dx.doi.org/10.1088/0741-3335/54/1/015015}{{\em Plasma
  Phys. Control. Fusion}} \textbf{54}, 015015 (2012).

\bibitem{Garbet16}
X.~Garbet, J.H. Ahn, S.~Breton, P.~Donnel, D.~Esteve, R.~Guirlet, H.~L\"utjens,
  T.~Nicolas, Y.~Sarazin, C.~Bourdelle, O.~F\'evrier, G.~Dif-Pradalier,
  P.~Ghendrih, V.~Grandgirard, G.~Latu, J.F. Luciani, P.~Maget, A.~Marx, and
  A.~Smolyakov.
\newblock \textit{Synergetic effects of collisions, turbulence and sawtooth
  crashes on impurity transport}.
\newblock
  \href{hal.archives-ouvertes.fr/hal-01367373}{hal.archives-ouvertes.fr/hal-01367373}
  (2016).

\bibitem{Ahn16}
J.H. Ahn, X.~Garbet, H.~L\"utjens, and R.~Guirlet.
\newblock \textit{Dynamics of heavy impurities in non-linear \text{MHD}S
  simulations of sawtoothing tokamak plasmas}.
\newblock
  \href{hal.archives-ouvertes.fr/hal-01340291}{hal.archives-ouvertes.fr/hal-01340291}
  (2016).

\bibitem{Nicolas12}
T.~Nicolas, R.~Sabot, X.~Garbet, H.~L\"utjens, J.F. Luciani,
  Z.~Guimaraes-Filho, J.~Decker, and A.~Merle.
\newblock Non-linear magnetohydrodynamic simulations of density evolution in
  \text{T}ore \text{S}upra sawtoothing plasmas.
\newblock \href{http://dx.doi.org/10.1063/1.4766893}{{\em Phys. Plasmas}}
  \textbf{19}, 112305 (2012).

\bibitem{Nicolas14}
T.~Nicolas, H.~L\"utjens, J.F. Luciani, X.~Garbet, and R.~Sabot.
\newblock Impurity behavior during sawtooth activity in tokamak plasmas.
\newblock \href{http://dx.doi.org/10.1063/1.4861859}{{\em Phys. Plasmas}}
  \textbf{21}, 012507 (2014).

\bibitem{Pereverzev02}
G.V. Pereverzev and P.N. Yushmanov.
\newblock {\em \text{ASTRA} - \text{A}utomated \text{S}ystem for
  \text{TR}ansport \text{A}nalysis}.
\newblock Max-Planck-Institut F\"ur Plasmaphysik, IPP-Report, IPP 5/98,
  Februray 2002.

\bibitem{Fable13}
E.~Fable, C.~Angioni, F.J. Casson, D.~Told, A.A. Ivanov, F.~Jenko, R.M.
  McDermott, S.Yu. Medvedev, G.V. Pereverzev, F.~Ryter, W.~Treutterer,
  E.~Viezzer, and the ASDEX Upgrade~Team.
\newblock Novel free-boundary equilibrium and transport solver with
  theory-based models and its validation against \text{ASDEX} \text{U}pgrade
  current ramp scenarios.
\newblock \href{http://dx.doi.org/10.1088/0741-3335/55/12/124028}{{\em Plasma
  Phys. Control. Fusion}} \textbf{55}, 124028 (2013).

\end{thebibliography}

	\end{document}